\documentclass[longauth]{aa}

\usepackage{graphicx}
\usepackage{txfonts}

\usepackage{multirow}

\usepackage[breaklinks=true,colorlinks=true,citecolor=blue,linkcolor=blue,urlcolor=blue]{hyperref}
\usepackage{tablefootnote}
\usepackage{natbib}
\usepackage{amsmath} 
\usepackage{amssymb}
\usepackage{mathtools}
\usepackage{amsmath} 
\usepackage{microtype}
\usepackage[dvipsnames]{xcolor}
\usepackage{array}
\usepackage{makecell}
\usepackage{multirow}
\usepackage{siunitx}
\usepackage{tabularx}
\usepackage{caption}
\usepackage{float}
\usepackage[caption=false]{subfig}
\usepackage{verbatim}

\newcommand{\fmp}{\mathrm{FMP}}
\newcommand{\MOKA}{\ensuremath{\mathrm{MOKA^{3D}}}}

\usepackage{hyperref}

\begin{document} 

\title{Cosmic Duets }
\subtitle{I. High-spatial resolution spectroscopy of dual and lensed active galactic nuclei with MUSE-NFM}

\titlerunning{Cosmic Duets I}
\authorrunning{Scialpi et al.}

\author{
M.~Scialpi\inst{\ref{iUNITN},\ref{iUNIFI},\ref{Inaf-oaa}}
\and F.~Mannucci\inst{\ref{Inaf-oaa}}
\and Q.~D'Amato\inst{\ref{Inaf-oaa}}
\and C.~Marconcini\inst{\ref{iUNIFI},\ref{Inaf-oaa}}
\and G.~Cresci\inst{\ref{Inaf-oaa}}
\and A.~Marconi\inst{\ref{iUNIFI},\ref{Inaf-oaa}}
\and L.~Ulivi\inst{\ref{iUNITN},\ref{iUNIFI},\ref{Madrid}}
\and M.~Fumagalli \inst{\ref{bicocca},\ref{inaf_trieste}}
\and P.~Rosati\inst{\ref{iUNIFE},\ref{OASBO}}
\and G.~Tozzi\inst{\ref{MPE}}
\and M.V.~Zanchettin\inst{\ref{Inaf-oaa}}
\and L.~Battistini \inst{\ref{roma3}}
\and E.~Bertola\inst{\ref{Inaf-oaa}}
\and C.~Bracci\inst{\ref{iUNIFI},\ref{Inaf-oaa}}
\and S.~Carniani\inst{\ref{Pisa}}
\and E.~Cataldi\inst{\ref{iUNIFI},\ref{Inaf-oaa}}
\and M.~Ceci\inst{\ref{iUNIFI},\ref{Inaf-oaa}}
\and A.~Chakraborty \inst{\ref{Inaf-oaa}, \ref{india}}
\and C.~Cicone\inst{\ref{Oslo}}
\and A.~Ciurlo\inst{\ref{UCLA}}
\and A.~De~Rosa \inst{\ref{Inaf-roma}}
\and G.~Di~Rosa \inst{\ref{iUNIFE}}
\and A.~Feltre\inst{\ref{Inaf-oaa}}
\and M.~Ginolfi\inst{\ref{iUNIFI},\ref{Inaf-oaa}}
\and I.~Lamperti\inst{\ref{iUNIFI}, \ref{Inaf-oaa}}
\and E.~Lusso\inst{\ref{iUNIFI},\ref{Inaf-oaa}}
\and B.~Moreschini\inst{\ref{iUNIFI},\ref{Inaf-oaa}}
\and E.~Nardini\inst{\ref{Inaf-oaa}}
\and M.~Parvatikar \inst{\ref{torvergata}, \ref{Inaf-roma}}
\and M.~Perna\inst{\ref{Madrid}}
\and K.~Rubinur\inst{\ref{Oslo}}
\and P.~Severgnini\inst{\ref{Inaf-brera}}
\and J.~Singh\inst{\ref{Inaf-brera}}
\and C.~Spingola\inst{\ref{Inaf-bologna}}
\and G.~Venturi\inst{\ref{Pisa}}
\and C.~Vignali\inst{\ref{iUNIBO},\ref{OASBO}}
\and M.~Volonteri\inst{\ref{Paris}}
}

\institute{
University of Trento, Via Sommarive 14, I-38123 Trento, Italy
\label{iUNITN}
\and 
Università di Firenze, Dipartimento di Fisica e Astronomia, via G. Sansone 1, 50019 Sesto F.no, Firenze, Italy\\
\email{martina.scialpi@unifi.it}
\label{iUNIFI}
\and 
INAF – Osservatorio Astrofisico di Arcetri, Largo E. Fermi 5, 50125 Firenze, Italy
\label{Inaf-oaa}
\and
Centro de Astrobiología (CAB), CSIC--INTA, Ctra. de Ajalvir Km 4, 28850 Torrejón de Ardoz, Madrid, Spain\label{Madrid}
\and
Universit\`a degli Studi di Milano-Bicocca, Piazza della Scienza 3, 20126 Milano, Italy \label{bicocca}
\and  
INAF - Osservatorio Astronomico di Trieste, via G.B. Tiepolo 11, I-34143 Trieste, Italy\label{inaf_trieste}
\and
Dipartimento di Fisica e Scienze della Terra, Università di Ferrara, Via G. Saragat 21, 44122 Ferrara, Italy
\label{iUNIFE}
\and 
INAF -- Osservatorio di Astrofisica e Scienza dello Spazio di Bologna, Via Gobetti 93/3, I-40129 Bologna, Italy\label{OASBO}
\and
Max Planck Institute for extraterrestrial Physics, Giessenbachstraße 1, 85748 Garching, Germany \label{MPE}
\and
Dipartimento di Matematica e Fisica, Università degli Studi Roma Tre, via della Vasca Navale 84, 00146 Roma\label{roma3} 
\and
Scuola Normale Superiore, Piazza dei Cavalieri 7, 56126 Pisa, Italy
\label{Pisa}
\and
Indian Institute Of Astrophysics, 100 Feet Rd, Santhosapuram, 2nd Block, Koramangala, Bengaluru, Karnataka 560034 \label{india}
\and
Institute of Theoretical Astrophysics, University of Oslo, P.O Box 1029, Blindern, 0315 Oslo, Norway
\label{Oslo}
\and 
Department of Physics and Astronomy, University of California Los Angeles, 430 Portola Plaza, Los Angeles, CA 90095, USA
\label{UCLA}
\and
INAF - Istituto di Astrofisica e Planetologia Spaziali, Via Fosso del Cavaliere 100, 00133 Rome, Italy
\label{Inaf-roma}
\and 
Dipartimento di Fisica, Università di Roma Tor Vergata, Via della Ricerca Scientifica, I-00133, Roma, Italy \label{torvergata}
\and 
INAF – Osservatorio Astronomico di Brera, via Brera 28, 20121 Milano, Italy
\label{Inaf-brera}
\and
INAF -- Istituto di Radioastronomia, Via Piero Gobetti 101, 40129 Bologna, Italy
\label{Inaf-bologna}
\and 
Physics and Astronomy Department ``Augusto Righi'', Università di Bologna, Via Gobetti 93/2, 40129 Bologna, Italy
\label{iUNIBO}
\and
Institut d'Astrophysique de Paris, 98bis Bd Arago, 75014 Paris, France\label{Paris}
}

   \date{}

\abstract{
We present the first-year results of the MUSE Large Program ``Cosmic Duets,'' whose goal is to obtain adaptive-optics assisted MUSE observations with a typical angular resolution of $0.1''-0.2''$ in order to provide integral-field spectroscopy of sub-arcsec separation dual and lensed active galactic nucleus (AGN) candidates. These observations reveal previously unexplored properties of dual and lensed systems that are key to understanding galaxy evolution, supermassive black hole mergers, and strong-lensing modeling.

Targets were efficiently selected using the \textit{Gaia} multipeak (GMP) technique, which identifies pairs of point-like sources with angular separations below $0.8''$ in the \textit{Gaia} catalog. MUSE spatially resolved spectroscopy provides accurate redshifts, ionization diagnostics, and identification of absorption systems along the line of sight.

We report results for 30 GMP-selected targets at $z = 0.5$–3.5. All systems show at least two spatially resolved components. Nineteen objects are confirmed as AGN multiplets, including six dual AGN, ten doubly lensed quasars, and three quadruply lensed systems, while the remaining 11 correspond to chance alignments with foreground stars. Among all the spectroscopically confirmed dual AGN in the literature, 27 pairs have projected separations below 7 kpc in this redshift regime, and our sample accounts for 22\% of the total.
We studied dual and lensed AGN distributions as a function of redshift, magnitude, and projected separation while accounting for selection effects, and we find that bright systems ($J \lesssim 16.5$) are dominated by lensed quasars, whereas the relative fraction of dual AGN increases at fainter magnitudes.

This first-year sample demonstrates the high efficiency of GMP preselection combined with MUSE spectroscopy for identifying sub-arcsec dual and lensed AGN. The full program, targeting $\sim$150 systems, will enable statistical studies of dual AGN incidence as a function of redshift, separation, and luminosity and allow for precise characterization of the mass composition and distribution in strong lensing galaxies.}


   \maketitle

\section{Introduction}
According to the $\Lambda$ Cold Dark Matter ($\Lambda$CDM) cosmological paradigm, structure forms hierarchically, with smaller systems such as galaxies merging to build more massive ones \citep{White78,White91}. The most massive galaxies host a central supermassive black hole (SMBH), which grows through gas accretion and mergers and ultimately regulates star formation and shapes the structure of the host galaxy through feedback processes \citep{dimatteo05, Croton06}.

During galaxy mergers, the SMBHs involved undergo orbital decay via dynamical friction and can form bound pairs \citep{Begelman80,volonteri03,Colpi14}. When the two SMBHs are actively accreting material, the system becomes a dual active galactic nucleus \citep[AGN;][]{koss12, Rosas-Guevara2019, Volonteri22}, with projected separations from tens of kiloparsecs (kpc) down to sub-kpc scales \citep{derosa20, Bigmac25}. Dual AGN represent the direct precursors of gravitationally bound binary SMBHs at parsec separations \citep[e.g.,][]{Kelley17, Volonteri22}, which may eventually coalesce and emit gravitational waves (GWs). The GW signals from SMBH binaries are expected to be detectable with pulsar timing arrays (\citealt{Arzoumanian18, Agazie2023a, EPTACollaboration2023}) and, at higher frequencies, with the upcoming Laser Interferometer Space Antenna (LISA; \citealt{AmaroSeoane2023,Colpi2024arxiv}).

Dual AGN therefore offer a unique laboratory to investigate gas accretion and feedback in merging galaxies and the connection between black hole growth and hierarchical structure formation. Characterizing their observed separations, accretion rates, BH masses, and host-galaxy properties provide key constraints for models of SMBH coalescence and for the prediction of the GW detection rate of future missions such as LISA, which will be sensitive to merging SMBHs with masses of $10^{4-8}\,M_\odot$ up to $z\sim20$ \citep[e.g.,][]{AmaroSeoane2023}. Quantifying the abundance and demographics of dual AGN is therefore essential to linking galaxy evolution with the emerging GW landscape \citep{ Sedda23,Perna2025dual}.

It is important to clarify the terminology used in this work. In the literature, the term "dual AGN" generally refers to systems hosting two actively accreting SMBHs over a wide range of projected separations, from $\sim 100$~kpc down to sub-kpc scales \citep[e.g.,][]{Liu11_sdsspair, derosa20, Bigmac25}. 
The Cosmic Duets systems targeted here represent a specific subset of close dual AGN selected to probe the small projected separation regime ($<7$ kpc), where the merging galaxies are expected to be in an advanced stage of interaction and the two SMBHs reside within a common merger remnant. While wider dual AGN systems trace interacting galaxies that are still distinct, Cosmic Duets is focused on the regime most directly connected to the late stages of SMBH orbital decay, merger-driven accretion, and the onset of feedback within a single galactic potential.

The direct identification of dual AGN at approximately kpc separations remains challenging due to their intrinsic rarity, the high angular resolution required to resolve two active nuclei, and the potential increase in obscuration associated with late-stage mergers \citep{Koss2018, derosa20}. Simulations and observations indicate that column densities and obscuring material tend to increase as mergers progress \citep[e.g.,][]{Capelo17, Blecha18, Chen23b, Ricci17}. Several discoveries have in fact been serendipitous \citep[e.g.,][]{Junkkarinen01, Glikman23}, including cases revealed with spatially resolved Integral Field Unit (IFU) spectroscopy using JWST \citep{Ubler2024dual, Matsuoka2024, Perna2025dual} and candidates identified from complex broad-line structures in spatially integrated spectra \citep{Maiolino2024dual}, highlighting the difficulty of systematically uncovering such systems.

Initial targeted searches relied on indirect or resource-intensive techniques, such as the selection of double-peaked emission lines \citep{Zhou2004, Comerford2009, Rubinur19}. Other systematic approaches have targeted dual AGN through spectroscopic galaxy pairs \citep{Liu11_sdsspair,Jing25_desidual}, radio-selected samples \citep{Fu15_a,Fu15_b, Fu18_sdss}, mid-infrared preselection \citep{Satyapal17, Pfeifle19_triple}, quasar clustering studies \citep{Hennawi06, Hennawi10}, probabilistic mid-IR pairs \citep{Barrows23}, and serendipitous discoveries in lens surveys \citep{Inada08,Inada12}.
However, many of these surveys primarily probe systems at projected separations of several tens of kiloparsec, corresponding to earlier merger stages or physically associated pairs rather than the compact ($\sim$ few kpc) regime where both SMBHs are embedded within a common merger remnant. 
While a subset of studies (e.g., \citealt{Fu15_a, Fu18_sdss, Satyapal17, Pfeifle19_triple}) have focused on AGN pairs with projected separations $<10$ kpc, the majority of candidates identified across these surveys exhibit projected separations on scales of tens of kiloparsecs, largely driven by large low-redshift samples such as \citet{Liu11_sdsspair}. In addition, these works predominantly probe redshift regimes that are distinct from those targeted here and are based on selection techniques sensitive to different AGN populations (e.g., radio-selected or narrow-line AGN in \citealt{Fu15_a, Fu18_sdss} and mid-infrared-selected AGN in \citealt{Satyapal17, Pfeifle19_triple}). 
As a result, they often explore a different region of parameter space in separation, dynamical state, redshift, and gas obscuration compared to the sub-arcsecond dual AGN targeted here.

A major breakthrough came with the \textit{Gaia} mission \citep{GaiaMission} of the European Space Agency (ESA), which enabled systematic all-sky searches for closely separated AGN through multiple complementary techniques. Dual AGN candidates can be identified as multiple \textit{Gaia} detections within a few arcseconds \citep{Lemon17, Chen22a, Shen23b, Lemon23}, via characteristic variability \citep{Kochanek06, KroneMartins19}, or through astrometric jitter in the photocenters of \textit{Gaia} quasars (“vastrometry”; \citealt{Shen19, Hwang20, Chen22a, Schwartzman24_vastrometry,Schwartzman25_vastrometry, gross25_vodka}).
High-resolution imaging from ESA/\textit{Euclid} provides a powerful complement, enabling the identification of closely separated AGN candidates at sub-arcsec scales. Machine-learning methods applied to \textit{Euclid} images allow for systematic searches for dual photometric AGN in order to uncover sources that are often too faint for \textit{Gaia} detection \citep{Ulivi2025arXiv}.

These developments have enabled recent large-scale efforts to expand the census of close dual AGN, particularly at $z > 0.5$ and projected separations below $1''$ (a few kiloparsecs at cosmic noon). An example is the compilation presented by \citet{Bigmac25}, which brings together candidate and confirmed systems identified with heterogeneous selection techniques, although the number of spectroscopically confirmed dual AGN at these separations remains limited.
This parameter space, in which simulations place SMBHs at separations consistent with a common merger remnant \citep{Volonteri22}, has been targeted by innovative \textit{Gaia}-based techniques. In particular, the innovative \textit{Gaia} multipeak (GMP) method identifies sub-arcsecond multiple AGN candidates from \textit{Gaia} light-profile peaks \citep{Mannucci22}. Typically, GMP candidates are brighter than \textit{Euclid}-selected sources, which reach a depth of $I_{\rm AB} \sim 24.5$ in the VIS filter, whereas \textit{Gaia} detects sources down to $G \sim 20.5$~mag (Vega-like), enabling efficient IFU follow-up to spectroscopically distinguish dual AGN, lensed quasars, or AGN–star projections \citep{Mannucci22, Mannucci23, Ciurlo23, Scialpi24}.
However, as an optical selection technique, GMP is primarily sensitive to unobscured, optically bright AGN and is therefore likely biased against heavily obscured dual AGN, particularly at higher redshift, a limitation shared by most current high-$z$ dual AGN searches.

A fraction of GMP-selected sources are expected to be gravitationally lensed quasars by foreground galaxies, producing multiple images at sub-arcsec separations. 
Although rare, these compact lensed systems are powerful probes of the central regions of lens galaxies, where stellar mass dominates \citep{Treu10b}. In these regions, they enable precise measurements of the mass enclosed within the Einstein radius (total mass), stellar mass-to-light ratios ($M_{\star}/L$), and the stellar initial mass function (\citealt{Treu10a, Shajib24}), while also constraining the relative contributions of baryons and dark matter and the overall mass profile of the lens galaxy \citep{Sonnenfeld15, Massey10, Newman17}.
Moreover, strongly lensed quasars provide independent constraints on cosmological parameters, such as the Hubble constant \citep{Wong20, Lemon23}, and have been extensively used in the literature to test models of structure formation and lens galaxy evolution \citep[e.g.,][]{Kochanek06, treu15}.

Beyond their role in probing lens galaxies and cosmology, strongly lensed AGN are also powerful tools to study the quasars themselves.  The magnification and amplification produced by lensing acts as a natural "boost" of the AGN signal, allowing detailed studies of high-redshift quasars that would otherwise be too faint to observe. Lensing can also induce microlensing variability, which probes the internal structure of the AGN, including the accretion disk and broad-line region \citep[e.g.,][]{Sluse12,Mediavilla21}. By combining lens modeling with high-resolution imaging and spectroscopy, lensed AGN provide a unique laboratory to investigate the physical properties of quasars across cosmic time.

Several large surveys have systematically targeted lensed AGN to build statistical samples, providing insights into both lens galaxies and background quasars. The Sloan Digital Sky Survey (SDSS) Quasar Lens Search (\citealt{Inada05, Inada12}) produced a total of about 62 strongly lensed quasars in the SDSS Data Release 7 (DR7) quasar catalog, 26 in the well-defined statistical sample, and 36 additional lenses identified with various techniques, providing one of the largest homogeneous samples for statistical studies of strong lensing. The Sloan Lens ACS Survey (\citealt{Bolton08}) used spectroscopic selection from SDSS to find 131 strong galaxy-galaxy lenses at $z\sim0.05$–0.5, providing detailed constraints on the mass, light, and kinematics of massive early-type galaxies. The SL2S survey \citep{Sonnenfeld13, Sonnenfeld15} and the COSMOS HST survey \citep{Faure08} identified tens of strong lens candidates, allowing the study of stellar-to-dark matter fractions, radial mass profiles, and magnification properties in a larger but still moderate redshift range. The Grism Lens-Amplified Survey from Space (\citealt{treu15}) extended lens spectroscopy to cluster fields at $z\sim0.3$–0.7, delivering near-IR spectra and emission-line redshifts for both lenses and background sources, with a focus on faint systems.

While these surveys provide valuable statistical samples, they typically target lenses with image separations $\geq1''$ and redshifts below two, limiting the study of very compact and higher-redshift systems. In contrast, the GMP method \citep{Mannucci22} identifies subarcsec multiple AGN candidates, including some of the most tightly separated lensed systems known, with quasar (QSO) redshifts in the range $0.5<z<3.5$ and projected separations between 0.15$''$ and 0.8$''$. These compact lenses enable high-resolution imaging and spectroscopic follow-up, allowing us to extend detailed analyses to a population of lensed AGN previously inaccessible to systematic statistical studies, while simultaneously probing both the structure of the lens galaxy and the properties of the AGN. Moreover, our technique targets Einstein radii smaller than the effective radius of the lens galaxies, enabling us to probe their inner regions even at high redshift \citep{Sonnenfeld19,DAmato26_quad}.

Here we present the European Southern Observatory (ESO) Large Program 114.27BY (PI: M. Scialpi) consisting of adaptive-optics (AO) assisted Multi Unit Spectroscopic Explorer (MUSE, \citealt{Bacon10}) narrow-field mode (NFM) observations at ESO’s Very Large Telescope (VLT) targeting 150 GMP-selected systems at sub-arcsec separations and $z>0.5$.  
In this paper, we describe the dataset of 30 systems collected during the first year (2025) as well as the observing strategy, data reduction, and calibration procedures, and we outline the analysis used to classify each system as dual AGN, lensed AGN, or foreground stellar alignment.

The paper is structured as follows. 
In Sect.~\ref{sec:target} we describe the GMP-based target selection and the use of integrated spectra to filter out AGN+star contaminants. Section~\ref{sec:obs} presents the MUSE observations, the data reduction, and the resulting dataset of images and spectra. Section~\ref{sec:analysis} details the spectral modeling and analysis procedures used to classify the systems and study their properties, while Sect.~\ref{sec:results} reports the spectroscopic properties and final classification of each observed system. Section~\ref{sec:distribution} examines the distributions of dual and lensed AGN as a function of separation, magnitude, and redshift while taking into account selection effects. 
Finally, in Sect.~\ref{sec:summary} we summarize the results, discuss the implications for the population of dual and lensed AGN at subarcsec separations, and outline the prospects for the full MUSE Large Program.
All magnitudes we report are in the Vega system. We adopted a flat $\Lambda$CDM cosmology with parameters from \cite{Planck2020}, with $H_{\rm 0}=67.7\ {\rm km\ s^{-1}\ Mpc^{-1}}$ and $\Omega_{\rm m,0}=0.31$.

\vspace{-2mm}

\section{Target selection}
\label{sec:target}
We selected our AGN candidates using the GMP technique \citep{Mannucci22, Mannucci23} 
applied to \textit{Gaia} DR3 light profiles \citep{Gaia_EDR3_1}, as briefly summarized in the following, 
while full details can be found in \citet{Mannucci22}. The fraction of \textit{Gaia} scans displaying a 
double peak is quantified by the $\fmp$ parameter \citep{Gaia_EDR3_1, Gaia_EDR3_2}, 
ranging from 0 (never detected) to 100 (always detected). We adopted a selection threshold of 
$\fmp>8$, shown by \citet{Mannucci22, Mannucci23} to be a reliable criterion for identifying 
multiple systems.

The primary source must be brighter than $G < 20.5$, corresponding to the efficiency limit of the 
GMP method \citep{Mannucci23}. The effective selection window for GMP dual-AGN candidates is 
therefore $\sim 0.15''$–$0.8''$, as detectability depends on both angular separation and luminosity ratio. 
Systems at separations smaller than the \textit{Gaia} point spread function (PSF) ($\sim 0.15''$) or with large luminosity contrasts 
($L_A/L_B \gtrsim 15$) may fail to produce a measurable multipeak signature, while sources with 
separations $\gtrsim 0.8''$ are typically resolved as distinct detections in the \textit{Gaia} archive. 
This upper limit naturally selects dual AGN that are physically close enough to reside within the same host galaxy, while also ensuring that the GMP method is still sensitive to multiplicity. 
In Table~\ref{tab:targets} we list the properties of all systems, and for those resolved by \textit{Gaia} 
we report the individual magnitudes of both components (G1 for the primary and G2 for the secondary).

To minimize stellar contamination, we restricted the selection to regions at Galactic latitudes 
$|b|>20^\circ$ and excluded areas surrounding large Local Group galaxies (M31, LMC, SMC). 
We also excluded over-crowded regions of the sky, defined as having more than 12 \textit{Gaia} detections 
per square arcminute within 3 arcmin of the target. 
Priority was given to systems at $z > 0.5$, both because galaxy merger activity peaks in this 
redshift range \citep{Chen23b} and because at lower redshift the brightness of the host galaxy can 
significantly reduce the GMP detection efficiency \citep{Mannucci22}. This ensures an optimal balance 
between astrophysical relevance and selection completeness.

\smallskip
The GMP selection was applied to a parent sample of AGN drawn from two complementary datasets: a catalog of spectroscopically confirmed AGN and a catalog of photometric AGN candidates, which are described below.

(i) \emph{Spectroscopic AGN.} This sample includes sources with available spectra and secure redshift measurements, drawn from large spectroscopic surveys such as the Sloan Digital Sky Survey (SDSS; \citealt{Lyke20_SDSS}) and the Dark Energy Spectroscopic Instrument (DESI; \citealt{DESI25_arxive}). For these sources, we adopt version 8 of the Milliquas compilation \citep{Flesch23}.

(ii) \emph{Photometric AGN.} This sample consists of AGN candidates selected from recent photometric catalogs \citep{Delchambre22, quaia, Flesch21, wise_ps1, Fu24, Salvato25}, which lack spectroscopic confirmation but significantly extend the accessible parameter space.

The spectroscopic sample provides robust AGN classifications and accurate redshift measurements for the primary component up to $z \sim 4$. In contrast, the photometric catalogs extend the GMP selection to regions not yet covered by large spectroscopic surveys, increasing the sample size at the expense of lower purity.

Together, these datasets enable an efficient identification of multiple-peak AGN candidates across a broad range of magnitudes and redshifts, approaching full-sky coverage.

\subsection{Spectral confirmation of photometric candidates}
\label{sec:phot_candidates}
To confirm the AGN nature of photometric candidates and obtain robust spectroscopic redshifts, we carried out follow-up observations through several programs using multiple instruments. Specifically, we employed the ESO Faint Object Spectrograph and Camera (v.2, EFOSC2) mounted at the Nasmyth B focus of the 3.58\,m ESO New Technology Telescope (NTT; \citealt{efosc2_ntt}; Program IDs: 109.22W4, 112.25CT, PI: Mannucci; 113.26TB, PI: Scialpi), the Device Optimized for the LOw RESolution (DOLORES) on the Telescopio Nazionale \textit{Galileo} (TNG; Program IDs: \texttt{A47TAC\_28}, \texttt{A48TAC\_67}, PI: Scialpi), and the FOcal Reducer/low dispersion Spectrograph 2 (FORS2) on the VLT in Long-Slit Spectroscopy (LSS) mode (Program ID: \texttt{115.28CQ}, PI: Scialpi).
These observations allowed us to confirm the AGN classification of the primary sources and to determine their spectroscopic redshifts. The systems in which the presence of an AGN as the primary source is confirmed are further analyzed to reliably exclude the presence of secondary stellar components, thereby removing systems that correspond to AGN–star pairs through the analysis described in Sect.~\ref{sec:deconv}.
Observations were conducted under varying atmospheric conditions, with a typical seeing better than $1.1''$. A $1''$ slit was used by default and widened when necessary to ensure that both components remained within the slit. Exposure times were optimized using the ESO and TNG Exposure Time Calculators to achieve a continuum S/N of $\sim10$–15 per spectral resolution element for targets with $G \simeq 20$. Each exposure sequence included additional overheads for acquisition and calibration frames, resulting in total integrations of 20–60\,min per target depending on the instrument and conditions.

The redshifts of all targets observed with MUSE are listed in Table~\ref{tab:targets}, while the full catalog will be presented in a future paper (Scialpi \& Marconcini, in prep.).

\subsection{Spectral decomposition}
\label{sec:deconv}
To remove stellar contaminants as secondary components in our sample, we built a Python routine that quantitatively estimates the likelihood of contamination via  spectral decomposition of the unresolved spectra (see Sect. \ref{sec:phot_candidates}).
The spectral decomposition used here is un updated version of the one described in \citet{Scialpi24} and will be fully presented in Marconcini \& Scialpi in prep.
We applied our custom spectral decomposition to all targets with available integrated spectra, including AGN from DESI and SDSS and photometric candidates confirmed through our follow-up observations, for a total of 650 sources. 
All these spectra lack the spatial resolution to disentangle the single components, which may correspond to a second AGN (either an independent AGN or multiple images of the same lensed AGN) or a foreground aligned star. 
This decomposition process has proven crucial to remove a large fraction of AGN+star contaminants prior to IFU follow-up observations \citep[for a similar approach see][]{Shen23_deconv}.

In brief, the observed spatially unresolved spectrum is fit using a Python implementation of the Penalized PiXel-Fitting algorithm (\texttt{pPXF}, \citealt{Cappellari2004}) with a linear combination of quasar (from a library covering different ionization levels; \citealt{Temple21}) and stellar (G, K, M types) spectral templates from the MaStar library \citep{Yan19}, covering the wavelength range $3622-10354\ \AA$ at $R\sim1800$. 

First, we convolve the quasar and stellar templates with the data spectral resolution. Then, we create a single template grid composed of 26 stellar spectra (G0-G9, K0--K7, M0--M8)  and 66 quasar spectra. 
Quasar spectra are created using the \textit{qsosed} package, with intrinsic visual extinction in the range [0, 1] with spacing 0.1 and spanning different levels of ionization.

Then, we used \texttt{pPXF} to infer the best-fit combination of stellar and quasar templates to reproduce the data. In doing so, the free parameters are the stellar spectral type, the QSO extinction ($A_V$; $E(B-V)$), the redshift ($z_{\rm QSO}$), and the flux normalization between templates ($F_{\rm star}$, $F_{\rm QSO}$). Moreover, we allowed for a spectral shift of the stellar component along the line of sight to account for its proper motion of $\pm$ 300 km/s.

The routine first degrades the spectral resolution of the templates to match the observed data, then finds the linear combination of stellar and QSO templates that best fits the spectrum, accounting for dust extinction, redshift, and radial velocity shifts. 
While the formal statistical uncertainties on the redshift derived from the pPXF fit are typically smaller than the spectral resolution, we adopt a conservative estimate based on the spectral resolution of the data. Given the resolving power of our observing setups, this corresponds to a redshift uncertainty of $\Delta z \sim 0.001$ over the redshift range of our targets ($z \sim 0.7-3.2$). This accuracy is sufficient for secure line identification and for planning IFU follow-up observations.

To quantify the probability of stellar contamination in our spectra, we defined seven of the brightest stellar absorption features for each stellar type and evaluated the $\chi^2$ in a spectral window of $\pm 300$~km/s around each feature to estimate the improvement relative to a pure quasar template. Based on this, we assign a probability of stellar contamination when the improvement in $\chi^2_R$ exceeds 10\%, or alternatively when $\Delta\mathrm{BIC} > 6$.
This procedure removes most AGN+star contaminants prior to IFU follow-up. \citet{Mannucci22} found that 30\% of GMP systems at $0.5\arcsec$ are expected to be AGN+star, increasing to 60\% at $0.8 \arcsec$ (scaling roughly with the square of the separation at small separations; see the discussion in Sect.~\ref{sec:star_fraction}). 

For this first year, we also targeted some objects that had a non-zero probability of containing a star based on the spectroscopic decomposition, in order to test the method. The preselection procedure removed roughly 75\% of the expected stellar contaminants, implying that about 30 AGN+star systems were filtered out prior to the MUSE follow-up. Among the 30 objects observed in year one, only 11 were confirmed to be AGN+star systems (see Table~\ref{tab:targets}), confirming the method’s effectiveness.

\subsection{Targets for MUSE follow-up}
From the subset of GMP-selected sources with available spectroscopic data (both fiber and single-slit spectra), and after applying all the selection cuts described above, we retained only those targets whose spectral decomposition showed no evidence of a stellar secondary component. We then applied additional constraints to ensure that the selected systems were observable from Paranal (Dec~$< +15^{\circ}$) and sufficiently bright in the near-infrared to serve as tip-tilt reference stars for the MUSE AO system ($J<17.5$). 
J magnitudes were obtained from 2MASS \citep{Skrutskie06}, VHS \citep{McMahon13}, and other surveys. In cases where the objects were too faint to be detected, we estimated their magnitudes from a linear combination of the observed $R_p$ and $B_p$ magnitudes from \textit{Gaia} \citep{Gaia_EDR3_1}, with a typical uncertainty of $\sim0.27$\,mag \citep[see the appendix of][]{Scialpi24}:
\begin{equation}
J = R_p - 0.82 \,(B_p-R_p) - 0.26,  \quad \sigma = 0.27
\label{eq:J}.
\end{equation}

Applying this procedure to the full GMP-selected AGN sample, we identified the best 150 targets suitable for MUSE-NFM follow-up observations. These sources span the redshift range $0.5 \lesssim z \lesssim 3.5$, as higher-redshift AGN are rarely detected by \textit{Gaia}, and exhibit angular separations between $\sim0.15\arcsec$ and $\sim0.8\arcsec$ (corresponding to projected physical separations of approximately $1$–$7$ kpc across this redshift range), consistent with close AGN pairs in interacting or merging systems. Simulations suggest that at such separations the SMBHs may often reside within a common merger remnant \citep[e.g.,][]{Volonteri22,Chen23b}.

These sources span the redshift range $0.5 \lesssim z \lesssim 3.5$, as higher-redshift AGN are rarely detected by \textit{Gaia}, and exhibit angular separations between $\sim0.15\arcsec$ and $\sim0.8\arcsec$ (corresponding to projected physical separations of approximately $1$–$7$ kpc across this redshift range), consistent with close physical AGN pairs in merging systems \citep[e.g.,][]{Volonteri22,Chen23b}.

\section{The Cosmic Duets large program}
\label{sec:obs}
In this work, we discuss the observing strategy, data reduction, and analysis procedures of our Cosmic Duets ESO Large Program 114.27BY (PI: Scialpi), whose ultimate goal is to obtain AO assisted MUSE-NFM observations to characterize dual and lensed AGN at subarcsec separations in the redshift range $0.5<z<3.5$. Here we present the results for the objects observed during the first year of the program, from 1 October 2024 to 30 September 2025. The program, motivated by the identification of 150 GMP-selected dual and lensed AGN candidates observable with MUSE/NFM (Sect.~\ref{sec:target}), has a total allocation time of 137 hours. The same observing strategy and analysis procedures described here will be applied to the rest of the sample, which will be observed in cycles P116 and P117.

\begin{table*}[t!] \footnotesize
\caption[]{Main information on the 30 observed targets, including exposure times ($T_{\rm exp}$), number of exposures ($N_{\rm exp}$), projected separation (Sep), and spectroscopic classification.}
\label{tab:targets}

\begin{tabular}{c@{\hspace{3.5mm}}c@{\hspace{3.5mm}}cc@{\hspace{2.5mm}}c@{\hspace{2.5mm}}c@{\hspace{3.5mm}}c@{\hspace{3.5mm}}c@{\hspace{3.5mm}}c@{\hspace{3.5mm}}c@{\hspace{3.5mm}}c@{\hspace{1.5mm}}c@{\hspace{1.5mm}}c}

\hline
Target & Coords (J2000) & z & $\fmp$ & \textit{J} & \textit{G1} & \textit{G2} & $T_{\rm exp} \times N_{\rm exp}$ & Obs. & Sep & PA & Classification \\  
 & & & & & & & (sec) & date & (\arcsec) &$^\circ$ (N→E)& \\
\hline
\\
J0032--1053  & 00:32:50.12 --10:53:57.92 & 2.439 & 12 & 17.68 & 19.02 & 20.13 & 600x3 & 09/17/25 & 0.662 & 79&Dual \\
J0034--1623 & 00:34:55.85 --16:23:28.90 & 1.994 & 11 & 17.82 & 18.96 & - & 600x3 & 09/17/25 & 0.590 & --54 &Lensed  \\
J0120--0542 & 01:20:18.64 --05:42:42.37 & 2.292 & 20 & 17.41 & 19.74 & 20.11 & 600x3 & 07/22/25 & 0.485& --12&AGN+star \\
J0144--5745 & 01:44:09.19 --57:45:27.31 & 0.793 & 8 & 17.03 & 19.59 & - & 600x3 & 10/29/24 & 0.437 & 114 &Dual \\
J0230--3201 & 02:30:48.03 --32:01:24.46 & 2.872 & 17 & 17.25 & 19.35 & - & 600x3 & 07/22/25 & 0.369& 118 &AGN+star \\
J0259--0901 & 02:59:26.08 --09:01:36.39 & 1.156 & 11 & 17.41 & 18.87 & - & 600x3 & 12/04/24 & 0.325 &90 &Lensed \\
J0315--4608 & 03:15:48.36 --46:08:43.09 & 2.698 & 11 & 17.45 &19.17 &20.46& 600x3 & 12/01/24 & 0.707&100 &AGN+star \\
J0317--5604 & 03:17:43.01 --56:04:54.66 & 1.932 & 12 & 17.54 & 19.88 & 19.61 & 600x2 & 12/23/24 & 0.500 & 143 &Lensed \\
J0404--2836 & 04:04:46.13 --28:36:29.67 & 1.890 & 27 & 17.49 & 19.74 & 19.91 & 600x6 & 12/24/24 & 0.548 & 47 &Lensed \\
J0408--3503 & 04:08:46.51 --35:03:13.31 & 0.883 & 9 & 17.07 & 19.27 & 20.97 & 600x3 & 12/22/24 & 0.825 & 125&AGN+star \\
J0437--3711 & 04:37:05.14 --37:11:19.45 & 2.146 & 53 & 17.53 & 19.14 & - & 600x5 & 12/25/24 & 0.461& --139&AGN+star \\
J0528--2838 & 05:28:09.23 --28:38:57.28 & 2.850 & 9 & 16.94 & 19.09 & 20.46 & 600x3 & 12/21/24 &0.656 & 130&AGN+star \\
J0530--3730 & 05:30:36.98 --37:30:10.54 & 2.858 & 38 & 15.31 & 17.33 & 20.32 & 600x3 & 12/21/24 & 0.531* & - &Quad\\
J0551--4629 & 05:51:47.84 --46:29:53.40 & 1.215 & 23 & 17.36 & 19.04 & 20.23 & 600x3 & 12/25/24 & 0.740 & --168&Lensed \\
J0637--4137 & 06:37:01.38 --41:37:41.67 & 2.743 & 18 & 16.63 & 18.94 & - & 600x3 & 01/22/25 & 0.414 & --155&Dual \\
J0802+1944 & 08:02:43.93 +19:44:23.42 & 1.778 & 37 & 17.09 & 19.33 & - & 600x3 & 12/22/24 & 0.340 & 107&Lensed \\
J0957--2242 & 09:57:52.71 --22:42:02.96 & 2.061 & 12 & 16.36 & 18.54 & 18.55 & 600x4 & 12/25/24 & 0.335* & -&Quad \\
J1113--1621 & 11:13:04.93 --16:21:31.00 & 0.727 & 22 & 17.51 & 19.69 & - & 600x3 & 04/24/25 & 0.280&--117 &AGN+star \\
J1116--0657 & 11:16:23.53 --06:57:38.46 & 1.236 & 29 & 16.37 & 17.90 & 17.90 & 600x3 & 04/30/25 & 0.341* & -&Quad \\
J1447--0501 & 14:47:08.72 --05:01:13.31 & 1.490 & 50 & 16.21 & 18.26 & - & 600x3 & 01/22/25 & 0.309 & --76&Lensed  \\
J1616--2149 & 16:16:46.43 --21:49:26.65 & 1.484 & 12 & 17.32 & 18.95 & - & 600x3 & 04/27/25 & 0.403&--30 &AGN+star \\
J1739+1847 & 17:39:49.29 +18:47:54.23 & 1.201 & 16 & 17.46 & 19.17 & 20.54 & 600x3 & 05/31/25 & 0.643& --13&AGN+star \\
J1837--6711 & 18:37:00.67 --67:11:57.28 & 1.124 & 24 & 16.95 & 18.51 & - & 600x3 & 04/07/25 & 0.412 & 166 &Dual \\
J2050--6657 & 20:50:52.38 --66:57:08.11 & 2.230 & 33 & 17.03 & 18.77 & - & 600x3 & 05/30/25 & 0.364& 164&AGN+star\\
J2056+1013 & 20:56:20.16 +10:13:22.46 & 0.533 & 18 & 17.62 & 19.40 & 20.78 & 600x3 & 08/28/25 & 0.610& 35&AGN+star \\
J2100+0612 & 21:00:39.18 +06:12:30.54 & 0.785 & 9 & 17.63 & 18.72 & - & 600x3 & 05/29/25 & 0.375 & 127 &Dual  \\
J2204--6530 & 22:04:40.09 --65:30:00.09 & 1.880 & 11 & 17.57 & 19.00 & - & 600x3 & 05/31/25 & 0.416 & --123 &Dual \\
J2257--6555 & 22:57:36.14 --65:55:08.84 & 1.633 & 17 & 16.33 & 17.69 & - & 600x3 & 07/18/25 & 0.407 & 43 &Lensed \\
J2344--4259 & 23:44:07.11 --42:59:42.40 & 3.164 & 20 & 17.64 & 20.28 & - & 600x3 & 09/17/25 & 0.246 & --66 &Lensed \\
J2353--0655 & 23:53:27.64 --06:55:17.24 & 1.463 & 22 & 17.08 & 18.20 & - & 600x3 & 10/11/24 & 0.424 & 45 &Lensed  \\
\\
\hline
\end{tabular}
\\
\footnotesize{
Notes: \textit{Gaia} G-band magnitudes of the primary (G1) and secondary (G2, if multiple components are present) sources, as well as $\fmp$, are present in the \emph{\textit{Gaia}} archive. 
J-band magnitudes are either observed or derived from \emph{\textit{Gaia}} BP and RP photometry, as described in the appendix of \citet{Scialpi24}. 
$T_{\rm exp} \times N_{\rm exp}$ indicates the exposure time in seconds multiplied by the number of exposures for the MUSE observations. Separation and classification are derived from our analysis of the data. The separation (Sep) corresponds to the distance between the two components in case of a double system (lensed, dual or AGN+star), and to the radius of the Einstein ring in the case of a quadruple lensed system (indicated with *). The PA is the angle measured clockwise from north toward east, calculated from components A to B.  }
\end{table*}

\subsection{Observations}
The observations were carried out with MUSE in NFM assisted by the GALACSI AO system at VLT \citep{Muse_bacon, museAO_strobele, museAO_Arsenault}. GALACSI operates in closed-loop at 1000 Hz, delivering near-diffraction-limited images. The instrument field of view (FoV) is about $7"\times7"$. The pixel size is $0.025'' \times 0.025''$, covering the observed-frame wavelength range 4750–9350~$\AA$ with a spectral resolving power $R \sim 1700$–3400, corresponding to a velocity resolution of $\Delta v \sim 180$–90 km s$^{-1}$. The wavelength range 5780–6050~$\AA$ is blocked by a sodium notch filter to avoid contamination from the laser guide stars (LGS). It uses four sodium LGS to correct for high-order turbulence and a natural guide star (NGS) for tip-tilt correction. 
Data were collected over multiple nights from October 2024 to September 2025 under excellent seeing conditions (FWHM $<0.85''$) and with lunar illumination $<50\%$. Each target was observed with three dithered exposures offset by $0.2''$ to improve PSF sampling, for a total exposure time of 30–45~min depending on target brightness. This strategy ensures sufficient signal-to-noise (S/N), effective sky subtraction, and mitigation of instrumental artifacts.

In all cases, the target itself was used as the AO reference source. In some observations, closing the AO loop on these double systems proved challenging, as the seeing conditions varied and only the total system magnitude (typically brighter than $J \sim 18$) was known, while the individual magnitudes of the two components were not. Apart from these occasional issues, the AO system consistently delivered excellent correction, with measured PSF FWHMs at $\sim7000\ \AA$ between $0.14''$ and $0.17''$. The multiple components of all targets are well separated (see Fig. \ref{fig:cubes_AGN}) and enabled detailed analysis of each source.

In contrast to the common practice of rotating the field between exposures (e.g.\ by 90$^\circ$), we maintained a fixed position angle (PA) for all observations. This choice is driven by the nature of our targets: pairs of point-like sources separated by $\lesssim 0.5\arcsec$ and of comparable brightness, with the brightest used as the NGS for AO correction. Although the AO loop remains closed during individual exposures, rotating the field between them increases the risk that the NGS temporarily falls outside the control region or that the deformable secondary mirror returns to the wrong configuration when the PA is reset, potentially leading to aborted exposures and increased overheads. 
Moreover, field rotation is not required for our science case. Unlike deep observations of extended, low–surface-brightness objects, where rotation mitigates systematic background structures, our compact point sources allow robust background estimation directly from multiple regions of the FoV. Rotating the field therefore provides no practical advantage while introducing operational risks, while a fixed PA ensures stable AO performance and observational efficiency.

\subsection{Data reduction}
We reduced our data using the MUSE pipeline \citep{Weilbacher20} within the \texttt{EsoRex} (ESO Recipe Execution Tool) environment. The standard reduction involved bias and dark subtraction to remove instrumental noise, flat-fielding to correct for detector response variations, wavelength calibration using arc lamp exposures, characterization of the line spread function to account for spectral resolution, sky subtraction and flux scaling. 
The sky was estimated from the faintest 50\% of the field, avoiding regions containing the science target, which occupies only a small portion of the FoV. 
Flux calibration was performed using a standard star observed during the same night with the same instrumental setup. Circular aperture extraction ensured that the star's total flux was captured across the full wavelength range.
Finally, the individual exposures were astrometrically aligned to correct for coordinate offsets and combined by averaging, resulting in the final calibrated datacube.
\vspace{0.3cm}

\begin{figure}[!h]
    \centering
    \includegraphics[width=1\linewidth]{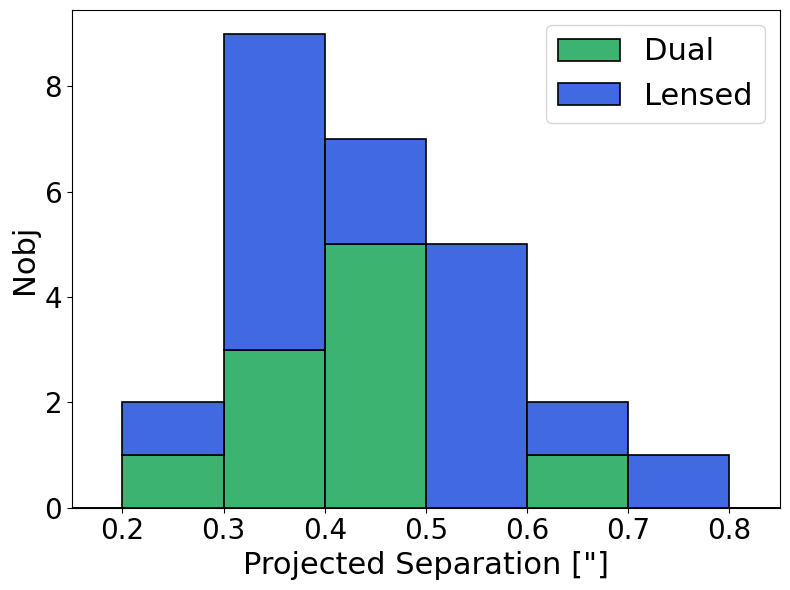}
    \includegraphics[width=1\linewidth]{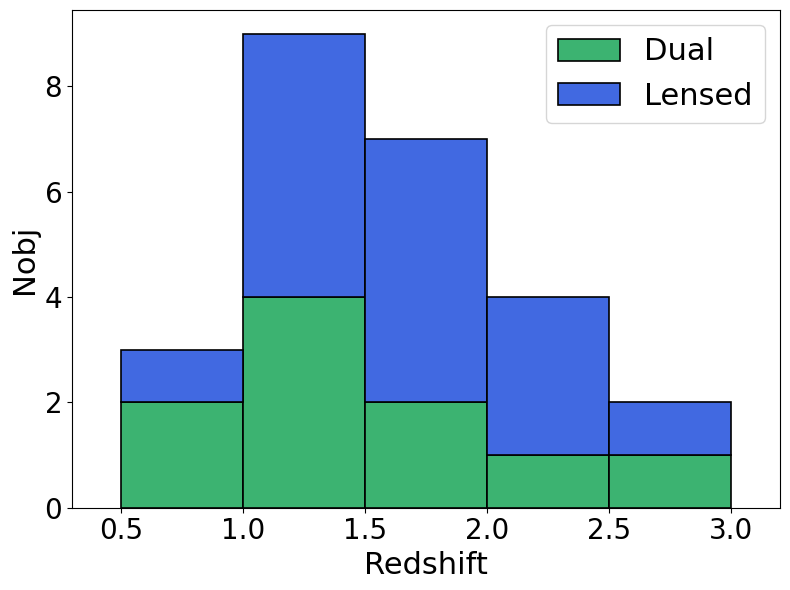}
    \caption{Distribution of dual (green) and lensed (blue) AGN as a function of projected separation (top panel) and redshift (bottom panel). The sample, observed with MUSE–NFM, includes sources presented in this work and by \citet{Scialpi24}. For quads, the Einstein radius was used to characterize the separation.}
    \label{fig:dual_lensed_sep}
\end{figure}

\subsection{Datasets}
We present the first 30 systems observed during the first year of the Cosmic Duets Large Program (ESO 114.27BY; PI: Scialpi). 
Table~\ref{tab:targets} summarizes the main information of these targets, including exposure times ($T_{\rm exp}$), number of exposures ($N_{\rm exp}$), projected separations among the components (Sep), and preliminary classification. 
The following subsections describe the extraction of images and spectra from the calibrated MUSE-NFM datacubes.

\subsubsection{Image extraction}

\begin{figure*}[h!]
    \centering
    \includegraphics[width=1\linewidth]{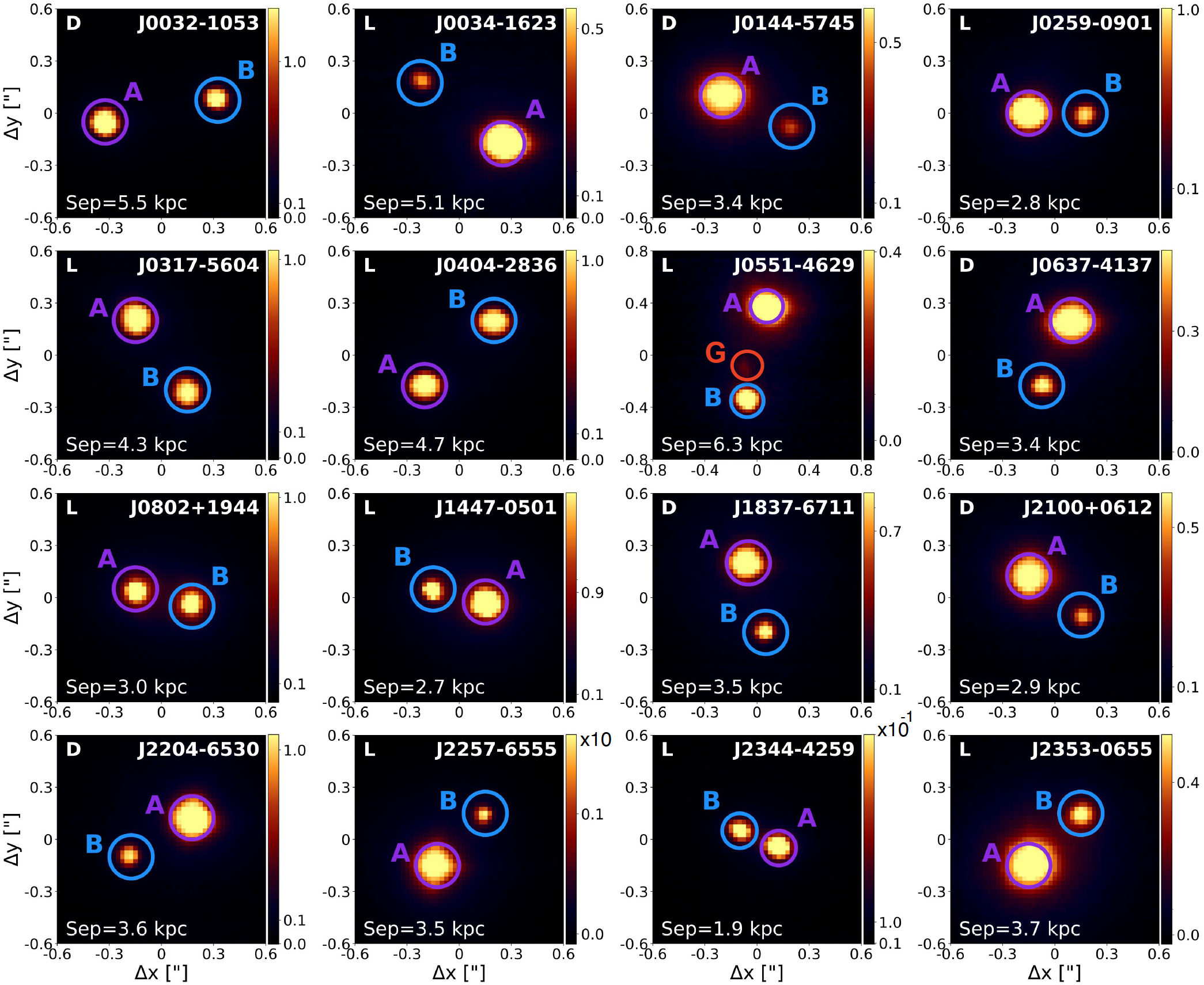}
    \caption{White-light images of the double AGN systems (lensed or dual) in units of 10$^{-15}$ erg/s/cm$^{-2}$. The target name and the separation between the two components are indicated in each panel. The spectra are extracted from the circular apertures shown on the maps, with purple and blue apertures corresponding to the A (brightest) and B (faintest) components, respectively. Dual systems are marked with the flag D, lensed systems with L. For systems identified as gravitational lenses, if an additional lensed image is detected, it is shown inside the red aperture and labeled as component G.}
    \label{fig:cubes_AGN}
\end{figure*}

\begin{figure*}[]
    \centering
    \includegraphics[width=1\linewidth]{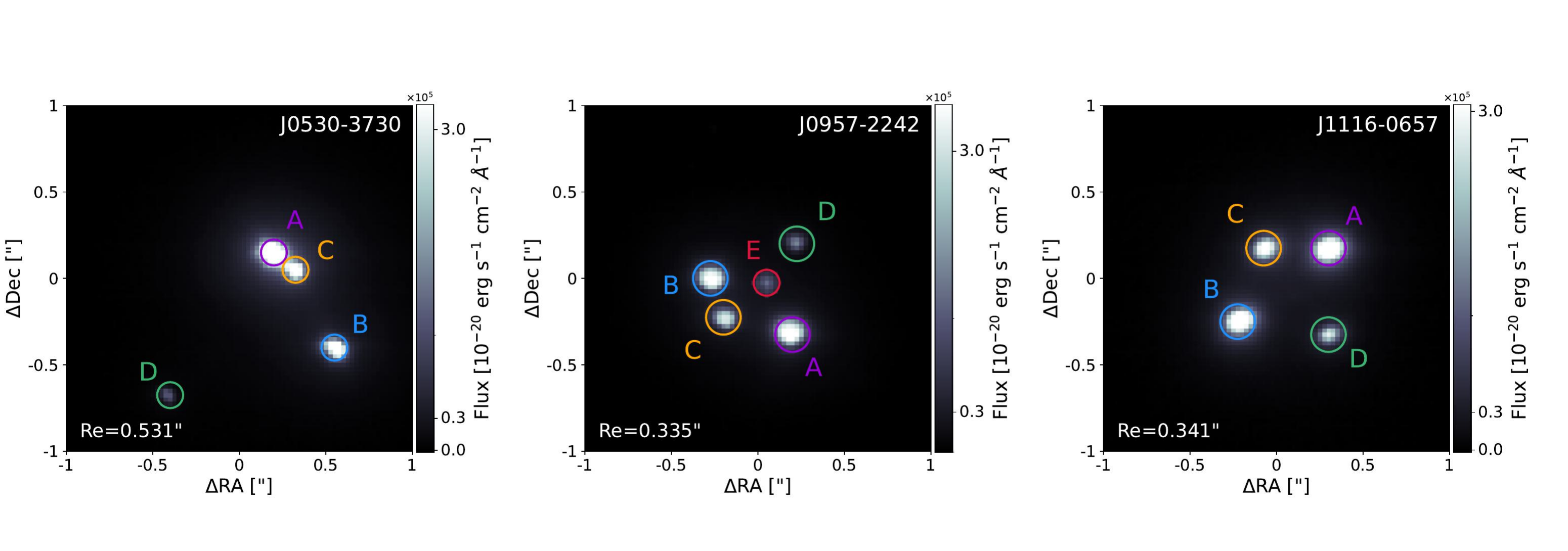}
    \caption{White-light images of the three quadruply lensed AGN systems. The target name and the value of the Einstein radius ($R_e$) is indicated in each panel. The spectra are extracted from the colored circular apertures. The colors of the apertures follow the order of brightness: A (purple, brightest), B (blue), C (orange), and D (green, faintest). For J0957, we also detect the emission-line lensing source (E) in red.}
    \label{fig:cubes_quads}
\end{figure*}

For each target, we extracted a white-light image by summing up all spectral channels. 
Figure~\ref{fig:cubes_AGN} shows cut-outs of the 16 double systems (lensed or dual AGN) with logarithmic scaling to enhance faint sources. 
The separation between the components is indicated in each panel. 
Figure~\ref{fig:cubes_quads} shows the three quadruply imaged systems, with the radius of the smallest circle encompassing AGN images (Einstein radius, $R_E$) indicated. 
All 30 observed systems are resolved into multiple point sources with angular separations ranging from $0.2''$ to $0.8''$, confirming the success of the GMP selection method. Figure~\ref{fig:dual_lensed_sep} shows the distribution of dual AGN and lensed systems from this sample and from \citet{Scialpi24}, all observed with MUSE–NFM, as a function of projected separation and redshift.

\subsubsection{Spectral extraction}
All spectra were extracted using a fixed 5~pixel-radius (0.125$''$) circular aperture, chosen to match the typical FWHM of the PSF across all wavelengths. For~J2344--4259~and~J0530--3730, a smaller aperture of 4~pix~was used because the components are closer than $0.25''$. This choice optimizes the trade-off between flux recovery and contamination from the companion sources.~The extracted spectra are independent, with flux uncertainties of~$\sim10\%$.

To assess whether aperture extraction is affected by cross-contamination \citep{Husemann18, pfeifle23_dual}, we performed an independent test using profile-fitting extraction with Gaussian and Moffat models. A white-light image was created by summing all spectral channels, and fit with two Gaussians to obtain initial estimates for the centroids and widths ($\sigma_x$, $\sigma_y$). Because the sources are very close, we fixed their relative separation and imposed common widths, as PSF variations at this scale are negligible compared to flux uncertainties.

Using these initial parameters, we then fit two Gaussians independently in each spectral channel. The resulting fit parameters, the widths ($\sigma_{x,\mathrm{fit}}, \sigma_{y,\mathrm{fit}}$) and amplitude ($A_{\mathrm{fit}}$), were smoothed as a function of wavelength by fitting a polynomial of degree 3, in order to remove outliers caused by noisy or bad channels. The flux of each component was then computed as
\begin{equation}
F_A = A_{\mathrm{fit}} \, 2 \pi \, \sigma_{x,\mathrm{fit}} \, \sigma_{y,\mathrm{fit}},
\end{equation}
where $A_{\mathrm{fit}}$ is the fit amplitude of the Gaussian for component A, and $\sigma_{x,\mathrm{fit}}$ and $\sigma_{y,\mathrm{fit}}$ are the fit Gaussian widths along the $x$ and $y$ directions, respectively. This procedure ensures that the extracted flux accounts for the PSF shape and minimizes contamination from the nearby source.

This test allowed us to evaluate whether contamination or wavelength–dependent PSF variations systematically alter the flux recovered by fixed apertures.
At blue wavelengths, however, the PSF broadens significantly due to the reduced efficiency of AO correction, causing the fit widths to increase and the profile–fitting technique to overestimate the true flux. Tests using Moffat profiles or mixed models (two Gaussians or two Moffats per component) did not improve the situation: despite their greater flexibility, they proved highly sensitive to noise and channel–to–channel fluctuations, leading to spurious variations in the extracted flux at low S/N.

By comparison, circular-aperture extraction yields stable measurements at all wavelengths, with flux variations below 5\% and negligible cross-contamination (<1\%) thanks to the small separation and the excellent MUSE–NFM spatial resolution. Because AO assisted optical observations have modest Strehl ratios, the extracted spectra contain only a fraction of the total flux; however, this fraction is identical for both components in each pair. We therefore normalized the spectra to their \textit{Gaia} magnitudes \citep{Gaia_EDR3_1}. When \textit{Gaia} resolves both components, each spectrum was normalized to its own magnitude; when only a combined magnitude is available, we first measured the flux ratio between components and normalized the sum of their spectra to the system magnitude.

These normalization factors reflect not only the AO Strehl ratio, but also the absolute flux–calibration accuracy of MUSE, which contributes significantly to the overall uncertainty. Assuming similar Strehl ratios and PSF for all components, we applied these normalization factors to extract the final spectra from each datacube. The resulting normalized spectra are shown in Figs.~\ref{fig:spectra_AGN} (dual AGN) and~\ref{fig:spectra_AGN_lensed} (double lensed AGN). Figure~\ref{fig:cubes_AGN} shows the corresponding circular apertures used for extraction. In all the figures, the brightest component is shown in purple and the faintest in blue. For the three quadruple-lensed AGN (Figs.~\ref{fig:cubes_quads} and \ref{fig:spectra_quads}), the third and fourth images are shown in orange and green, respectively, while the lens galaxy appears in red when detected.

\section{Data analysis}
\label{sec:analysis}
To analyze the data, we followed the methodology applied in a previous~MUSE-NFM~pilot program adopting the same observational strategy, and target selection described in \cite{Scialpi24}. Details on the main steps of the analysis are provided in the following. 

\subsection{Spectral modeling}
For systems classified as AGN+AGN, the extracted spectra of each component were modeled independently using AGN templates from \cite{Vandenberk01} and \cite{Temple21},~including interstellar dust extinction following the empirical quasar attenuation law of \cite{Temple21}. This fitting aims to derive the reddening~($E(B-V)$)~and redshift of each AGN individually.~Because the spectra were extracted from circular apertures and the components are well separated, we treat each spectrum as originating from a single source; potential differences in emission-line profiles between the two AGN are thus captured within the respective template fits. Galactic extinction is neglected, as its effect is minimal at the high Galactic latitudes of~our targets~($|b|>20^\circ$).

\vspace{-0.3mm}
For systems composed of an AGN and a foreground star, we modeled the AGN emission using the same template-fitting procedure described above, while the stellar component was fit with templates from the \citet{Yan19} library, including G-, K-, and M-type stars to determine its spectral type.

\subsection{Narrow absorption lines}
\label{sec:NAL}
In addition to continuum and emission-line diagnostics, we identified and analyzed narrow absorption lines (NALs) in our MUSE spectra, which in some cases provide complementary information to distinguish lensed from dual AGN systems.
Intrinsic absorbers are close in redshift to the quasar, typically with velocity shifts of $<3000~{\rm km\,s^{-1}}$ relative to the systemic redshift. These absorbers, often showing blueshifted velocities, consist of gas associated with the quasar or its host galaxy, and intrinsic NALs trace the quasar environment \citep[e.g.,][]{Hamann11, Misawa07, WeiJian19}. 
In lensed systems, if one image exhibits intrinsic NALs, the other image should generally show the same features. Differences in equivalent width (EW) can arise due to microlensing (see Sect. \ref{sec:classification}) or the geometry of the lens, where each light path intersects different regions of the lens galaxy \citep[e.g.,][]{Chartas09, Green06}. Conversely, if intrinsic absorbers differ significantly between components in redshift, depth, or profile, this indicates that the two sight lines probe physically distinct AGN, favoring a dual AGN scenario.

Intervening absorbers consist of gas clouds or galaxies along the line of sight, unrelated to the quasar itself. These features appear at redshifts different from that of the quasar, but can be present at the same redshift in both components of a system if both sight lines intersect the same intervening structure \citep{Martin+10}. While identical intervening NALs in both spectra are consistent with lensing, this is not by itself conclusive, and must be interpreted alongside other diagnostics. Even simple measurements of absorber redshifts provide valuable information on the intervening structures along closely spaced sight lines (separations of a few kiloparsecs at the absorber redshift), offering a first step toward probing the small-scale properties of the circumgalactic and intergalactic medium \citep[e.g.,][]{Dutta24}. 
To search for NALs in our sample, each spectrum is normalized by a locally fit continuum and inspected for narrow features. All absorber redshifts have been measured, providing a comprehensive catalog of both intrinsic and intervening NALs in our sample. For intervening absorbers, a line is considered common (from the same absorber) if it is present in both components with a velocity difference $\Delta v < 50$ km s$^{-1}$ and consistent depth $\Delta EW / EW < 20\%$. For intrinsic absorbers, differences between components can indicate dual AGN or quasar-driven outflows.

\subsection{Classification of dual and lensed systems}
\label{sec:classification}
To discriminate between lensed QSOs and dual AGN, we compared the two components in terms of redshift, EWs, emission-line flux ratios, line profiles, continuum shape and NALs. 
Before classifying a system as a dual AGN, it is important to verify that observed differences in continuum or broad-line EWs cannot be explained by gravitational microlensing. Microlensing occurs when compact objects (e.g., stars) in the lens galaxy preferentially magnify the most compact emission regions, such as the accretion disk and the high-ionization portion of the broad-line region \citep[e.g.,][]{Wambsganss06, mosquera2013, jimenezvicente2015}. Macro- and microlensing magnifications ($M$ and $\mu$) were estimated via the Macro--Micro Decomposition (MmD) method \citep{Sluse07}. Deviations caused by microlensing are recognized by their wavelength-dependent signatures and by the absence of corresponding changes in narrow lines, ensuring they are not misinterpreted as evidence of dual AGN.

\vspace{1mm}
\noindent Below we summarize the criteria adopted for classification:
\begin{itemize}
\item Redshift ($\Delta z$). Since lensed images originate from the same background AGN, they must share the same intrinsic redshift. Gravitational lensing cannot produce differential line-of-sight velocity shifts larger than the small, apparent differences due to microlensing on the broad line shape. Therefore any measurable offset ($\Delta z \geq 0.001$, i.e., $\Delta v \gtrsim 300$ km s$^{-1}$) cannot be produced by lensing and indicates two physically distinct AGN.
    
\item Equivalent widths. For each rest-frame spectrum extracted from circular apertures, line fluxes were measured via trapezoidal integration after estimating the local continuum from the edges of a velocity window. Uncertainties were propagated from individual flux errors and continuum estimates. The EWs were then compared between components. Significant differences not compatible with microlensing, variability, or differential extinction, support a dual AGN interpretation.

\item Emission-line ratios. Line flux ratios, especially for narrow emission lines, are a robust diagnostic. Significant deviations between components indicate physically distinct AGN, since narrow-line regions are spatially extended and largely unaffected by microlensing.

\item Continuum shape and line profiles. Differences in continuum or line profiles were quantified via a cross-correlation procedure, progressively modifying one spectrum to match the other and optimizing parameters via $\chi^2$ minimization. Initial fits considered only a redshift offset and flux scaling; further refinements allowed independent variations in continuum and emission lines, including wavelength-dependent extinction modeled with a second-degree polynomial. Minor deviations consistent with these effects do not indicate a dual AGN.

\item Narrow absorption lines. Narrow absorption lines provide complementary information for classification (See Sect \ref{sec:NAL}). We identify both intrinsic and intervening absorbers. Each spectrum is normalized by a locally fit continuum, and absorber redshifts are measured. For intervening absorbers, a line is considered \emph{common} if present in both components with $\Delta v < 50$ km s$^{-1}$. For intrinsic absorbers, we consider differences between the two components to be significant if the absorber centroids differ by $\Delta v > 300\ \mathrm{km\ s^{-1}}$. In these cases, gravitational lensing (including differential potential and time-delay effects) cannot produce systematic centroid shifts and they are naturally explained by two distinct sight lines intersecting different absorbing regions (outflows/host-galaxy kinematics).
\end{itemize}
Overall, a system is classified as a lensed AGN if all diagnostics are consistent with the same source (allowing for microlensing and extinction effects), and as a dual AGN if one or more diagnostics show significant differences.
\vspace{-4mm}
\section{Results}
\label{sec:results}
Table~\ref{tab:targets} summarizes the classifications obtained for the observed systems. 
The cross-correlation analysis described in Sect.~\ref{sec:analysis} was used as a quantitative diagnostic to assess spectral similarity between components, particularly for identifying lensed AGN candidates. However, the final classification of each system as dual AGN, lensed AGN, or AGN+star projection is based on a combination of spectral, kinematic, and morphological criteria, including emission-line profiles and ratios, velocity offsets, flux ratios, and the presence of lensing features in the datacube. 
Applying this multicriteria approach, we identified six systems as dual AGN, ten as doubly lensed AGN candidates, three as quadruply lensed systems, and 11 AGN+star.
In the following, we describe the observed systems in detail, highlighting their spectral properties and the procedures adopted for their final classification.

\subsection{AGN+star systems}
\label{sec:star_fraction}
The spectral decomposition used during the target selection stage (see Sect. \ref{sec:deconv}) is designed to remove stellar contaminants prior to IFU follow-up and succeeds in excluding the majority of such systems; see Sect.~\ref{sec:analysis}. 
Nevertheless, in our sample of 30 observed systems, 11 are revealed by the MUSE data to be AGN+star projections that were not rejected during the initial selection. 
These cases correspond to configurations in which the stellar component is significantly fainter than the AGN,
so that its spectral features are too faint to be detected with sufficient confidence in the integrated spectrum used for preselection.

In addition, several spatially unresolved observations were obtained under atmospheric conditions that reduced the effective S/N (seeing of $0.8$--$1.2$ arcsec), further limiting the detectability of stellar absorption features. 
Once the spatial and spectral information of the IFU is taken into account, the nature of the secondary point source becomes evident, and the stellar component can be isolated and fit, allowing a robust AGN+star classification. The spectra of these systems are reported in Fig.~\ref{fig:AGN_STAR} for completeness.

\begin{figure}[h!]
    \centering
    \includegraphics[width=0.95\linewidth]{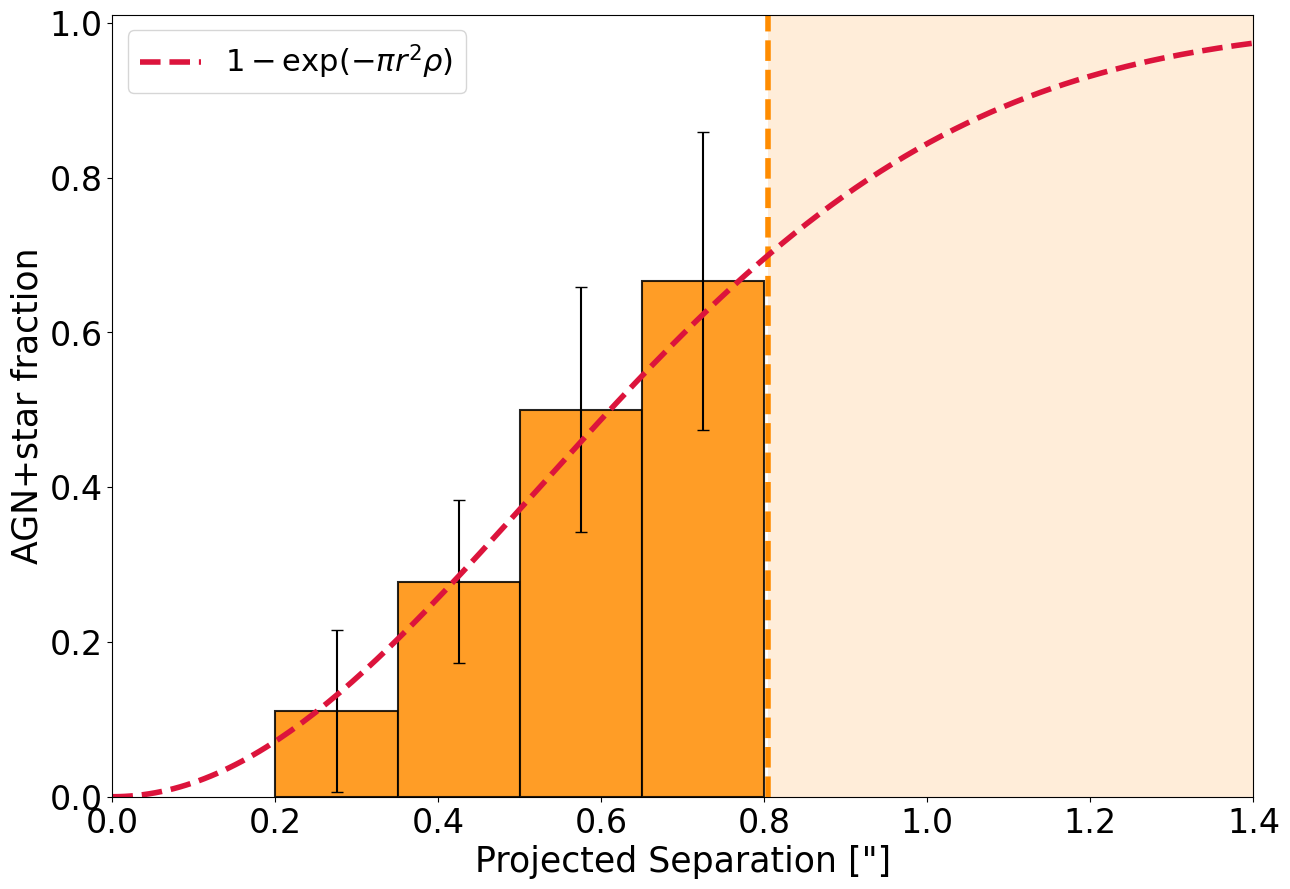}
    \caption{Fraction of AGN+star projections as a function of projected separation.  
    The dotted red line shows the expected contamination probability function (Eq.~\ref{eq:contam_prob} with $D=0.65''$), representing the separation distribution of random stellar contaminants. The shaded area shows the separation range excluded by the GMP selection.}
    \label{fig:star_sep}
\end{figure}

Considering all~GMP~systems observed with~MUSE~(this work and \citealt{Scialpi24}),~we~find~that~the~occurrence~of AGN+star projections strongly~depends~on the projected separation between the two point sources.
Figure~\ref{fig:star_sep} shows the fraction of AGN+star systems,~$N_{\mathrm{AGN+star}}/N_{\mathrm{tot}}$,~in~bins~of~separation.

\noindent The probability of stellar contamination can be described by
\begin{equation}
P(r) = 1 - \exp\left(-\frac{\pi r^2}{4 D^2}\right)=1 - \exp\left(-{\pi r^2 \rho}\right),
\label{eq:contam_prob}
\end{equation}
where $D$ is the mean distance to the nearest contaminant star and $\rho =1/4 D^2$ is the surface density of contaminants.
For $r \ll D$, $P(r)$ behaves approximately as a parabola \citep{Banerjee21}, while it asymptotically approaches one for $r \gg D$.
For illustration, the curve with $D = 0.65''$ is overplotted in Fig.~\ref{fig:star_sep} in red.
The contamination fraction rises significantly for separations $\gtrsim 0.6''$, where AGN+star projections become more common than double AGN candidates.
At separations $>0.8''$, contamination exceeds 80\%, and the curve’s asymptotic approach to 1 indicating that dual AGN searches at these scales are expected to have low efficiency. These pairs are therefore excluded from our GMP sample (dashed area).
These levels agree with estimates by \citet[$P(0.5)\sim0.3$]{Mannucci22} and \citet{Ulivi2025arXiv}, who found a lower critical separation (0.4") due to deeper sensitivity, resulting in a smaller value of D and therefore higher contamination.

\subsection{Dual AGN systems}
\label{sec:Dual AGNystems}

Six of the AGN+AGN systems in our sample have been classified as dual AGN following our spectroscopic and spatially resolved analysis.
Only one of these sources, J0032–1053, was also selected by an independent method and previously identified as a dual AGN candidate by \citet{Hwang20}, who flagged the object due to significant non-zero proper motions in \textit{Gaia} DR2.
In the following, we describe the analysis and properties of each individual system.

\subsubsection{J0032--1053} 
\vspace{-1mm}
J0032--1053 is confirmed as a dual AGN system at redshift $z=2.439 \pm 0.001$, with the two nuclear components separated by a projected distance of $5.5 \pm 0.1$~kpc and a relative velocity of $\sim 200 \pm 130 $~km~s$^{-1}$. Spectroscopic analysis reveals notable differences between the two nuclei, labeled AGN~A and AGN~B (purple and blue, respectively, in Fig.~\ref{fig:spectra_AGN}). AGN~A is more luminous, with its observed continuum flux requiring a downward scaling factor of $\sim 3.2$ for direct spectral comparison. The line ratios also differ: the lower-ionization species, such as C\,\textsc{ii}] $\lambda2326$, are proportionally stronger in the AGN~B spectrum relative to the high-ionization C\,\textsc{iv} $\lambda1549$ emission.

\begin{figure}[]
    \centering
    \includegraphics[width=1\linewidth]{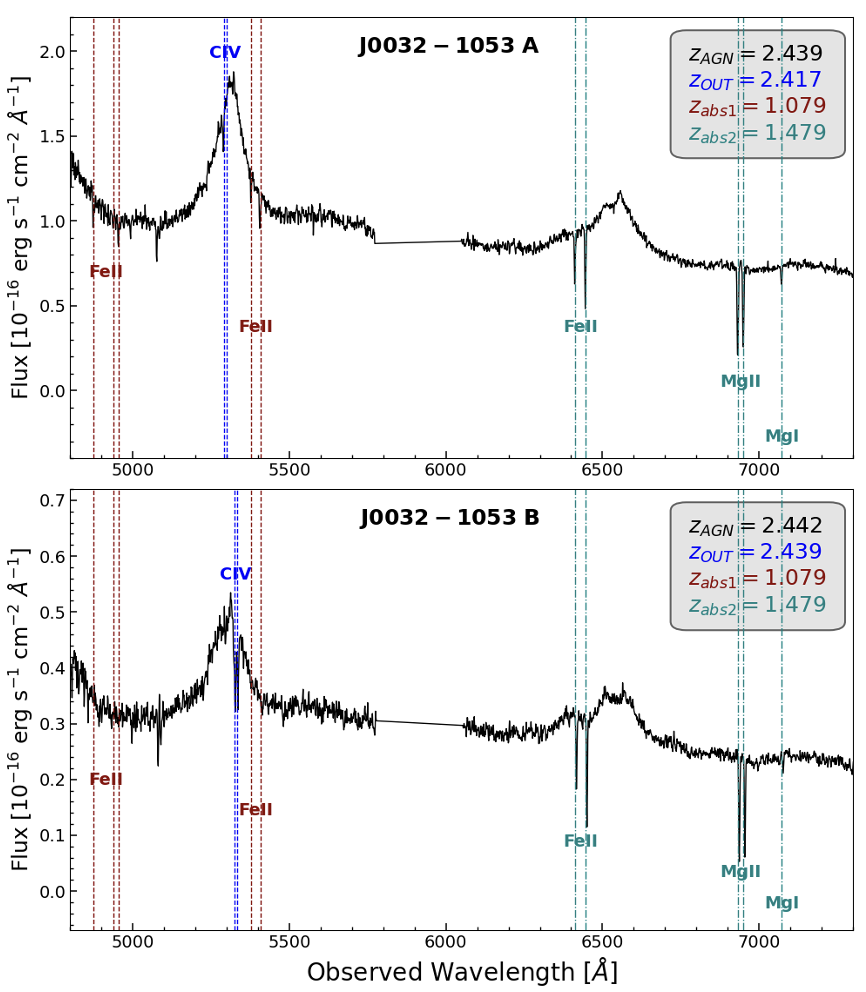}
    \caption{Absorption systems of the dual AGN J0032--1053. The upper panel shows component~A, while the bottom panel shows component~B. The two intervening absorption systems are indicated: the system at $z = 1.079$ is shown in dark red and the system at $z = 1.479$ in teal. Intrinsic \ion{C}{iv} absorption at the AGN redshifts is shown in blue.}
    \label{fig:J0032}
\end{figure}

Two intervening absorption systems are detected in both components (see Fig.~\ref{fig:J0032}). The first system is located at $z = 1.0794 \pm 0.0004$, where the Mg\,\textsc{ii} $\lambda\lambda2796.35, 2803.53$ doublet is observed at $\sim 6943$~\AA, along with Fe\,\textsc{ii} $\lambda\lambda2586.65, 2600.17$ transitions around 6448~\AA\ in the observed frame. A second intervening absorber is detected at $z = 1.4791 \pm 0.0006$, showing multiple Fe\,\textsc{ii} transitions ($\lambda2344$, $\lambda2374$, $\lambda2382$, $\lambda2586$, $\lambda2600$) spanning $\sim 4800$–5400~\AA, as illustrated in Fig.~\ref{fig:J0032}.

In both AGN~A and AGN~B, C\,\textsc{iv} absorption is detected, but with markedly different velocity shifts relative to their systemic redshifts.
For AGN~B, adopting a systemic redshift of $z_{\rm sys,B} = 2.442$, the C\,\textsc{iv} $\lambda\lambda1548,1550$ doublet is detected at $z_{\rm out,B} = 2.4388$, corresponding to a blueshift of $v_{B\rightarrow B} \simeq -170~\mathrm{km\,s^{-1}}$, well within the typical range of AGN outflows \citep[e.g., a few hundred km\,s$^{-1}$;][]{Hamann11}. The line ratio is consistent with an intrinsic narrow absorption line (NAL), indicating absorbing gas associated with AGN~B, but without a fast outflow.
For AGN~A, with systemic redshift $z_{\rm sys,A} = 2.43967$, a weaker doublet-like feature is detected at $z_{\rm out,A} = 2.4175$. Interpreted as intrinsic C\,\textsc{iv} absorption, this implies a much larger blueshift of $v_{A\rightarrow A} \simeq -1935~\mathrm{km\,s^{-1}}$, consistent with a fast quasar outflow. Compared to the systemic redshift of AGN~B, the offset becomes $v_{A\rightarrow B} \simeq -2140~\mathrm{km\,s^{-1}}$. The shallow profile and non-unity doublet ratio suggest that the sight line to AGN~A intercepts a lower-density, possibly more diffuse region of the same outflow traced in AGN~B, but at lower optical depth. 

Alternatively, the absorption feature in AGN~A may originate from a distinct intervening system unrelated to the intrinsic outflow seen in AGN~B.
In this case, the absence of C\,\textsc{iv} absorption at the systemic redshift of AGN~A supports the interpretation that only AGN~B exhibits an intrinsic outflow, while the two sight lines do not intersect the same absorbing structure, despite their projected 5.5~kpc separation.

\subsubsection{J0144--5745 and J2100+0612} 
These two systems are confirmed dual AGN at redshifts $z=0.791$ and $z=0.779$, with projected separations of 3.4~$\pm$ 0.1~kpc and 3.9~$\pm$~0.1~kpc, respectively. In the MUSE spectra at these redshifts, we detect the Mg\,\textsc{ii} $\lambda\lambda 2796,2803$ doublet at the bluest wavelength, H$\beta$ and [O\,\textsc{iii}] $\lambda\lambda 4959,5007$ at the reddest wavelength, and additional lines at intermediate wavelengths. In both systems, the brightest AGN (purple in Fig.~\ref{fig:spectra_AGN}) exhibits a broader H$\beta$ line, and in J2100+0612 a broader Mg\,\textsc{ii} is also observed. Additionally, the line ratios between the narrow lines differ between the two nuclei, with measured ratios of [O III] $\lambda$5007/H$\beta = 4.2 \pm 0.2$ for AGN A and $7.8 \pm 0.2$ for AGN B. No narrow absorption line systems are detected in either component of these two dual AGN. These differences suggest varying ionization conditions or metallicities between the nuclei

\subsubsection{J0637--4137} 
The system J0637–4137 at $z = 2.743 \pm 0.001$ is a strong candidate for a dual AGN, with the two components separated by a projected distance of $3.3\pm0.1$ kpc. The shape and EW of the C\textsc{iii}] line in the fainter component (light blue in Fig.~\ref{fig:spectra_AGN}) differ from those of the brighter nucleus, providing evidence that the system hosts two distinct AGN.

Multiple absorption systems are detected in both spectra (Fig.~\ref{fig:spectra_AGN}). All of these are intervening and thus not physically associated with the quasars themselves. We identify three such systems at $z_{\rm abs1} = 1.2523$, $z_{\rm abs2} = 2.1352$, and $z_{\rm abs3} = 2.3924$, each showing consistent metal absorption features (e.g., \ion{Mg}{ii}, \ion{Fe}{ii}, \ion{C}{iv}, \ion{Si}{ii}) in both components. The close match in absorption redshifts along the two sight-lines provides independent confirmation that the quasar pair is physically associated, complementing the evidence from the nearly identical emission-line redshifts and reducing the likelihood of a chance line-of-sight alignment.

\subsubsection{J1837--6711} 
The system J1837--6711 is a dual AGN. First, the narrow emission lines differ between the two components, with [O\,\textsc{ii}], [Ne\,\textsc{iii}], H$\epsilon$, H$\delta$, and H$\gamma$ showing distinct flux ratios. Since the NLR is spatially extended, microlensing cannot alter its flux, and these differences therefore require two separate narrow-line regions.
Second, the Fe\,\textsc{ii} emission complexes around Mg\,\textsc{ii} also differ. The blue side of Mg\,\textsc{ii} shows stronger absorption near $\sim$2700 \AA\ in the purple component, and higher bumps at $\sim$2900 \AA\ and $\sim$3200\ \AA. Fe\,\textsc{ii} originates in the outer BLR, which is only weakly and smoothly affected by microlensing. Thus, the observed Fe\,\textsc{ii} variations cannot be produced by microlensing of a single BLR, and instead imply intrinsically distinct BLR conditions.
Taken together, these differences in the spectral properties cannot be explained by microlensing of a single quasar. They instead strongly indicate the presence of two active nuclei in the same system, consistent with an ongoing galaxy merger at $z=1.120$.

\begin{figure}
    \centering
    \includegraphics[width=0.95\linewidth]{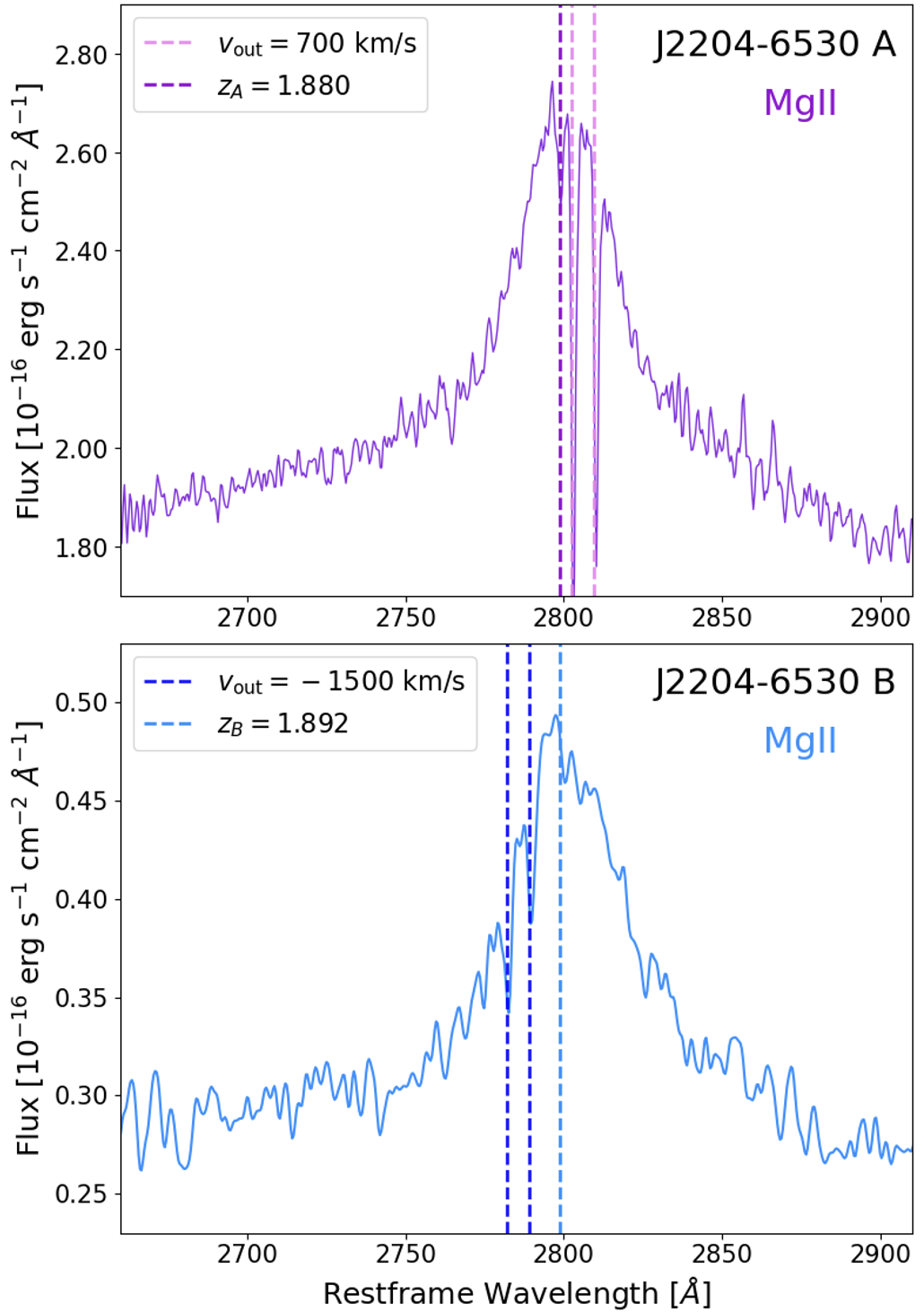}
    \caption{Enlargement of the Mg\,\textsc{ii} spectral region of J2204--6530 (AGN A in the top panel and AGN B in the bottom one). Both AGN show the  Mg\,\textsc{ii} doublet absorption lines at rest-frame wavelengths of 2796~$\AA$ and 2803~$\AA$. In each panel, the redshift of each QSO and the NAL velocity relative to its systemic redshift are reported, assuming that the two outflows are independent.}
    
    \label{fig:J2204_zoom}
\end{figure}

\subsubsection{J2204--6530} 
J2204--6530 is confirmed as a dual AGN system with a projected separation of $3.6$~kpc. 
The two components show clear differences in velocity, line centroids, and line profiles, consistent with the presence of two distinct active nuclei.

Using the spectral decomposition and fitting procedure described in Sect.~\ref{sec:analysis}, we measured redshifts of $z_A = 1.880$ and $z_B = 1.892$. These values are based on broad emission lines (C III] and Mg II; the latter is partially affected by absorption), as no narrow lines suitable for precise systemic measurements are available in the spectral range (See Fig.~\ref{fig:spectra_AGN}). Accounting for these limitations, we estimate a relative velocity of $\sim 1236 \pm 250$ km s$^{-1}$ between the two components.\\
This velocity shift cannot be explained within a gravitational lensing scenario. In lensed systems, the spectra of multiple images are expected to be nearly identical, as the typical time delays (of order days; \citealt{Treu10b}) are too short to produce such large differences in emission-line centroids and profiles through intrinsic variability.

Although the measured offset lies at the high end of the velocity distribution observed for close dual AGN, it remains within the range reported in the literature (e.g., \citealt{Bigmac25}), particularly when accounting for the large uncertainties associated.

As shown in Fig.~\ref{fig:J2204_zoom}, both AGN display intrinsic NALs at different velocities. In both spectra, the \ion{Mg}{ii} $\lambda\lambda 2796,2803$ doublet is detected in absorption, although at different observed wavelengths. 
For AGN~A, the \ion{Mg}{ii} NALs are redshifted by $700 \pm 45$~km~s$^{-1}$ relative to its systemic redshift, while in AGN~B the absorption appears at $-1500 \pm 300$~km~s$^{-1}$ relative to AGN~B, and at $-250 \pm 50$~km~s$^{-1}$ when referred to the systemic redshift of AGN~A.\\
This configuration can be interpreted in two ways: each AGN hosts its own independent outflow with distinct velocities; a single outflow (likely associated with one AGN) intersects the line of sight of both nuclei, producing absorption at different projected velocities depending on geometry and relative orientation. While these velocities are consistent with outflowing gas, the current data do not allow a precise characterization of the outflow properties. Distinguishing between alternative scenarios requires spatially resolved kinematic analysis of the absorbing gas, which is beyond the scope of this work and will be addressed in future follow-up studies.  
We note that inflowing gas is unlikely to explain the observed absorption, as the measured velocity offsets are mostly blueshifted relative to the systemic redshifts, and redshifted absorption expected from inflows is not observed. The line profiles and partial coverage signatures also favor an interpretation as AGN-driven outflows rather than accreting material.

\subsection{Double lensed AGN}
\label{sec:Lensed AGNystems}
In this section we present the systems classified as doubly lensed AGN from our data and analysis.
Out of the sixteen double AGN systems analyzed, ten have consistent redshifts between their components and show minor spectral differences that are fully consistent with microlensing effects (see Fig. \ref{fig:spectra_AGN_lensed}). In these cases, variations appear primarily in the continuum flux and, in some instances, in the equivalent widths of broad emission lines, while narrow lines remain unaffected. No intrinsic differences between the images are required to explain the observations.
These signatures are expected for gravitational microlensing, where compact objects in the lensing galaxy (e.g., stars) preferentially magnify the most compact emission regions, such as the accretion disk and the high-ionization portion of the broad line region \citep{Wambsganss06}. Consequently, the blue continuum and high-ionization lines can be more strongly magnified than the red continuum or low-ionization lines \citep[e.g.,][]{mosquera2013,jimenezvicente2015}. Narrow emission lines, emitted on scales of hundreds to thousands of parsecs, are essentially immune to microlensing effects.
Using the macro--micro decomposition (MmD) method \citep{Sluse07} on all available emission lines, we obtained detailed estimates of $M$ and $\mu$ for nearly all sources in the sample. Systems with values consistent with literature expectations ($\mu \sim 1.1$--1.5 for microlensed continuum regions, while narrow-line regions remain unaffected) are classified as lensed AGN. The observed differences in these ten systems are therefore fully explained by microlensing alone, supporting their classification as lensed AGN candidates.

The systems J0259--0901, J0551--4629, J1447--0501 and J2257--6555 show nearly identical spectra from both images, with no significant differences detected. This strongly supports their classification as lensed AGN, with each pair of components corresponding to multiple images of the same source. In the MUSE datacube of J0551--4629, the lensing galaxy is directly detected and is highlighted in red as component C in Fig.~\ref{fig:cubes_AGN}, providing a definitive confirmation of the lensing nature of the system.

J0802+1944, J2344--4259, and J2353--0655 show some differences in the continuum shape, with marginally significant effects on the broad or narrow emission lines within the measurement flux errors. In these cases, microlensing can primarily account for the observed variations in the continuum, driving the lensed classification of the systems.

While J0034--1623, J0317--5604, and J0404--2836 exhibit differences in their Mg\,\textsc{ii} emission line profiles, these variations are quantified by the differential microlensing factor $\mu = A/M$, where $A$ is the continuum flux ratio and $M$ the Mg\,\textsc{ii} line flux ratio between the images. The resulting $\mu$ values agree with differential amplification expected from the literature \citep{Sluse07,Sluse12}.
For J0034--1623, $A \approx 5.60$ and $M \approx 6.21$, yielding $\mu \approx 0.90$, indicating that Mg\,\textsc{ii} is de-magnified relative to the compact continuum source and explaining the observed differences in line shape and EW.
For J0317--5604 ($A \approx 1.57$, $M \approx 1.46$, $\mu \approx 1.07$) and J0404--2836 ($A \approx 1.82$, $M \approx 1.73$, $\mu \approx 1.05$), the Mg\,\textsc{ii} BLR is only marginally affected. While physically the same in both images, microlensing induces small variations in the EWs due to the different amplification of the continuum versus the more extended BLR. Narrow lines in J0404--2836 remain unaffected ($\mu \approx 1$), confirming that the observed differences in broad lines are due to microlensing.
Although dual AGN could, in principle, produce differences in Mg\,\textsc{ii}, this analysis strongly supports microlensing as the main cause.

Several of the systems in our sample show narrow absorption features at redshifts lower than the quasar systemic redshift. The fact that these absorbers lie at lower redshifts than the background AGN indicates that they are produced by foreground structures along the line of sight rather than by intrinsic outflows from the quasar.
For instance, J0034--1623 shows a strong absorber at $z = 1.6339$, while J0317--5604 hosts three intervening systems at $z = 1.4204$, $1.6146$, and
$1.6372$. Similar features are found in J0404--2836 ($z = 1.1231$, $1.2510$,
$1.3038$), and in J0551--4629 ($z = 0.7903$, $1.0151$), among others. The presence of the same absorption systems in both images confirms that the light paths of the two images cross the same foreground structures, supporting the gravitational lens interpretation.

In contrast, two systems show absorption features that are not intervening but intrinsic to the quasar. In J1447--0501, the Mg\,\textsc{ii} doublet
($\lambda_\mathrm{rest}=2796, 2803$\,\AA) is observed in absorption, with a velocity of $\sim 200 \pm 50$\,km\,s$^{-1}$ in both images, indicating a moderate quasar-driven outflow. 
In J2344--4259, several high-ionization lines show stronger blueshifted absorption, including the N\,\textsc{v} doublet ($\lambda_\mathrm{rest}=1238.82, 1242.80$\,\AA) and the C\,\textsc{iv} doublet ($\lambda_\mathrm{rest}=1548.20, 1550.77$\,\AA), with velocities of $-2748 \pm 40$\,km\,s$^{-1}$. These features are also detected in both components, confirming their intrinsic origin and therefore the lensed nature of the system, rather than indicating distinct physical AGN.

\subsection{Quadruple lensed AGN}
\label{sec:Quads}
From our GMP selection procedure (Sect. \ref{sec:target}), we identified three systems that each exhibit four AGN images (quads; see Fig. \ref{fig:cubes_quads}). For each system, we extracted both the images and the spectra of the four AGN. In one case, we were also able to isolate the spectrum of the lensing galaxy.

We decided to observe even those systems previously known as quads from imaging in order to obtain MUSE spatially resolved spectroscopy. When the lens is sufficiently bright, this allowed us to directly detect and resolve both the lensing galaxy and the multiple AGN components. The high angular resolution provided by AO assisted MUSE observations ($\sim 0.1''$) is critical for accurate modeling of such small-scale lens systems. Moreover, this approach ensures that the measurement of lensing prevalence among spectroscopic GMP candidates is independent of prior assumptions.

\subsubsection{J0530--3730}
This system was identified as a quads candidate by \cite{Delchambre19} based on a search over \textit{Gaia} DR2 \citep{GaiaDR2}. It was also selected using the method described by \cite{Ostrovski17} and \cite{Lemon17} as a triple detection around a photometric quasar candidate, and through the GMP technique \citep{Mannucci22}, because AGN A and C are separated by 0.16$''$ and AGN A and B by 0.67$''$ (see Fig. \ref{fig:cubes_quads}). It was confirmed as a quasar at $z = 2.838$ from NTT spectra \citep{Anguita18}.\\
This system was modeled using the automated time-delay cosmography pipeline by \cite{Schmidt23}, based on high-resolution HST multi-band imaging and informative priors. The method provides accurate strong-lens mass modeling, deriving key lens parameters, magnifications, and time-delay predictions. 

Using our data, we measured a circle enclosing all four AGN images with a radius of $0.531 \pm 0.009''$, consistent with previous estimates \citep{Schmidt23}. Thanks to the high spatial resolution of our observations, we are able, for the first time, to simultaneously spatially and spectrally resolve all four AGN, detecting in all of them the same absorption systems at $z=1.0054 \pm 0.0002$ and $z=1.2641 \pm 0.0002$.

\subsubsection{J0957--2242}
J0957--2242 is a newly discovered quadruple lensed system observed in this Large Program , as it is exclusively GMP-selected. MUSE data reveal the four images of the AGN at $z = 2.070 \pm 0.005$, as well as the foreground lens galaxy, which is detected in emission (component E in Fig. \ref{fig:cubes_quads}). 
We extract the spectra of the AGN from circular apertures of 5 px (0.125$''$) radius and that of the lens galaxy from a smaller aperture of 3 px (0.075$''$) radius (see the bottom panels of Fig.~\ref{fig:spectra_quads}). The smaller aperture for the lens galaxy minimizes contamination from the much brighter AGN emission, while still capturing the central light of the deflector.
We estimate an Einstein radius of $0.32''$ ($\sim 1.7$ kpc), making this one of the most compact quadruply imaged AGN currently known, with an emission-line galaxy acting as the lens (D'Amato in prep.).

\begin{figure}
    \centering
    \includegraphics[width=1\linewidth]{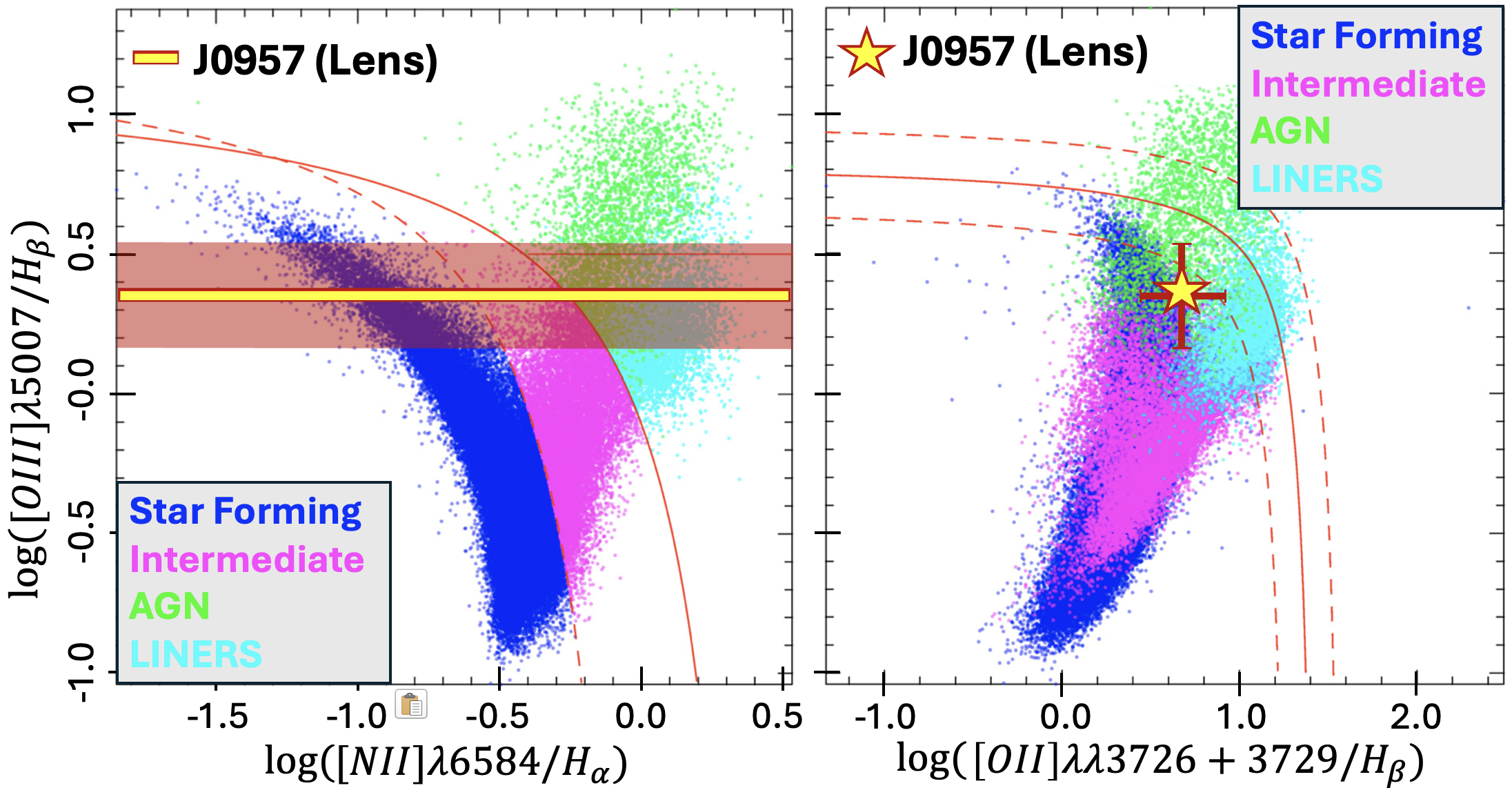}
    \caption{BPT diagram adapted from \cite{Lamareille10}  showing the positions of the local SDSS galaxies. In the left panel, the yellow strip shows the range of [N,\textsc{ii}]$\lambda$6584/H$\alpha$ values corresponding to the observed [O,\textsc{iii}]/H$\beta$ ratio (the dashed red area indicates the measurement uncertainty). In the right panel, the yellow star marks the position of the lens galaxy in the [O,\textsc{iii}]/H$\beta$ versus [O,\textsc{ii}]/H$\beta$ diagram, which lies very close to the separation line between star-forming galaxies (blue) and AGN (green and cyan).} 
    \label{fig:BPT}
\end{figure}

From the MUSE spectra, we estimate the redshift of the lens source to be $z=0.492 \pm 0.001$, we detect H$\beta$, the [O\,\textsc{iii}]~$\lambda\lambda4959,5007$ doublet, and the [O\,\textsc{ii}]~$\lambda\lambda3726,3729$ doublet (Fig.~\ref{fig:spectra_quads}), and we measured the individual line fluxes.
These line fluxes can be used to discriminate between photoionization by star formation and AGN activity. 
However, as shown by Baldwin–Phillips–Terlevich (BPT) emission-line diagnostic diagrams \citep{Baldwin81} in Fig.~\ref{fig:BPT} adapted from \cite{Lamareille10}, the available lines are insufficient to confirm or exclude the presence of an AGN. 
The $\log_{10}$[O\,\textsc{iii}]/H$\beta$ = $0.36\pm0.2$, consistent with both star-forming and AGN-dominated ionization, 
while the $\log_{10}[\mathrm{O\,II}]/\mathrm{H}\beta = 0.65 \pm 0.2$, placing the galaxy near the separation line. 
Accurate classification would require the [N\,\textsc{ii}]~$\lambda6584$/H$\alpha$ ratio, which is not covered by the MUSE spectra, following the diagnostic scheme of \citet{Lamareille10}.

The ratio of the [O,\textsc{ii}] lines, O,\textsc{ii}~$\lambda 3726/\lambda 3729 = 0.85 \pm 0.10$, implies an electron density of $n_e \simeq 263 \pm 50$~cm$^{-3}$, assuming an electron temperature of $T_e = 10^4$~K \citep{Osterbrock06}. These relations are valid for ionized gas typical of star-forming galaxies, where collisional excitation dominates. Considering a plausible range of electron temperatures for low-$z$ star-forming galaxies, $T_e = 7000$–$18000$~K, the inferred density varies only moderately, from $n_e \simeq 223$~cm$^{-3}$ to $n_e \simeq 313$~cm$^{-3}$, remaining within the quoted uncertainties.

We computed the oxygen abundance using the $R23$ metallicity indicator, defined as $R23 = (\mathrm{[O\,\textsc{ii}]}\,\lambda\lambda3726,3729 + \mathrm{[O\,\textsc{iii}]}\,\lambda4959,\lambda5007)/\mathrm{H\beta}$ \citep{Curti20}. From the measured fluxes we obtain $\log_{10}(R23)=0.91 \pm 0.20$. Using the empirical calibration of \citet{Maiolino08}, this corresponds to two possible oxygen abundances: $12 + \log(\mathrm{O/H}) = 7.85 \pm 0.20$ (low--metallicity branch) and $12 + \log(\mathrm{O/H}) = 8.44 \pm 0.20$ (high--metallicity branch). To break the degeneracy we computed the $R2$ and $R3$ ratios, obtaining $\log R2 = 0.65 \pm 0.15$ and $\log R3 = 0.55 \pm 0.15$. Both values place the source on the high-metallicity branch according to the trends of \citet{Maiolino08}. The agreement between $R2$ (dust-sensitive) and $R3$ (dust-insensitive) also indicates minimal nebular extinction. We therefore adopted the high-metallicity solution $12 + \log(\mathrm{O/H}) = 8.44 \pm 0.20$ consistent with moderately metal-rich ionized gas. 

Using the observed H$\beta$ luminosity and the \citet{Kennicutt98} calibration we estimate the star formation rate (SFR) $\mathrm{SFR}_{\mathrm{H}\beta} = 0.016 \pm 0.002\ M_\odot\ \mathrm{yr^{-1}}$ (uncorrected for dust).
Because $R2$ and $R3$ are in agreement, suggesting minimal dust extinction, this value should be regarded as a robust lower limit and is likely close to the true SFR.

Overall, the combination of these emission-line ratios and diagnostics indicates highly ionized gas 
with moderate electron density, providing meaningful constraints on the excitation mechanism 
and chemical enrichment even in the absence of H$\alpha$ and [N\,\textsc{ii}] coverage.

\begin{figure}
\centering
    \includegraphics[width=\linewidth]{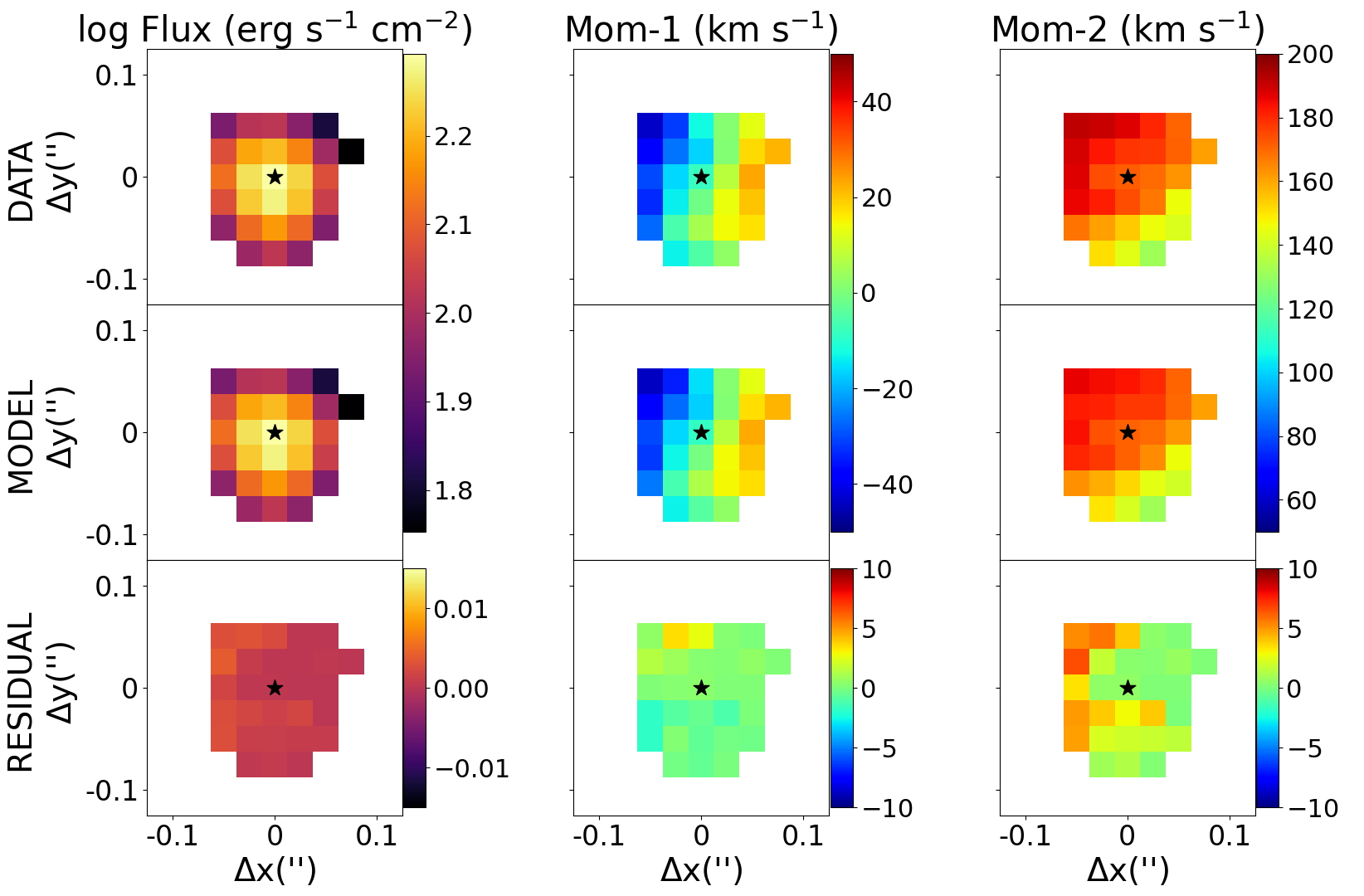}
    \caption{\MOKA \ rotating disk best-fit model of J0957-2242 traced by the [O\,\textsc{iii}]5007 transition. Panels show the observed (top), best-fit (center), and residual (bottom) moment maps. The residual maps were obtained by subtracting the model from the observed moment maps. The star marks the kinematic center, inferred as the peak flux emission. Pixels are masked at SN $\leq$ 4.}
    \label{fig:moka}

\end{figure}

The ionized gas kinematics traced by the [O\,\textsc{iii}]5007 transition reveal a tentative rotation pattern, as shown by moment maps in the upper panel in Fig. \ref{fig:moka}. To test whether a pure rotating disk is suited to reproduce the observed kinematic features we adopted the \MOKA \ kinematic model \citep{Marconcini23}. In particular, we assumed a circular thin disk with inner and outer radius of 0--0.15\arcsec ($\sim$ 0--0.6 kpc) and a disk height comparable with the instrumental PSF (i.e., 0.1\arcsec). In doing so, we assumed that the gas kinematics can be described as a purely rotating disk with constant circular velocity, with no radial motions. The free parameters of the fit are the inclination, the circular velocity and the disk PA. Our best fit model returns an inclination of 55$^{\circ} \pm$ 3$^{\circ}$, an intrinsic circular velocity of 380 $\pm$ 11 km s$^{-1}$ and a PA of 18$^{\circ} \pm$ 4$^{\circ}$. \\ As shown in Fig.~\ref{fig:moka} the inferred best-fit model is able to reproduce the observed ionized gas features with extremely high-accuracy and residuals $\leq$ 10 km s$^{-1}$ both in the line-of-sight velocity and velocity dispersion maps. Then we derived the dynamical mass profile from the deprojected circular velocity while assuming circular motion in a spherically symmetric potential:
\begin{equation}
    M = \frac{V_\mathrm{circ}^2\,R}{G},
\end{equation}
where $R$ is the galactocentric radius in kpc, and $G$ is is the gravitational constant. We computed the uncertainties on the dynamical mass propagating the uncertainties of the circular velocity. This procedure provides a direct estimate of the enclosed dynamical mass, under the assumption that the gas trace circular motion in the plane of the disk and that non-circular motions are negligible. As a result, we estimate a dynamical mass within the disk maximum extension (0.15\arcsec) of $M$ = 3.1 $\pm$0.2 $\times$ 10$^{10} M_{\odot} $.

\subsubsection{J1116--0657}

The quasar J1116--0657 was identified as a quadruple gravitational lens by \cite{blackburne08}, with a source redshift of $z_S = 1.235$. They detect a faint lensing galaxy estimated at $z_L \sim 0.7$. A simple lens model reproduces the observed image positions but fails to match the flux ratios, possibly due to substructure or microlensing. The estimated time delays between the images are on the order of one day.

Our measurements are in agreement with \cite{blackburne08}, with an Einstein radius of $0.341 \pm 0.005''$ (see Fig. \ref{fig:cubes_quads}). Importantly, we were able to extract spectra for all four AGN images individually, providing the first resolved spectral information for this system and enabling the identification of spectral features in each component.

\section{Distribution of dual and lensed systems}
\label{sec:distribution}

Building on the classifications derived for each target, we next analyzed the population properties of the GMP-selected systems observed with MUSE–NFM. For this purpose we consider the full sample with available MUSE follow-up, which includes the 19 systems presented in this work together with the seven additional sources from \citealt{Scialpi24}. This consolidated sample provides a uniform basis to investigate the redshift, magnitude, and separation distributions of dual and lensed AGN, and understand the selection effects introduced by both the GMP method and the MUSE observations.

The sample is limited to projected separations $0.15'' < \rm sep < 0.8''$. The lower limit corresponds to the sensitivity of the GMP method, while the upper limit reflects our focus on systems in the regime of a few kiloparsecs, where simulations predict that the SMBHs are likely embedded within a common merger remnant or in the late stages of galaxy coalescence \citep{Volonteri22}.
At the median redshift of the sample ($z \sim 1.5$), these values correspond to projected physical separations of $\sim$1.7--6.8~kpc.  \\ Only systems with $z>0.5$ are included, ensuring that the GMP technique identifies spatially distinct components and not single extended sources.
\begin{figure}[h!]
    \centering
    \includegraphics[width=1\linewidth]{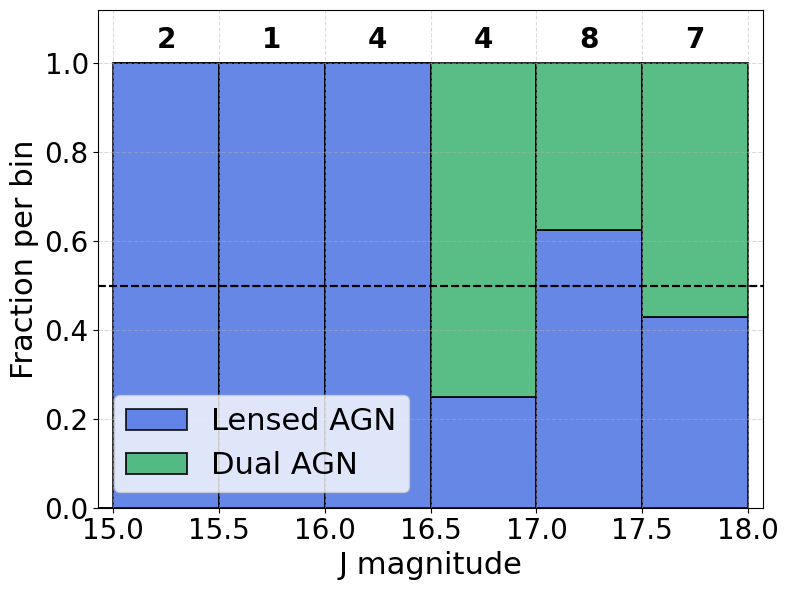}
    \includegraphics[width=1\linewidth]{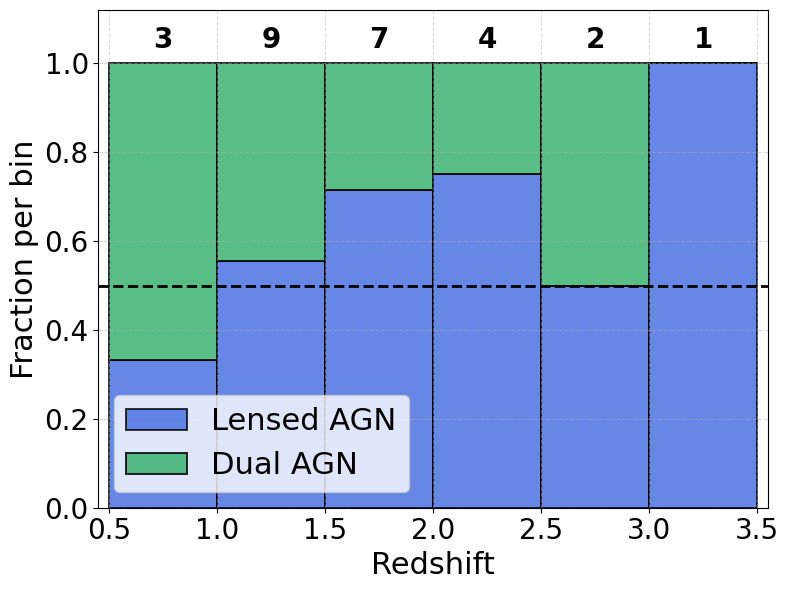}

    \caption{Top panel: Fraction of lensed systems (blue) and dual AGN (green) as a function of $J$-band magnitude for the MUSE sample (this work and \citealt{Scialpi24}). Bottom panel: Same but as a function of redshift. Data are binned in intervals of 0.5~mag (top) and 0.5 (bottom). The number above each column indicates the total number of objects in that bin. The horizontal dashed line marks the 50\% level for reference.} 
    \label{fig:hist_J}
\end{figure}

Figure~\ref{fig:hist_J} shows the fraction of lensed systems (blue) and dual AGN (green) as a function of $J$-band magnitude (top panel) and redshift (bottom panel) for the MUSE sample (this work and \citealt{Scialpi24}). In both cases, the fractions are computed per bin so that the sum of lensed and dual systems equals one, providing a normalized view of the relative contributions removing the selection effects.
Lensed systems clearly dominate the bright end of the sample ($J \lesssim 16.5$), while dual AGN become increasingly prevalent at fainter magnitudes, highlighting the strong influence of magnification bias.
The lensing fraction peaks around $z \sim 2.1$, reflecting the combination of survey selection effects (also relevant for dual AGN), the increasing lensing efficiency due to the ratio of angular diameter distances of lenses and sources, and the quasar luminosity function.

\smallskip
\noindent
These distributions can be interpreted in the broader context of dual AGN searches. 
Relative to previous GMP studies \citep{Mannucci22, Mannucci23,Ciurlo23, Scialpi24}, the present MUSE–NFM sample confirms similar candidates over comparable magnitude, separation, and redshift ranges, but increases the number of spectroscopically classified systems, enabling a more robust characterization of the population properties.

Compared to varstrometry-based selections (e.g., VODKA and VADAR; \citealt{Shen19, Hwang20, Schwartzman24_vastrometry, Schwartzman25_vastrometry, gross25_vodka}), the GMP method relies on a different observational signature in \textit{Gaia} data. Varstrometry is sensitive to unresolved systems through astrometric jitter due to flux variability \citep{Hwang20}, whereas GMP identifies multiple peaks in the \textit{Gaia} light profile, consistent with multiple components \citep{Mannucci22}.
Both techniques primarily select unobscured, optically bright quasars and, in practice, probe broadly similar regimes in redshift and physical separation.
GMP candidates are generally found at $z \gtrsim 0.5$ and at kpc-scale separations. Varstrometry, in principle, is sensitive to even smaller separations and lower-redshift systems \citep{Hwang20}; however, current observational results indicate that confirmed dual AGN from varstrometry largely occupy similar redshift ranges and comparable -- or in some cases larger -- projected separations than those identified by GMP.
Available data suggest that the contamination rates reported in this work and by \cite{Gross2024} are comparable for both techniques, with star–quasar projections affecting each sample at similar levels. Notably, only two sources from the GMP sample also are varstrometry candidates. On the other hand, all currently known varstrometry-selected systems at $0.5 \lesssim z \lesssim 3.5$ are also identified by GMP, provided the secondary AGN has $G < 20.5$, which is the completeness limit of the GMP selection \citet{Mannucci23}.
Finally, the redshift distribution of the dual AGN identified here is broadly consistent with expectations from clustered quasar surveys \citep[e.g.,][]{Hennawi06, Hennawi10}, which find an increasing incidence of close quasar pairs at $z \gtrsim 1$, although typically at larger projected separations.

\section{Conclusions}
\label{sec:summary}

We have presented the first-year results of our MUSE Large Program targeting dual and lensed AGN candidates selected via the GMP technique \citep{Mannucci22}. The first observed dataset comprises 30 targets spanning redshifts $0.7 \lesssim z \lesssim 3.2$ and projected separations of $\sim0.24\arcsec$–$0.82\arcsec$, corresponding to $\sim 1.5$–7~kpc at typical redshifts.
Our observations have led to the discovery of six new dual AGN systems in the redshift range $z \sim 0.9$–$2.5$ and with projected separations of $<7$~kpc as well as 13 lensed AGN (including three quadruply imaged systems) in the redshift range $z \sim 0.7$–$3.2$. \\
Based on literature compilations (e.g., \citealt{Mannucci22, Chen22a, Bigmac25}), only $\sim 27$ dual AGN are currently confirmed at such close separations across the redshift range $0.5 \lesssim z \lesssim 3.5$, including the six systems presented here. The dual AGN in this work therefore represent $\sim 22\%$ of the known population in this separation and redshift regime.

Combining this work with previous GMP-based confirmations \citep{Mannucci22, Mannucci23, Ciurlo23, Scialpi24},~15~out~of the~27~currently known dual AGN at $0.5<z<3.5$ with separations $<7$ kpc were originally identified through GMP selection (56\% of the sample).~A~larger number of systems~(18~out~of~27)~satisfy the GMP criteria,~although~three~of these were first observed or identified through other selection methods \citep{Junkkarinen01,Scheckter17,Chen23a}.
A similar consideration applies to strongly lensed AGN at subarcsec separations.~While several hundred lensed quasars are known overall, only a small fraction have image separations below~$\sim0.7''$~with spatially resolved spectroscopic characterization.~The~13~lensed systems identified here therefore represent a substantial addition to the sample of compact-separation~lensed AGN at cosmic noon accessible~to~IFU-based~studies.

Using the MUSE data, we systematically classified all targets. The six dual AGN exhibit clear spectral differences between the two nuclei in emission-line profiles and line ratios, with projected separations of~3–6~kpc and relative velocities up to $\sim 1200$~km~s$^{-1}$. Lensed AGN were identified through flux ratios, spectral similarity, and the detection of lensing features in the MUSE datacube. Eleven systems correspond to AGN+star projections, reflecting a low contamination fraction and the reliability of the GMP selection.

We also systematically measured redshifts for both intrinsic and intervening NALs in all spectra.
Intervening NALs provide a unique probe of the small-scale structure of the CGM and IGM along closely spaced sight lines ($\lesssim 10$ kpc), tracing the distribution, kinematics, metallicity, and ionization state of cold gas clumps \citep{mandalker19,Dutta24}. Intrinsic NALs instead offer diagnostics of AGN outflows and dual activity (e.g., \citealt{Hamann12}).

The distributions of targets in angular separation, $J$-band and \textit{Gaia} $G$ magnitudes, and redshift highlight the effectiveness of the GMP selection and the capabilities of the MUSE AO system. Lensed systems are predominantly detected at the bright end of the $J$-band distribution, while dual AGN are increasingly identified at fainter magnitudes.

This first-year dataset demonstrates the power of combining GMP preselection with MUSE IFU spectroscopy to systematically identify and characterize dual and lensed AGN at subarcsec separations. As an optical~\textit{Gaia}-based selection, the GMP method is primarily sensitive to unobscured, optically bright AGN. Spatially resolved spectroscopy allows us to disentangle close pairs, measure redshifts and line properties, and detect both intrinsic and intervening absorption systems. 
The full~MUSE~Large~Program, which will expand the sample to~$\sim 150$~targets,~will enable statistically robust studies of dual AGN occurrence as well as their physical properties, including black hole masses and~luminosities, in relation to separation, redshift, and magnitude.~
Future~analyses will include detailed modeling of outflows, absorber kinematics, and lensing configurations, providing unprecedented insights into the dynamics and environments of AGN at cosmic~noon.

In summary, the combination of GMP selection and MUSE IFU observations offers a highly efficient strategy for exploring the sub-arcsec AGN population. This approach bridges the gap between photometric identification and detailed spectroscopic characterization, and it establishes a solid foundation for future large-scale studies of dual and strongly lensed AGN.

\section*{Data availability}The observational data used in this work are available from the European Southern Observatory archive: {\color{blue}https://archive.eso.org.}

\begin{acknowledgements}

This publication  was produced while attending the PhD program in Space Science and Technology at the University of Trento, Cycle XXXIX, with the support of a scholarship financed by the Ministerial Decree no. 118 of 2nd march 2023, based on the NRRP - funded by the European Union - NextGenerationEU - Mission 4 "Education and Research," Component 1 "Enhancement of the offer of educational services: from nurseries to universities” - Investment 4.1 “Extension of the number of research doctorates and innovative doctorates for public administration and cultural heritage”.    
\\
Based on observations collected at the European Southern Observatory (ESO) programs 109.22WA, 110.23SM, 112.25ET, and 114.27BY with VLT/MUSE,
and 109.22W4, 112.25CT, 113.26TB, 113.26D7, 114.27FC, and 115.28CQ with NTT/EFOSC2 and VLT/FORS2.
The authors sincerely thank the ESO staff in Garching and at Paranal for their availability, patience, and support during both the preparation and execution of the observations. In particular, we are grateful to the ESO instrument experts for their guidance on the MUSE AO system, which was invaluable in planning the observing strategy and ensuring optimal data quality.

Based on observations made with the Italian Telescopio Nazionale \textit{Galileo} (TNG) operated on the island of La Palma by the Fundación Galileo Galilei of the INAF (Istituto Nazionale di Astrofisica) at the Spanish Observatorio del Roque de los Muchachos of the Instituto de Astrofisica de Canarias.

This work has made use of data from the European Space Agency (ESA) mission \textit{Gaia} (https://www.cosmos.esa. int/gaia), processed by the \textit{Gaia} Data Processing and Analysis Consortium (DPAC, https://www.cosmos.esa.int/ web/gaia/dpac/consortium). Funding for the DPAC has been provided by national institutions, in particular the institutions participating in the \textit{Gaia} Multilateral Agreement.
 We thank Elena Pancino for her valuable support regarding the \textit{Gaia} Archive and properties.\\ 
We acknowledge financial contribution from INAF Large Grant “Dual AGN and binary supermassive black holes in the multimessenger era: from galaxy mergers to gravitational waves” (Bando Ricerca Fondamentale INAF 2022), from the INAF project "VLT-MOONS" CRAM 1.05.03.07, from the French National Research Agency (grant ANR-21-CE31-0026, project MBH\_waves), from the INAF Large Grant 2022 "The metal circle: a new sharp view of the baryon cycle up to Cosmic Dawn with the latest generation IFU facilities," from INAF Large Grant "The Quest for dual and binary massive black
holes in the gravitational wave era," (Bando Ricerca Fondamentale INAF 2024), the MIRACLE INAF 2024 GO grant "A JWST/MIRI MIRACLE: Mid-IR Activity of Circumnuclear Line Emission"

We also acknowledge financial support by the PRIN-MUR project ``PROMETEUS''  202223XDPZM financed by the European Union -  Next Generation EU, Mission 4 Component 1, CUP B53D23004750006 and C53D2300080006.

AF and EB acknowledge financial support from the Ricerca Fondamentale INAF 2024 under project 1.05.24.07.01 MiniGrant RSN1.

MP acknowledges support through the grants PID2021-127718NB-I00, PID2024-159902NA-I00, and RYC2023-044853-I, funded by the Spain Ministry of Science and Innovation/State Agency of Research MCIN/AEI/10.13039/501100011033 and El Fondo Social Europeo Plus FSE+.

SC and GV acknowledge support from European Union's HE ERC Starting Grant No. 101040227 - WINGS.\\

We acknowledge financial support from the ASI-INAF agreements n. 2024-36-HH.1-2025  and 2025-29-HH.0.

Finally, we thank the referees for their helpful comments and careful review, which improved the clarity of the paper.
\end{acknowledgements}

\bibliographystyle{aa}
\bibliography{references}

\begin{thebibliography}{136}
\expandafter\ifx\csname natexlab\endcsname\relax\def\natexlab#1{#1}\fi

\bibitem[{{Agazie} {et~al.}(2023){Agazie}, {Anumarlapudi}, {Archibald}, {Arzoumanian}, {Baker}, {B{\'e}csy}, {Blecha}, {Brazier}, {Brook}, {Burke-Spolaor}, {Burnette}, {Case}, {Charisi}, {Chatterjee}, {Chatziioannou}, {Cheeseboro}, {Chen}, {Cohen}, {Cordes}, {Cornish}, {Crawford}, {Cromartie}, {Crowter}, {Cutler}, {Decesar}, {Degan}, {Demorest}, {Deng}, {Dolch}, {Drachler}, {Ellis}, {Ferrara}, {Fiore}, {Fonseca}, {Freedman}, {Garver-Daniels}, {Gentile}, {Gersbach}, {Glaser}, {Good}, {G{\"u}ltekin}, {Hazboun}, {Hourihane}, {Islo}, {Jennings}, {Johnson}, {Jones}, {Kaiser}, {Kaplan}, {Kelley}, {Kerr}, {Key}, {Klein}, {Laal}, {Lam}, {Lamb}, {Lazio}, {Lewandowska}, {Littenberg}, {Liu}, {Lommen}, {Lorimer}, {Luo}, {Lynch}, {Ma}, {Madison}, {Mattson}, {McEwen}, {McKee}, {McLaughlin}, {McMann}, {Meyers}, {Meyers}, {Mingarelli}, {Mitridate}, {Natarajan}, {Ng}, {Nice}, {Ocker}, {Olum}, {Pennucci}, {Perera}, {Petrov}, {Pol}, {Radovan}, {Ransom}, {Ray}, {Romano}, {Sardesai}, {Schmiedekamp}, {Schmiedekamp}, {Schmitz},
  {Schult}, {Shapiro-Albert}, {Siemens}, {Simon}, {Siwek}, {Stairs}, {Stinebring}, {Stovall}, {Sun}, {Susobhanan}, {Swiggum}, {Taylor}, {Taylor}, {Turner}, {Unal}, {Vallisneri}, {van Haasteren}, {Vigeland}, {Wahl}, {Wang}, {Witt}, {Young}, \& {Nanograv Collaboration}}]{Agazie2023a}
{Agazie}, G., {Anumarlapudi}, A., {Archibald}, A.~M., {et~al.} 2023, \apjl, 951, L8

\bibitem[{{Amaro-Seoane} {et~al.}(2023){Amaro-Seoane}, {Andrews}, {Arca Sedda}, {Askar}, {Baghi}, {Balasov}, {Bartos}, {Bavera}, {Bellovary}, {Berry}, {Berti}, {Bianchi}, {Blecha}, {Blondin}, {Bogdanovi{\'c}}, {Boissier}, {Bonetti}, {Bonoli}, {Bortolas}, {Breivik}, {Capelo}, {Caramete}, {Cattorini}, {Charisi}, {Chaty}, {Chen}, {Chru{\'s}li{\'n}ska}, {Chua}, {Church}, {Colpi}, {D'Orazio}, {Danielski}, {Davies}, {Dayal}, {De Rosa}, {Derdzinski}, {Destounis}, {Dotti}, {Du{\c{t}}an}, {Dvorkin}, {Fabj}, {Foglizzo}, {Ford}, {Fouvry}, {Franchini}, {Fragos}, {Fryer}, {Gaspari}, {Gerosa}, {Graziani}, {Groot}, {Habouzit}, {Haggard}, {Haiman}, {Han}, {Istrate}, {Johansson}, {Khan}, {Kimpson}, {Kokkotas}, {Kong}, {Korol}, {Kremer}, {Kupfer}, {Lamberts}, {Larson}, {Lau}, {Liu}, {Lloyd-Ronning}, {Lodato}, {Lupi}, {Ma}, {Maccarone}, {Mandel}, {Mangiagli}, {Mapelli}, {Mathis}, {Mayer}, {McGee}, {McKernan}, {Miller}, {Mota}, {Mumpower}, {Nasim}, {Nelemans}, {Noble}, {Pacucci}, {Panessa}, {Paschalidis}, {Pfister}, {Porquet},
  {Quenby}, {Ricarte}, {R{\"o}pke}, {Regan}, {Rosswog}, {Ruiter}, {Ruiz}, {Runnoe}, {Schneider}, {Schnittman}, {Secunda}, {Sesana}, {Seto}, {Shao}, {Shapiro}, {Sopuerta}, {Stone}, {Suvorov}, {Tamanini}, {Tamfal}, {Tauris}, {Temmink}, {Tomsick}, {Toonen}, {Torres-Orjuela}, {Toscani}, {Tsokaros}, {Unal}, {V{\'a}zquez-Aceves}, {Valiante}, {van Putten}, {van Roestel}, {Vignali}, {Volonteri}, {Wu}, {Younsi}, {Yu}, {Zane}, {Zwick}, {Antonini}, {Baibhav}, {Barausse}, {Bonilla Rivera}, {Branchesi}, {Branduardi-Raymont}, {Burdge}, {Chakraborty}, {Cuadra}, {Dage}, {Davis}, {de Mink}, {Decarli}, {Doneva}, {Escoffier}, {Gandhi}, {Haardt}, {Lousto}, {Nissanke}, {Nordhaus}, {O'Shaughnessy}, {Portegies Zwart}, {Pound}, {Schussler}, {Sergijenko}, {Spallicci}, {Vernieri}, \& {Vigna-G{\'o}mez}}]{AmaroSeoane2023}
{Amaro-Seoane}, P., {Andrews}, J., {Arca Sedda}, M., {et~al.} 2023, Living Reviews in Relativity, 26, 2

\bibitem[{{Anguita} {et~al.}(2018){Anguita}, {Schechter}, {Kuropatkin}, {Morgan}, {Ostrovski}, {Abramson}, {Agnello}, {Apostolovski}, {Fassnacht}, {Hsueh}, {Motta}, {Rojas}, {Rusu}, {Treu}, {Williams}, {Auger}, {Buckley-Geer}, {Lin}, {McMahon}, {Abbott}, {Allam}, {Annis}, {Bernstein}, {Bertin}, {Brooks}, {Burke}, {Carnero Rosell}, {Carrasco-Kind}, {Carretero}, {Cunha}, {D'Andrea}, {De Vicente}, {DePoy}, {Desai}, {Diehl}, {Doel}, {Flaugher}, {Garc{\'\i}a-Bellido}, {Gerdes}, {Gruen}, {Gruendl}, {Gschwend}, {Hartley}, {Hollowood}, {Honscheid}, {James}, {Kuehn}, {Lima}, {Maia}, {Miquel}, {Plazas}, {Sanchez}, {Scarpine}, {Smith}, {Soares-Santos}, {Sobreira}, {Suchyta}, {Tarle}, \& {Walker}}]{Anguita18}
{Anguita}, T., {Schechter}, P.~L., {Kuropatkin}, N., {et~al.} 2018, \mnras, 480, 5017

\bibitem[{{Arsenault} {et~al.}(2008){Arsenault}, {Madec}, {Hubin}, {Paufique}, {Stroebele}, {Soenke}, {Donaldson}, {Fedrigo}, {Oberti}, {Tordo}, {Downing}, {Kiekebusch}, {Conzelmann}, {Duchateau}, {Jost}, {Hackenberg}, {Bonaccini Calia}, {Delabre}, {Stuik}, {Biasi}, {Gallieni}, {Lazzarini}, {Lelouarn}, \& {Glindeman}}]{museAO_Arsenault}
{Arsenault}, R., {Madec}, P.~Y., {Hubin}, N., {et~al.} 2008, in Society of Photo-Optical Instrumentation Engineers (SPIE) Conference Series, Vol. 7015, Adaptive Optics Systems, ed. N.~{Hubin}, C.~E. {Max}, \& P.~L. {Wizinowich}, 701524

\bibitem[{{Arzoumanian} {et~al.}(2018){Arzoumanian}, {Baker}, {Brazier}, {Burke-Spolaor}, {Chamberlin}, {Chatterjee}, {Christy}, {Cordes}, {Cornish}, {Crawford}, {Thankful Cromartie}, {Crowter}, {DeCesar}, {Demorest}, {Dolch}, {Ellis}, {Ferdman}, {Ferrara}, {Folkner}, {Fonseca}, {Garver-Daniels}, {Gentile}, {Haas}, {Hazboun}, {Huerta}, {Islo}, {Jones}, {Jones}, {Kaplan}, {Kaspi}, {Lam}, {Lazio}, {Levin}, {Lommen}, {Lorimer}, {Luo}, {Lynch}, {Madison}, {McLaughlin}, {McWilliams}, {Mingarelli}, {Ng}, {Nice}, {Park}, {Pennucci}, {Pol}, {Ransom}, {Ray}, {Rasskazov}, {Siemens}, {Simon}, {Spiewak}, {Stairs}, {Stinebring}, {Stovall}, {Swiggum}, {Taylor}, {Vallisneri}, {van Haasteren}, {Vigeland}, {Zhu}, \& {NANOGrav Collaboration}}]{Arzoumanian18}
{Arzoumanian}, Z., {Baker}, P.~T., {Brazier}, A., {et~al.} 2018, \apj, 859, 47

\bibitem[{{Bacon} {et~al.}(2010){Bacon}, {Accardo}, {Adjali}, {Anwand}, {Bauer}, {Biswas}, {Blaizot}, {Boudon}, {Brau-Nogue}, {Brinchmann}, {Caillier}, {Capoani}, {Carollo}, {Contini}, {Couderc}, {Daguis{\'e}}, {Deiries}, {Delabre}, {Dreizler}, {Dubois}, {Dupieux}, {Dupuy}, {Emsellem}, {Fechner}, {Fleischmann}, {Fran{\c{c}}ois}, {Gallou}, {Gharsa}, {Glindemann}, {Gojak}, {Guiderdoni}, {Hansali}, {Hahn}, {Jarno}, {Kelz}, {Koehler}, {Kosmalski}, {Laurent}, {Le Floch}, {Lilly}, {Lizon}, {Loupias}, {Manescau}, {Monstein}, {Nicklas}, {Olaya}, {Pares}, {Pasquini}, {P{\'e}contal-Rousset}, {Pell{\'o}}, {Petit}, {Popow}, {Reiss}, {Remillieux}, {Renault}, {Roth}, {Rupprecht}, {Serre}, {Schaye}, {Soucail}, {Steinmetz}, {Streicher}, {Stuik}, {Valentin}, {Vernet}, {Weilbacher}, {Wisotzki}, \& {Yerle}}]{Muse_bacon}
{Bacon}, R., {Accardo}, M., {Adjali}, L., {et~al.} 2010, in Society of Photo-Optical Instrumentation Engineers (SPIE) Conference Series, Vol. 7735, Ground-based and Airborne Instrumentation for Astronomy III, ed. I.~S. {McLean}, S.~K. {Ramsay}, \& H.~{Takami}, 773508

\bibitem[{Bacon {et~al.}(2010)Bacon, Accardo, Adjali, Anwand, Bauer, Biswas, Blaizot, Boudon, Brau-Nogue, Brinchmann, Caillier, Capoani, Carollo, Contini, Couderc, Daguis{\'{e}}, Deiries, Delabre, Dreizler, Dubois, Dupieux, Dupuy, Emsellem, Fechner, Fleischmann, Fran{\c{c}}ois, Gallou, Gharsa, Glindemann, Gojak, Guiderdoni, Hansali, Hahn, Jarno, Kelz, Koehler, Kosmalski, Laurent, Floch, Lilly, Lizon, Loupias, Manescau, Monstein, Nicklas, Olaya, Pares, Pasquini, P{\'{e}}contal-Rousset, Pell{\'{o}}, Petit, Popow, Reiss, Remillieux, Renault, Roth, Rupprecht, Serre, Schaye, Soucail, Steinmetz, Streicher, Stuik, H, Vernet, Weilbacher, Wisotzki, \& Yerle}]{Bacon10}
Bacon, R., Accardo, M., Adjali, L., {et~al.} 2010, in {SPIE} Proceedings, ed. I.~S. McLean, S.~K. Ramsay, \& H.~Takami ({SPIE})

\bibitem[{{Baldwin} {et~al.}(1981){Baldwin}, {Phillips}, \& {Terlevich}}]{Baldwin81}
{Baldwin}, J.~A., {Phillips}, M.~M., \& {Terlevich}, R. 1981, \pasp, 93, 5

\bibitem[{{Banerjee} \& {Abel}(2021)}]{Banerjee21}
{Banerjee}, A. \& {Abel}, T. 2021, \mnras, 500, 5479

\bibitem[{{Barrows}(2023)}]{Barrows23}
{Barrows}, S. 2023, in American Astronomical Society Meeting Abstracts, Vol. 242, American Astronomical Society Meeting Abstracts \#242, 406.06

\bibitem[{{Begelman} {et~al.}(1980){Begelman}, {Blandford}, \& {Rees}}]{Begelman80}
{Begelman}, M.~C., {Blandford}, R.~D., \& {Rees}, M.~J. 1980, \nat, 287, 307

\bibitem[{{Blackburne} {et~al.}(2008){Blackburne}, {Wisotzki}, \& {Schechter}}]{blackburne08}
{Blackburne}, J.~A., {Wisotzki}, L., \& {Schechter}, P.~L. 2008, \aj, 135, 374

\bibitem[{{Blecha} {et~al.}(2018){Blecha}, {Snyder}, {Satyapal}, \& {Ellison}}]{Blecha18}
{Blecha}, L., {Snyder}, G.~F., {Satyapal}, S., \& {Ellison}, S.~L. 2018, \mnras, 478, 3056

\bibitem[{{Bolton} {et~al.}(2008){Bolton}, {Burles}, {Koopmans}, {Treu}, {Gavazzi}, {Moustakas}, {Wayth}, \& {Schlegel}}]{Bolton08}
{Bolton}, A.~S., {Burles}, S., {Koopmans}, L. V.~E., {et~al.} 2008, \apj, 682, 964

\bibitem[{{Buzzoni} {et~al.}(1984){Buzzoni}, {Delabre}, {Dekker}, {Dodorico}, {Enard}, {Focardi}, {Gustafsson}, {Nees}, {Paureau}, \& {Reiss}}]{efosc2_ntt}
{Buzzoni}, B., {Delabre}, B., {Dekker}, H., {et~al.} 1984, The Messenger, 38, 9

\bibitem[{{Capelo} {et~al.}(2017){Capelo}, {Dotti}, {Volonteri}, {Mayer}, {Bellovary}, \& {Shen}}]{Capelo17}
{Capelo}, P.~R., {Dotti}, M., {Volonteri}, M., {et~al.} 2017, \mnras, 469, 4437

\bibitem[{{Cappellari} \& {Emsellem}(2004)}]{Cappellari2004}
{Cappellari}, M. \& {Emsellem}, E. 2004, \pasp, 116, 138

\bibitem[{{Chartas} {et~al.}(2009){Chartas}, {Charlton}, {Eracleous}, {Giustini}, {Hidalgo}, {Ganguly}, {Hamann}, {Misawa}, \& {Tytler}}]{Chartas09}
{Chartas}, G., {Charlton}, J., {Eracleous}, M., {et~al.} 2009, \nar, 53, 128

\bibitem[{{Chen} {et~al.}(2023{\natexlab{a}}){Chen}, {Di Matteo}, {Ni}, {Tremmel}, {DeGraf}, {Shen}, {Holgado}, {Bird}, {Croft}, \& {Feng}}]{Chen23b}
{Chen}, N., {Di Matteo}, T., {Ni}, Y., {et~al.} 2023{\natexlab{a}}, \mnras, 522, 1895

\bibitem[{{Chen} {et~al.}(2025){Chen}, {Gross}, {Liu}, {Shen}, {Zakamska}, {Hwang}, \& {Zhuang}}]{Chen25_vastrometry}
{Chen}, Y.-C., {Gross}, A.~C., {Liu}, X., {et~al.} 2025, \apj, 988, 126

\bibitem[{{Chen} {et~al.}(2022){Chen}, {Hwang}, {Shen}, {Liu}, {Zakamska}, {Yang}, \& {Li}}]{Chen22a}
{Chen}, Y.-C., {Hwang}, H.-C., {Shen}, Y., {et~al.} 2022, \apj, 925, 162

\bibitem[{{Chen} {et~al.}(2023{\natexlab{b}}){Chen}, {Liu}, {Foord}, {Shen}, {Oguri}, {Chen}, {Di Matteo}, {Holgado}, {Hwang}, \& {Zakamska}}]{Chen23a}
{Chen}, Y.-C., {Liu}, X., {Foord}, A., {et~al.} 2023{\natexlab{b}}, \nat, 616, 45

\bibitem[{{Ciurlo} {et~al.}(2023){Ciurlo}, {Mannucci}, {Yeh}, {Amiri}, {Carniani}, {Cicone}, {Cresci}, {Lusso}, {Marasco}, {Marconcini}, {Marconi}, {Nardini}, {Pancino}, {Rosati}, {Rubinur}, {Severgnini}, {Scialpi}, {Tozzi}, {Venturi}, {Vignali}, \& {Volonteri}}]{Ciurlo23}
{Ciurlo}, A., {Mannucci}, F., {Yeh}, S., {et~al.} 2023, \aap, 671, L4

\bibitem[{{Colpi}(2014)}]{Colpi14}
{Colpi}, M. 2014, \ssr, 183, 189

\bibitem[{{Colpi} {et~al.}(2024){Colpi}, {Danzmann}, {Hewitson}, {Holley-Bockelmann}, {Jetzer}, {Nelemans}, {Petiteau}, {Shoemaker}, {Sopuerta}, {Stebbins}, {Tanvir}, {Ward}, {Weber}, {Thorpe}, {Daurskikh}, {Deep}, {Fern{\'a}ndez N{\'u}{\~n}ez}, {Garc{\'\i}a Marirrodriga}, {Gehler}, {Halain}, {Jennrich}, {Lammers}, {Larra{\~n}aga}, {Lieser}, {L{\"u}tzgendorf}, {Martens}, {Mondin}, {Piris Ni{\~n}o}, {Amaro-Seoane}, {Arca Sedda}, {Auclair}, {Babak}, {Baghi}, {Baibhav}, {Baker}, {Bayle}, {Berry}, {Berti}, {Boileau}, {Bonetti}, {Brito}, {Buscicchio}, {Calcagni}, {Capelo}, {Caprini}, {Caputo}, {Castelli}, {Chen}, {Chen}, {Chua}, {Davies}, {Derdzinski}, {Domcke}, {Doneva}, {Dvorkin}, {Mar{\'\i}a Ezquiaga}, {Gair}, {Haiman}, {Harry}, {Hartwig}, {Hees}, {Heffernan}, {Husa}, {Izquierdo-Villalba}, {Karnesis}, {Klein}, {Korol}, {Korsakova}, {Kupfer}, {Laghi}, {Lamberts}, {Larson}, {Le Jeune}, {Lewicki}, {Littenberg}, {Madge}, {Mangiagli}, {Marsat}, {Vilchez}, {Maselli}, {Mathews}, {van de Meent}, {Muratore}, {Nardini},
  {Pani}, {Peloso}, {Pieroni}, {Pound}, {Quelquejay-Leclere}, {Ricciardone}, {Rossi}, {Sartirana}, {Savalle}, {Sberna}, {Sesana}, {Shoemaker}, {Slutsky}, {Sotiriou}, {Speri}, {Staab}, {Steer}, {Tamanini}, {Tasinato}, {Torrado}, {Torres-Orjuela}, {Toubiana}, {Vallisneri}, {Vecchio}, {Volonteri}, {Yagi}, \& {Zwick}}]{Colpi2024arxiv}
{Colpi}, M., {Danzmann}, K., {Hewitson}, M., {et~al.} 2024, arXiv e-prints, arXiv:2402.07571

\bibitem[{{Comerford} {et~al.}(2009){Comerford}, {Gerke}, {Newman}, {Davis}, {Yan}, {Cooper}, {Faber}, {Koo}, {Coil}, {Rosario}, \& {Dutton}}]{Comerford2009}
{Comerford}, J.~M., {Gerke}, B.~F., {Newman}, J.~A., {et~al.} 2009, \apj, 698, 956

\bibitem[{{Croton} {et~al.}(2006){Croton}, {Springel}, {White}, {De Lucia}, {Frenk}, {Gao}, {Jenkins}, {Kauffmann}, {Navarro}, \& {Yoshida}}]{Croton06}
{Croton}, D.~J., {Springel}, V., {White}, S. D.~M., {et~al.} 2006, \mnras, 365, 11

\bibitem[{{Curti} {et~al.}(2020){Curti}, {Mannucci}, {Cresci}, \& {Maiolino}}]{Curti20}
{Curti}, M., {Mannucci}, F., {Cresci}, G., \& {Maiolino}, R. 2020, \mnras, 491, 944

\bibitem[{{De Rosa} {et~al.}(2019){De Rosa}, {Vignali}, {Bogdanovi{\'c}}, {Capelo}, {Charisi}, {Dotti}, {Husemann}, {Lusso}, {Mayer}, {Paragi}, {Runnoe}, {Sesana}, {Steinborn}, {Bianchi}, {Colpi}, {del Valle}, {Frey}, {Gab{\'a}nyi}, {Giustini}, {Guainazzi}, {Haiman}, {Herrera Ruiz}, {Herrero-Illana}, {Iwasawa}, {Komossa}, {Lena}, {Loiseau}, {Perez-Torres}, {Piconcelli}, \& {Volonteri}}]{derosa20}
{De Rosa}, A., {Vignali}, C., {Bogdanovi{\'c}}, T., {et~al.} 2019, \nar, 86, 101525

\bibitem[{{Delchambre} {et~al.}(2023){Delchambre}, {Bailer-Jones}, {Bellas-Velidis}, {Drimmel}, {Garabato}, {Carballo}, {Hatzidimitriou}, {Marshall}, {Andrae}, {Dafonte}, {Livanou}, {Fouesneau}, {Licata}, {Lindstr{\o}m}, {Manteiga}, {Robin}, {Silvelo}, {Abreu Aramburu}, {{\'A}lvarez}, {Bakker}, {Bijaoui}, {Brouillet}, {Brugaletta}, {Burlacu}, {Casamiquela}, {Chaoul}, {Chiavassa}, {Contursi}, {Cooper}, {Creevey}, {Dapergolas}, {de Laverny}, {Demouchy}, {Dharmawardena}, {Edvardsson}, {Fr{\'e}mat}, {Garc{\'\i}a-Lario}, {Garc{\'\i}a-Torres}, {Gavel}, {Gomez}, {Gonz{\'a}lez-Santamar{\'\i}a}, {Heiter}, {Jean-Antoine Piccolo}, {Kontizas}, {Kordopatis}, {Korn}, {Lanzafame}, {Lebreton}, {Lobel}, {Lorca}, {Magdaleno Romeo}, {Marocco}, {Mary}, {Nicolas}, {Ordenovic}, {Pailler}, {Palicio}, {Pallas-Quintela}, {Panem}, {Pichon}, {Poggio}, {Recio-Blanco}, {Riclet}, {Rybizki}, {Santove{\~n}a}, {Sarro}, {Schultheis}, {Segol}, {Slezak}, {Smart}, {Sordo}, {Soubiran}, {S{\"u}veges}, {Th{\'e}venin}, {Torralba Elipe}, {Ulla},
  {Utrilla}, {Vallenari}, {van Dillen}, {Zhao}, \& {Zorec}}]{Delchambre22}
{Delchambre}, L., {Bailer-Jones}, C.~A.~L., {Bellas-Velidis}, I., {et~al.} 2023, \aap, 674, A31

\bibitem[{{Delchambre} {et~al.}(2019){Delchambre}, {Krone-Martins}, {Wertz}, {Ducourant}, {Galluccio}, {Kl{\"u}ter}, {Mignard}, {Teixeira}, {Djorgovski}, {Stern}, {Graham}, {Surdej}, {Bastian}, {Wambsganss}, {Le Campion}, \& {Slezak}}]{Delchambre19}
{Delchambre}, L., {Krone-Martins}, A., {Wertz}, O., {et~al.} 2019, \aap, 622, A165

\bibitem[{{DESI Collaboration} {et~al.}(2025){DESI Collaboration}, {Abdul-Karim}, {Adame}, {Aguado}, {Aguilar}, {Ahlen}, {Alam}, {Aldering}, {Alexander}, {Alfarsy}, {Allen}, {Allende Prieto}, {Alves}, {Anand}, {Andrade}, {Armengaud}, {Avila}, {Aviles}, {Awan}, {Bailey}, {Baleato Lizancos}, {Ballester}, {Bault}, {Bautista}, {BenZvi}, {Beraldo e Silva}, {Bermejo-Climent}, {Beutler}, {Bianchi}, {Blake}, {Blum}, {Bolton}, {Bonici}, {Brieden}, {Brodzeller}, {Brooks}, {Buckley-Geer}, {Burtin}, {Canning}, {Carnero Rosell}, {Carr}, {Carrilho}, {Casas}, {Castander}, {Cereskaite}, {Cervantes-Cota}, {Chaussidon}, {Chaves-Montero}, {Chen}, {Chen}, {Claybaugh}, {Cole}, {Cooper}, {Cousinou}, {Cuceu}, {Davis}, {Dawson}, {de Belsunce}, {de la Cruz}, {de la Macorra}, {de Mattia}, {Deiosso}, {Della Costa}, {Demina}, {Demirbozan}, {DeRose}, {Dey}, {Dey}, {Ding}, {Ding}, {Doel}, {Douglass}, {Dowicz}, {Ebina}, {Edelstein}, {Eisenstein}, {Elbers}, {Emas}, {Escoffier}, {Fagrelius}, {Fan}, {Fanning}, {Fawcett},
  {Fern\'andez-Garc\'ia}, {Ferraro}, {Findlay}, {Font-Ribera}, {Forero-Romero}, {Forero-S\'anchez}, {Frenk}, {G\''ansicke}, {Galbany}, {Garc\'ia-Bellido}, {Garcia-Quintero}, {Garrison}, {Gazta\~naga}, {Gil-Mar\'in}, {Gnedin}, {Gontcho}, {Gonzalez-Morales}, {Gonzalez-Perez}, {Gordon}, {Graur}, {Green}, {Gruen}, {Gsponer}, {Guandalin}, {Gutierrez}, {Guy}, {Hahn}, {Han}, {Han}, {He}, {Herrera-Alcantar}, {Honscheid}, {Hou}, {Howlett}, {Huterer}, {Ir\v{s}i\v{c}}, {Ishak}, {Jacques}, {Jimenez}, {Jing}, {Joachimi}, {Joudaki}, {Joyce}, {Jullo}, {Juneau}, {Kara\c{c}ayl\{\i\}}, {Karim}, {Kehoe}, {Kent}, {Khederlarian}, {Kirkby}, {Kisner}, {Kitaura}, {Kizhuprakkat}, {Kong}, {Koposov}, {Kremin}, {Krolewski}, {Lahav}, {Lai}, {Lamman}, {Lan}, {Landriau}, {Lang}, {Lange}, {Lasker}, {Le Goff}, {Le Guillou}, {Leauthaud}, {Levi}, {Li}, {Li}, {Lodha}, {Lokken}, {Luo}, {Magneville}, {Manera}, {Manser}, {Margala}, {Martini}, {Maus}, {McCullough}, {McDonald}, {Medina}, {Medina-Varela}, {Meisner}, {Mena-Fern\'andez}, {Menegas},
  {Mezcua}, {Miquel}, {Montero-Camacho}, {Moon}, {Moustakas}, {Mu\~noz-Guti\'errez}, {Mu\~noz-Santos}, {Myers}, {Myles}, {Nadathur}, {Najita}, {Napolitano}, {Newman}, {Nikakhtar}, {Nikutta}, {Niz}, {Noriega}, {Padmanabhan}, {Paillas}, {Palanque-Delabrouille}, {Palmese}, {Pan}, {Pan}, {Parkinson}, {Peacock}, {Percival}, {P\'erez-Fern\'andez}, {P\'erez-R\`afols}, \& {Peterson}}]{DESI25_arxive}
{DESI Collaboration}, {Abdul-Karim}, M., {Adame}, A.~G., {et~al.} 2025, arXiv e-prints, arXiv:2503.14745

\bibitem[{{Di Matteo} {et~al.}(2005){Di Matteo}, {Springel}, \& {Hernquist}}]{dimatteo05}
{Di Matteo}, T., {Springel}, V., \& {Hernquist}, L. 2005, in Growing Black Holes: Accretion in a Cosmological Context, ed. A.~{Merloni}, S.~{Nayakshin}, \& R.~A. {Sunyaev}, 340--345

\bibitem[{{Dutta} {et~al.}(2024){Dutta}, {Acebron}, {Fumagalli}, {Grillo}, {Caminha}, \& {Fossati}}]{Dutta24}
{Dutta}, R., {Acebron}, A., {Fumagalli}, M., {et~al.} 2024, \mnras, 528, 1895

\bibitem[{{D’Amato} {et~al.}(2026){D’Amato}, {Mannucci}, {Sonnenfeld}, {Scialpi}, {Nightingale}, {Spingola}, {Zibetti}, {Marconi}, {Rosati}, {Marconcini}, {Agapito}, {Gallazzi}, {Di Teodoro}, {Andreuzzi}, {Belfiore}, {Bertola}, {Bracci}, {Carniani}, {Cataldi}, {Chakraborty}, {Ceci}, {Cicone}, {Ciurlo}, {Cresci}, {De Rosa}, {Di Carlo}, {Feltre}, {Ginolfi}, {Lamperti}, {Moreschini}, {Nardini}, {Perna}, {Portaluri}, {Rubinur}, {Saracco}, {Severgnini}, {Testa}, {Tozzi}, {Venturi}, {Ulivi}, \& {Vignoli}}]{DAmato26_quad}
{D’Amato}, Q., {Mannucci}, F., {Sonnenfeld}, A., {et~al.} 2026, Nature Astronomy, Accepted: 25 February, DOI: 10.1038/s41550

\bibitem[{{EPTA Collaboration} {et~al.}(2023){EPTA Collaboration}, {InPTA Collaboration}, {Antoniadis}, {Arumugam}, {Arumugam}, {Babak}, {Bagchi}, {Bak Nielsen}, {Bassa}, {Bathula}, {Berthereau}, {Bonetti}, {Bortolas}, {Brook}, {Burgay}, {Caballero}, {Chalumeau}, {Champion}, {Chanlaridis}, {Chen}, {Cognard}, {Dandapat}, {Deb}, {Desai}, {Desvignes}, {Dhanda-Batra}, {Dwivedi}, {Falxa}, {Ferdman}, {Franchini}, {Gair}, {Goncharov}, {Gopakumar}, {Graikou}, {Grie{\ss}meier}, {Guillemot}, {Guo}, {Gupta}, {Hisano}, {Hu}, {Iraci}, {Izquierdo-Villalba}, {Jang}, {Jawor}, {Janssen}, {Jessner}, {Joshi}, {Kareem}, {Karuppusamy}, {Keane}, {Keith}, {Kharbanda}, {Kikunaga}, {Kolhe}, {Kramer}, {Krishnakumar}, {Lackeos}, {Lee}, {Liu}, {Liu}, {Lyne}, {McKee}, {Maan}, {Main}, {Mickaliger}, {Ni{\c{t}}u}, {Nobleson}, {Paladi}, {Parthasarathy}, {Perera}, {Perrodin}, {Petiteau}, {Porayko}, {Possenti}, {Prabu}, {Quelquejay Leclere}, {Rana}, {Samajdar}, {Sanidas}, {Sesana}, {Shaifullah}, {Singha}, {Speri}, {Spiewak}, {Srivastava},
  {Stappers}, {Surnis}, {Susarla}, {Susobhanan}, {Takahashi}, {Tarafdar}, {Theureau}, {Tiburzi}, {van der Wateren}, {Vecchio}, {Venkatraman Krishnan}, {Verbiest}, {Wang}, {Wang}, \& {Wu}}]{EPTACollaboration2023}
{EPTA Collaboration}, {InPTA Collaboration}, {Antoniadis}, J., {et~al.} 2023, \aap, 678, A50

\bibitem[{{Faure} {et~al.}(2008){Faure}, {Kneib}, {Covone}, {Tasca}, {Leauthaud}, {Capak}, {Jahnke}, {Smolcic}, {de la Torre}, {Ellis}, {Finoguenov}, {Koekemoer}, {Le Fevre}, {Massey}, {Mellier}, {Refregier}, {Rhodes}, {Scoville}, {Schinnerer}, {Taylor}, {Van Waerbeke}, \& {Walcher}}]{Faure08}
{Faure}, C., {Kneib}, J.-P., {Covone}, G., {et~al.} 2008, \apjs, 176, 19

\bibitem[{{Flesch}(2021)}]{Flesch21}
{Flesch}, E.~W. 2021, arXiv e-prints, arXiv:2105.12985

\bibitem[{{Flesch}(2023)}]{Flesch23}
{Flesch}, E.~W. 2023, The Open Journal of Astrophysics, 6, 49

\bibitem[{{Fu} {et~al.}(2015{\natexlab{a}}){Fu}, {Myers}, {Djorgovski}, {Yan}, {Wrobel}, \& {Stockton}}]{Fu15_a}
{Fu}, H., {Myers}, A.~D., {Djorgovski}, S.~G., {et~al.} 2015{\natexlab{a}}, \apj, 799, 72

\bibitem[{{Fu} {et~al.}(2018){Fu}, {Steffen}, {Gross}, {Dai}, {Isbell}, {Lin}, {Wake}, {Xue}, {Bizyaev}, \& {Pan}}]{Fu18_sdss}
{Fu}, H., {Steffen}, J.~L., {Gross}, A.~C., {et~al.} 2018, \apj, 856, 93

\bibitem[{{Fu} {et~al.}(2015{\natexlab{b}}){Fu}, {Wrobel}, {Myers}, {Djorgovski}, \& {Yan}}]{Fu15_b}
{Fu}, H., {Wrobel}, J.~M., {Myers}, A.~D., {Djorgovski}, S.~G., \& {Yan}, L. 2015{\natexlab{b}}, \apjl, 815, L6

\bibitem[{{Fu} {et~al.}(2024){Fu}, {Wu}, {Li}, {Pang}, {Joshi}, {Zhang}, {Wang}, {Yang}, {Ng}, {Liu}, {Qiu}, {Zhu}, {Wang}, {Wolf}, {Zhang}, {Huo}, {Ai}, {Ma}, {Feng}, \& {Bouwens}}]{Fu24}
{Fu}, Y., {Wu}, X.-B., {Li}, Y., {et~al.} 2024, \apjs, 271, 54

\bibitem[{{Gaia Collaboration} {et~al.}(2018){Gaia Collaboration}, {Brown}, {Vallenari}, {Prusti}, {de Bruijne}, {Babusiaux}, {Bailer-Jones}, {Biermann}, {Evans}, {Eyer}, {Jansen}, {Jordi}, {Klioner}, {Lammers}, {Lindegren}, {Luri}, {Mignard}, {Panem}, {Pourbaix}, {Randich}, {Sartoretti}, {Siddiqui}, {Soubiran}, {van Leeuwen}, {Walton}, {Arenou}, {Bastian}, {Cropper}, {Drimmel}, {Katz}, {Lattanzi}, {Bakker}, {Cacciari}, {Casta{\~n}eda}, {Chaoul}, {Cheek}, {De Angeli}, {Fabricius}, {Guerra}, {Holl}, {Masana}, {Messineo}, {Mowlavi}, {Nienartowicz}, {Panuzzo}, {Portell}, {Riello}, {Seabroke}, {Tanga}, {Th{\'e}venin}, {Gracia-Abril}, {Comoretto}, {Garcia-Reinaldos}, {Teyssier}, {Altmann}, {Andrae}, {Audard}, {Bellas-Velidis}, {Benson}, {Berthier}, {Blomme}, {Burgess}, {Busso}, {Carry}, {Cellino}, {Clementini}, {Clotet}, {Creevey}, {Davidson}, {De Ridder}, {Delchambre}, {Dell'Oro}, {Ducourant}, {Fern{\'a}ndez-Hern{\'a}ndez}, {Fouesneau}, {Fr{\'e}mat}, {Galluccio}, {Garc{\'\i}a-Torres},
  {Gonz{\'a}lez-N{\'u}{\~n}ez}, {Gonz{\'a}lez-Vidal}, {Gosset}, {Guy}, {Halbwachs}, {Hambly}, {Harrison}, {Hern{\'a}ndez}, {Hestroffer}, {Hodgkin}, {Hutton}, {Jasniewicz}, {Jean-Antoine-Piccolo}, {Jordan}, {Korn}, {Krone-Martins}, {Lanzafame}, {Lebzelter}, {L{\"o}ffler}, {Manteiga}, {Marrese}, {Mart{\'\i}n-Fleitas}, {Moitinho}, {Mora}, {Muinonen}, {Osinde}, {Pancino}, {Pauwels}, {Petit}, {Recio-Blanco}, {Richards}, {Rimoldini}, {Robin}, {Sarro}, {Siopis}, {Smith}, {Sozzetti}, {S{\"u}veges}, {Torra}, {van Reeven}, {Abbas}, {Abreu Aramburu}, {Accart}, {Aerts}, {Altavilla}, {{\'A}lvarez}, {Alvarez}, {Alves}, {Anderson}, {Andrei}, {Anglada Varela}, {Antiche}, {Antoja}, {Arcay}, {Astraatmadja}, {Bach}, {Baker}, {Balaguer-N{\'u}{\~n}ez}, {Balm}, {Barache}, {Barata}, {Barbato}, {Barblan}, {Barklem}, {Barrado}, {Barros}, {Barstow}, {Bartholom{\'e} Mu{\~n}oz}, {Bassilana}, {Becciani}, {Bellazzini}, {Berihuete}, {Bertone}, {Bianchi}, {Bienaym{\'e}}, {Blanco-Cuaresma}, {Boch}, {Boeche}, {Bombrun}, {Borrachero},
  {Bossini}, {Bouquillon}, {Bourda}, {Bragaglia}, {Bramante}, {Breddels}, {Bressan}, {Brouillet}, {Br{\"u}semeister}, {Brugaletta}, {Bucciarelli}, {Burlacu}, {Busonero}, {Butkevich}, {Buzzi}, {Caffau}, {Cancelliere}, {Cannizzaro}, {Cantat-Gaudin}, {Carballo}, {Carlucci}, {Carrasco}, {Casamiquela}, {Castellani}, {Castro-Ginard}, {Charlot}, {Chemin}, {Chiavassa}, {Cocozza}, {Costigan}, {Cowell}, {Crifo}, {Crosta}, {Crowley}, {Cuypers}, {Dafonte}, {Damerdji}, {Dapergolas}, {David}, {David}, {de Laverny}, \& {De Luise}}]{GaiaDR2}
{Gaia Collaboration}, {Brown}, A.~G.~A., {Vallenari}, A., {et~al.} 2018, \aap, 616, A1

\bibitem[{{Gaia Collaboration} {et~al.}(2021){Gaia Collaboration}, {Brown}, {Vallenari}, {Prusti}, {de Bruijne}, {Babusiaux}, {Biermann}, {Creevey}, {Evans}, {Eyer}, {Hutton}, {Jansen}, {Jordi}, {Klioner}, {Lammers}, {Lindegren}, {Luri}, {Mignard}, {Panem}, {Pourbaix}, {Randich}, {Sartoretti}, {Soubiran}, {Walton}, {Arenou}, {Bailer-Jones}, {Bastian}, {Cropper}, {Drimmel}, {Katz}, {Lattanzi}, {van Leeuwen}, {Bakker}, {Cacciari}, {Casta{\~n}eda}, {De Angeli}, {Ducourant}, {Fabricius}, {Fouesneau}, {Fr{\'e}mat}, {Guerra}, {Guerrier}, {Guiraud}, {Jean-Antoine Piccolo}, {Masana}, {Messineo}, {Mowlavi}, {Nicolas}, {Nienartowicz}, {Pailler}, {Panuzzo}, {Riclet}, {Roux}, {Seabroke}, {Sordo}, {Tanga}, {Th{\'e}venin}, {Gracia-Abril}, {Portell}, {Teyssier}, {Altmann}, {Andrae}, {Bellas-Velidis}, {Benson}, {Berthier}, {Blomme}, {Brugaletta}, {Burgess}, {Busso}, {Carry}, {Cellino}, {Cheek}, {Clementini}, {Damerdji}, {Davidson}, {Delchambre}, {Dell'Oro}, {Fern{\'a}ndez-Hern{\'a}ndez}, {Galluccio}, {Garc{\'\i}a-Lario},
  {Garcia-Reinaldos}, {Gonz{\'a}lez-N{\'u}{\~n}ez}, {Gosset}, {Haigron}, {Halbwachs}, {Hambly}, {Harrison}, {Hatzidimitriou}, {Heiter}, {Hern{\'a}ndez}, {Hestroffer}, {Hodgkin}, {Holl}, {Jan{\ss}en}, {Jevardat de Fombelle}, {Jordan}, {Krone-Martins}, {Lanzafame}, {L{\"o}ffler}, {Lorca}, {Manteiga}, {Marchal}, {Marrese}, {Moitinho}, {Mora}, {Muinonen}, {Osborne}, {Pancino}, {Pauwels}, {Petit}, {Recio-Blanco}, {Richards}, {Riello}, {Rimoldini}, {Robin}, {Roegiers}, {Rybizki}, {Sarro}, {Siopis}, {Smith}, {Sozzetti}, {Ulla}, {Utrilla}, {van Leeuwen}, {van Reeven}, {Abbas}, {Abreu Aramburu}, {Accart}, {Aerts}, {Aguado}, {Ajaj}, {Altavilla}, {{\'A}lvarez}, {{\'A}lvarez Cid-Fuentes}, {Alves}, {Anderson}, {Anglada Varela}, {Antoja}, {Audard}, {Baines}, {Baker}, {Balaguer-N{\'u}{\~n}ez}, {Balbinot}, {Balog}, {Barache}, {Barbato}, {Barros}, {Barstow}, {Bartolom{\'e}}, {Bassilana}, {Bauchet}, {Baudesson-Stella}, {Becciani}, {Bellazzini}, {Bernet}, {Bertone}, {Bianchi}, {Blanco-Cuaresma}, {Boch}, {Bombrun}, {Bossini},
  {Bouquillon}, {Bragaglia}, {Bramante}, {Breedt}, {Bressan}, {Brouillet}, {Bucciarelli}, {Burlacu}, {Busonero}, {Butkevich}, {Buzzi}, {Caffau}, {Cancelliere}, {C{\'a}novas}, {Cantat-Gaudin}, {Carballo}, {Carlucci}, {Carnerero}, {Carrasco}, {Casamiquela}, {Castellani}, {Castro-Ginard}, {Castro Sampol}, {Chaoul}, {Charlot}, {Chemin}, {Chiavassa}, {Cioni}, {Comoretto}, {Cooper}, {Cornez}, {Cowell}, {Crifo}, {Crosta}, {Crowley}, {Dafonte}, {Dapergolas}, {David}, {David}, {de Laverny}, {De Luise}, {De March}, {De Ridder}, {de Souza}, {de Teodoro}, {de Torres}, {del Peloso}, {del Pozo}, {Delbo}, {Delgado}, {Delgado}, {Delisle}, {Di Matteo}, {Diakite}, {Diener}, {Distefano}, {Dolding}, {Eappachen}, {Edvardsson}, {Enke}, {Esquej}, {Fabre}, {Fabrizio}, {Faigler}, {Fedorets}, {Fernique}, {Fienga}, {Figueras}, {Fouron}, {Fragkoudi}, {Fraile}, {Franke}, {Gai}, {Garabato}, {Garcia-Gutierrez}, {Garc{\'\i}a-Torres}, {Garofalo}, {Gavras}, {Gerlach}, {Geyer}, {Giacobbe}, {Gilmore}, {Girona}, {Giuffrida}, {Gomel}, {Gomez},
  {Gonzalez-Santamaria}, {Gonz{\'a}lez-Vidal}, {Granvik}, {Guti{\'e}rrez-S{\'a}nchez}, {Guy}, {Hauser}, {Haywood}, {Helmi}, {Hidalgo}, {Hilger}, {H{\l}adczuk}, {Hobbs}, {Holland}, {Huckle}, {Jasniewicz}, {Jonker}, {Juaristi Campillo}, {Julbe}, {Karbevska}, {Kervella}, {Khanna}, {Kochoska}, {Kontizas}, {Kordopatis}, {Korn}, {Kostrzewa-Rutkowska}, {Kruszy{\'n}ska}, {Lambert}, {Lanza}, {Lasne}, {Le Campion}, {Le Fustec}, {Lebreton}, {Lebzelter}, {Leccia}, {Leclerc}, {Lecoeur-Taibi}, {Liao}, {Licata}, {Lindstr{\o}m}, {Lister}, {Livanou}, {Lobel}, {Madrero Pardo}, {Managau}, {Mann}, {Marchant}, {Marconi}, {Marcos Santos}, {Marinoni}, {Marocco}, {Marshall}, {Martin Polo}, {Mart{\'\i}n-Fleitas}, {Masip}, {Massari}, {Mastrobuono-Battisti}, {Mazeh}, {McMillan}, {Messina}, {Michalik}, {Millar}, {Mints}, {Molina}, {Molinaro}, {Moln{\'a}r}, {Montegriffo}, {Mor}, {Morbidelli}, {Morel}, {Morris}, {Mulone}, {Munoz}, {Muraveva}, {Murphy}, {Musella}, {Noval}, {Ord{\'e}novic}, {Orr{\`u}}, {Osinde}, {Pagani}, {Pagano},
  {Palaversa}, {Palicio}, {Panahi}, {Pawlak}, {Pe{\~n}alosa Esteller}, {Penttil{\"a}}, {Piersimoni}, {Pineau}, {Plachy}, {Plum}, {Poggio}, {Poretti}, {Poujoulet}, {Pr{\v{s}}a}, {Pulone}, {Racero}, {Ragaini}, {Rainer}, {Raiteri}, {Rambaux}, {Ramos}, {Ramos-Lerate}, {Re Fiorentin}, {Regibo}, {Reyl{\'e}}, {Ripepi}, {Riva}, {Rixon}, {Robichon}, {Robin}, {Roelens}, {Rohrbasser}, {Romero-G{\'o}mez}, {Rowell}, {Royer}, {Rybicki}, {Sadowski}, {Sagrist{\`a} Sell{\'e}s}, {Sahlmann}, {Salgado}, {Salguero}, {Samaras}, {Sanchez Gimenez}, {Sanna}, {Santove{\~n}a}, {Sarasso}, {Schultheis}, {Sciacca}, {Segol}, {Segovia}, {S{\'e}gransan}, {Semeux}, {Shahaf}, {Siddiqui}, {Siebert}, {Siltala}, {Slezak}, {Smart}, {Solano}, {Solitro}, {Souami}, {Souchay}, {Spagna}, {Spoto}, {Steele}, {Steidelm{\"u}ller}, {Stephenson}, {S{\"u}veges}, {Szabados}, {Szegedi-Elek}, {Taris}, {Tauran}, {Taylor}, {Teixeira}, {Thuillot}, {Tonello}, {Torra}, {Torra}, {Turon}, {Unger}, {Vaillant}, {van Dillen}, {Vanel}, {Vecchiato}, {Viala}, {Vicente},
  {Voutsinas}, {Weiler}, {Wevers}, {Wyrzykowski}, {Yoldas}, {Yvard}, {Zhao}, {Zorec}, {Zucker}, {Zurbach}, \& {Zwitter}}]{Gaia_EDR3_1}
{Gaia Collaboration}, {Brown}, A.~G.~A., {Vallenari}, A., {et~al.} 2021, \aap, 649, A1

\bibitem[{{Gaia Collaboration} {et~al.}(2016){Gaia Collaboration}, {Prusti}, {de Bruijne}, {Brown}, {Vallenari}, {Babusiaux}, {Bailer-Jones}, {Bastian}, {Biermann}, {Evans}, {Eyer}, {Jansen}, {Jordi}, {Klioner}, {Lammers}, {Lindegren}, {Luri}, {Mignard}, {Milligan}, {Panem}, {Poinsignon}, {Pourbaix}, {Randich}, {Sarri}, {Sartoretti}, {Siddiqui}, {Soubiran}, {Valette}, {van Leeuwen}, {Walton}, {Aerts}, {Arenou}, {Cropper}, {Drimmel}, {H{\o}g}, {Katz}, {Lattanzi}, {O'Mullane}, {Grebel}, {Holland}, {Huc}, {Passot}, {Bramante}, {Cacciari}, {Casta{\~n}eda}, {Chaoul}, {Cheek}, {De Angeli}, {Fabricius}, {Guerra}, {Hern{\'a}ndez}, {Jean-Antoine-Piccolo}, {Masana}, {Messineo}, {Mowlavi}, {Nienartowicz}, {Ord{\'o}{\~n}ez-Blanco}, {Panuzzo}, {Portell}, {Richards}, {Riello}, {Seabroke}, {Tanga}, {Th{\'e}venin}, {Torra}, {Els}, {Gracia-Abril}, {Comoretto}, {Garcia-Reinaldos}, {Lock}, {Mercier}, {Altmann}, {Andrae}, {Astraatmadja}, {Bellas-Velidis}, {Benson}, {Berthier}, {Blomme}, {Busso}, {Carry}, {Cellino}, {Clementini},
  {Cowell}, {Creevey}, {Cuypers}, {Davidson}, {De Ridder}, {de Torres}, {Delchambre}, {Dell'Oro}, {Ducourant}, {Fr{\'e}mat}, {Garc{\'\i}a-Torres}, {Gosset}, {Halbwachs}, {Hambly}, {Harrison}, {Hauser}, {Hestroffer}, {Hodgkin}, {Huckle}, {Hutton}, {Jasniewicz}, {Jordan}, {Kontizas}, {Korn}, {Lanzafame}, {Manteiga}, {Moitinho}, {Muinonen}, {Osinde}, {Pancino}, {Pauwels}, {Petit}, {Recio-Blanco}, {Robin}, {Sarro}, {Siopis}, {Smith}, {Smith}, {Sozzetti}, {Thuillot}, {van Reeven}, {Viala}, {Abbas}, {Abreu Aramburu}, {Accart}, {Aguado}, {Allan}, {Allasia}, {Altavilla}, {{\'A}lvarez}, {Alves}, {Anderson}, {Andrei}, {Anglada Varela}, {Antiche}, {Antoja}, {Ant{\'o}n}, {Arcay}, {Atzei}, {Ayache}, {Bach}, {Baker}, {Balaguer-N{\'u}{\~n}ez}, {Barache}, {Barata}, {Barbier}, {Barblan}, {Baroni}, {Barrado y Navascu{\'e}s}, {Barros}, {Barstow}, {Becciani}, {Bellazzini}, {Bellei}, {Bello Garc{\'\i}a}, {Belokurov}, {Bendjoya}, {Berihuete}, {Bianchi}, {Bienaym{\'e}}, {Billebaud}, {Blagorodnova}, {Blanco-Cuaresma}, {Boch},
  {Bombrun}, {Borrachero}, {Bouquillon}, {Bourda}, {Bouy}, {Bragaglia}, {Breddels}, {Brouillet}, {Br{\"u}semeister}, {Bucciarelli}, {Budnik}, {Burgess}, {Burgon}, {Burlacu}, {Busonero}, {Buzzi}, {Caffau}, {Cambras}, {Campbell}, {Cancelliere}, {Cantat-Gaudin}, {Carlucci}, {Carrasco}, {Castellani}, {Charlot}, {Charnas}, {Charvet}, {Chassat}, {Chiavassa}, {Clotet}, {Cocozza}, {Collins}, {Collins}, {Costigan}, {Crifo}, {Cross}, {Crosta}, {Crowley}, {Dafonte}, {Damerdji}, {Dapergolas}, {David}, {David}, {De Cat}, {de Felice}, {de Laverny}, {De Luise}, {De March}, {de Martino}, {de Souza}, {Debosscher}, {del Pozo}, {Delbo}, {Delgado}, {Delgado}, {di Marco}, {Di Matteo}, {Diakite}, {Distefano}, {Dolding}, {Dos Anjos}, {Drazinos}, {Dur{\'a}n}, {Dzigan}, {Ecale}, {Edvardsson}, {Enke}, {Erdmann}, {Escolar}, {Espina}, {Evans}, {Eynard Bontemps}, {Fabre}, {Fabrizio}, {Faigler}, {Falc{\~a}o}, {Farr{\`a}s Casas}, {Faye}, {Federici}, {Fedorets}, {Fern{\'a}ndez-Hern{\'a}ndez}, {Fernique}, {Fienga}, {Figueras}, {Filippi},
  {Findeisen}, {Fonti}, {Fouesneau}, {Fraile}, {Fraser}, {Fuchs}, {Furnell}, {Gai}, {Galleti}, {Galluccio}, {Garabato}, {Garc{\'\i}a-Sedano}, {Gar{\'e}}, {Garofalo}, {Garralda}, {Gavras}, {Gerssen}, {Geyer}, {Gilmore}, {Girona}, {Giuffrida}, {Gomes}, {Gonz{\'a}lez-Marcos}, {Gonz{\'a}lez-N{\'u}{\~n}ez}, {Gonz{\'a}lez-Vidal}, {Granvik}, {Guerrier}, {Guillout}, {Guiraud}, {G{\'u}rpide}, {Guti{\'e}rrez-S{\'a}nchez}, {Guy}, {Haigron}, {Hatzidimitriou}, {Haywood}, {Heiter}, {Helmi}, {Hobbs}, {Hofmann}, {Holl}, {Holland}, {Hunt}, {Hypki}, {Icardi}, {Irwin}, {Jevardat de Fombelle}, {Jofr{\'e}}, {Jonker}, {Jorissen}, {Julbe}, {Karampelas}, {Kochoska}, {Kohley}, {Kolenberg}, {Kontizas}, {Koposov}, {Kordopatis}, {Koubsky}, {Kowalczyk}, {Krone-Martins}, {Kudryashova}, {Kull}, {Bachchan}, {Lacoste-Seris}, {Lanza}, {Lavigne}, {Le Poncin-Lafitte}, {Lebreton}, {Lebzelter}, {Leccia}, {Leclerc}, {Lecoeur-Taibi}, {Lemaitre}, {Lenhardt}, {Leroux}, {Liao}, {Licata}, {Lindstr{\o}m}, {Lister}, {Livanou}, {Lobel}, {L{\"o}ffler},
  {L{\'o}pez}, {Lopez-Lozano}, {Lorenz}, {Loureiro}, {MacDonald}, {Magalh{\~a}es Fernandes}, {Managau}, {Mann}, {Mantelet}, {Marchal}, {Marchant}, {Marconi}, {Marie}, {Marinoni}, {Marrese}, {Marschalk{\'o}}, {Marshall}, {Mart{\'\i}n-Fleitas}, {Martino}, {Mary}, {Matijevi{\v{c}}}, {Mazeh}, {McMillan}, {Messina}, {Mestre}, {Michalik}, {Millar}, {Miranda}, {Molina}, {Molinaro}, {Molinaro}, {Moln{\'a}r}, {Moniez}, {Montegriffo}, {Monteiro}, {Mor}, {Mora}, {Morbidelli}, {Morel}, {Morgenthaler}, {Morley}, {Morris}, {Mulone}, {Muraveva}, {Musella}, {Narbonne}, {Nelemans}, {Nicastro}, {Noval}, {Ord{\'e}novic}, {Ordieres-Mer{\'e}}, {Osborne}, {Pagani}, {Pagano}, {Pailler}, {Palacin}, {Palaversa}, {Parsons}, {Paulsen}, {Pecoraro}, {Pedrosa}, {Pentik{\"a}inen}, {Pereira}, {Pichon}, {Piersimoni}, {Pineau}, {Plachy}, {Plum}, {Poujoulet}, {Pr{\v{s}}a}, {Pulone}, {Ragaini}, {Rago}, {Rambaux}, {Ramos-Lerate}, {Ranalli}, {Rauw}, {Read}, {Regibo}, {Renk}, {Reyl{\'e}}, {Ribeiro}, {Rimoldini}, {Ripepi}, {Riva}, {Rixon},
  {Roelens}, {Romero-G{\'o}mez}, {Rowell}, {Royer}, {Rudolph}, {Ruiz-Dern}, {Sadowski}, {Sagrist{\`a} Sell{\'e}s}, {Sahlmann}, {Salgado}, {Salguero}, {Sarasso}, {Savietto}, {Schnorhk}, {Schultheis}, {Sciacca}, {Segol}, {Segovia}, {Segransan}, {Serpell}, {Shih}, {Smareglia}, {Smart}, {Smith}, {Solano}, {Solitro}, {Sordo}, {Soria Nieto}, {Souchay}, {Spagna}, {Spoto}, {Stampa}, {Steele}, {Steidelm{\"u}ller}, {Stephenson}, {Stoev}, {Suess}, {S{\"u}veges}, {Surdej}, {Szabados}, {Szegedi-Elek}, {Tapiador}, {Taris}, {Tauran}, {Taylor}, {Teixeira}, {Terrett}, {Tingley}, {Trager}, {Turon}, {Ulla}, {Utrilla}, {Valentini}, {van Elteren}, {Van Hemelryck}, {van Leeuwen}, {Varadi}, {Vecchiato}, {Veljanoski}, {Via}, {Vicente}, {Vogt}, {Voss}, {Votruba}, {Voutsinas}, {Walmsley}, {Weiler}, {Weingrill}, {Werner}, {Wevers}, {Whitehead}, {Wyrzykowski}, {Yoldas}, {{\v{Z}}erjal}, {Zucker}, {Zurbach}, {Zwitter}, {Alecu}, {Allen}, {Allende Prieto}, {Amorim}, {Anglada-Escud{\'e}}, {Arsenijevic}, {Azaz}, {Balm}, {Beck}, {Bernstein},
  {Bigot}, {Bijaoui}, {Blasco}, {Bonfigli}, {Bono}, {Boudreault}, {Bressan}, {Brown}, {Brunet}, {Bunclark}, {Buonanno}, {Butkevich}, {Carret}, {Carrion}, {Chemin}, {Ch{\'e}reau}, {Corcione}, {Darmigny}, {de Boer}, {de Teodoro}, {de Zeeuw}, {Delle Luche}, {Domingues}, {Dubath}, {Fodor}, {Fr{\'e}zouls}, {Fries}, {Fustes}, {Fyfe}, {Gallardo}, {Gallegos}, {Gardiol}, {Gebran}, {Gomboc}, {G{\'o}mez}, {Grux}, {Gueguen}, {Heyrovsky}, {Hoar}, {Iannicola}, {Isasi Parache}, {Janotto}, {Joliet}, {Jonckheere}, {Keil}, {Kim}, {Klagyivik}, {Klar}, {Knude}, {Kochukhov}, {Kolka}, {Kos}, {Kutka}, {Lainey}, {LeBouquin}, {Liu}, {Loreggia}, {Makarov}, {Marseille}, {Martayan}, {Martinez-Rubi}, {Massart}, {Meynadier}, {Mignot}, {Munari}, {Nguyen}, {Nordlander}, {Ocvirk}, {O'Flaherty}, {Olias Sanz}, {Ortiz}, {Osorio}, {Oszkiewicz}, {Ouzounis}, {Palmer}, {Park}, {Pasquato}, {Peltzer}, {Peralta}, {P{\'e}turaud}, {Pieniluoma}, {Pigozzi}, {Poels}, {Prat}, {Prod'homme}, {Raison}, {Rebordao}, {Risquez}, {Rocca-Volmerange}, {Rosen},
  {Ruiz-Fuertes}, {Russo}, {Sembay}, {Serraller Vizcaino}, {Short}, {Siebert}, {Silva}, {Sinachopoulos}, {Slezak}, {Soffel}, {Sosnowska}, {Strai{\v{z}}ys}, {ter Linden}, {Terrell}, {Theil}, {Tiede}, {Troisi}, {Tsalmantza}, {Tur}, {Vaccari}, {Vachier}, {Valles}, {Van Hamme}, {Veltz}, {Virtanen}, {Wallut}, {Wichmann}, {Wilkinson}, {Ziaeepour}, \& {Zschocke}}]{GaiaMission}
{Gaia Collaboration}, {Prusti}, T., {de Bruijne}, J.~H.~J., {et~al.} 2016, \aap, 595, A1

\bibitem[{{Gaia Collaboration} {et~al.}(2023){Gaia Collaboration}, {Vallenari}, {Brown}, {Prusti}, {de Bruijne}, {Arenou}, {Babusiaux}, {Biermann}, {Creevey}, {Ducourant}, {Evans}, {Eyer}, {Guerra}, {Hutton}, {Jordi}, {Klioner}, {Lammers}, {Lindegren}, {Luri}, {Mignard}, {Panem}, {Pourbaix}, {Randich}, {Sartoretti}, {Soubiran}, {Tanga}, {Walton}, {Bailer-Jones}, {Bastian}, {Drimmel}, {Jansen}, {Katz}, {Lattanzi}, {van Leeuwen}, {Bakker}, {Cacciari}, {Casta{\~n}eda}, {De Angeli}, {Fabricius}, {Fouesneau}, {Fr{\'e}mat}, {Galluccio}, {Guerrier}, {Heiter}, {Masana}, {Messineo}, {Mowlavi}, {Nicolas}, {Nienartowicz}, {Pailler}, {Panuzzo}, {Riclet}, {Roux}, {Seabroke}, {Sordo}, {Th{\'e}venin}, {Gracia-Abril}, {Portell}, {Teyssier}, {Altmann}, {Andrae}, {Audard}, {Bellas-Velidis}, {Benson}, {Berthier}, {Blomme}, {Burgess}, {Busonero}, {Busso}, {C{\'a}novas}, {Carry}, {Cellino}, {Cheek}, {Clementini}, {Damerdji}, {Davidson}, {de Teodoro}, {Nu{\~n}ez Campos}, {Delchambre}, {Dell'Oro}, {Esquej},
  {Fern{\'a}ndez-Hern{\'a}ndez}, {Fraile}, {Garabato}, {Garc{\'\i}a-Lario}, {Gosset}, {Haigron}, {Halbwachs}, {Hambly}, {Harrison}, {Hern{\'a}ndez}, {Hestroffer}, {Hodgkin}, {Holl}, {Jan{\ss}en}, {Jevardat de Fombelle}, {Jordan}, {Krone-Martins}, {Lanzafame}, {L{\"o}ffler}, {Marchal}, {Marrese}, {Moitinho}, {Muinonen}, {Osborne}, {Pancino}, {Pauwels}, {Recio-Blanco}, {Reyl{\'e}}, {Riello}, {Rimoldini}, {Roegiers}, {Rybizki}, {Sarro}, {Siopis}, {Smith}, {Sozzetti}, {Utrilla}, {van Leeuwen}, {Abbas}, {{\'A}brah{\'a}m}, {Abreu Aramburu}, {Aerts}, {Aguado}, {Ajaj}, {Aldea-Montero}, {Altavilla}, {{\'A}lvarez}, {Alves}, {Anders}, {Anderson}, {Anglada Varela}, {Antoja}, {Baines}, {Baker}, {Balaguer-N{\'u}{\~n}ez}, {Balbinot}, {Balog}, {Barache}, {Barbato}, {Barros}, {Barstow}, {Bartolom{\'e}}, {Bassilana}, {Bauchet}, {Becciani}, {Bellazzini}, {Berihuete}, {Bernet}, {Bertone}, {Bianchi}, {Binnenfeld}, {Blanco-Cuaresma}, {Blazere}, {Boch}, {Bombrun}, {Bossini}, {Bouquillon}, {Bragaglia}, {Bramante}, {Breedt},
  {Bressan}, {Brouillet}, {Brugaletta}, {Bucciarelli}, {Burlacu}, {Butkevich}, {Buzzi}, {Caffau}, {Cancelliere}, {Cantat-Gaudin}, {Carballo}, {Carlucci}, {Carnerero}, {Carrasco}, {Casamiquela}, {Castellani}, {Castro-Ginard}, {Chaoul}, {Charlot}, {Chemin}, {Chiaramida}, {Chiavassa}, {Chornay}, {Comoretto}, {Contursi}, {Cooper}, {Cornez}, {Cowell}, {Crifo}, {Cropper}, {Crosta}, {Crowley}, {Dafonte}, {Dapergolas}, {David}, {David}, {de Laverny}, {De Luise}, \& {De March}}]{Gaia_EDR3_2}
{Gaia Collaboration}, {Vallenari}, A., {Brown}, A.~G.~A., {et~al.} 2023, \aap, 674, A1

\bibitem[{{Glikman} {et~al.}(2023){Glikman}, {Langgin}, {Johnstone}, {Yoon}, {Comerford}, {Simmons}, {Stacey}, {Lacy}, \& {O'Meara}}]{Glikman23}
{Glikman}, E., {Langgin}, R., {Johnstone}, M.~A., {et~al.} 2023, \apjl, 951, L18

\bibitem[{{Green}(2006)}]{Green06}
{Green}, P.~J. 2006, \apj, 644, 733

\bibitem[{{Gross} {et~al.}(2025{\natexlab{a}}){Gross}, {Chen}, {Oguri}, {Nolan}, {Liu}, {Shen}, {Zhuang}, {Li}, {Zakamska}, {Hwang}, \& {Ishikawa}}]{gross25_vodka}
{Gross}, A.~C., {Chen}, Y.-C., {Oguri}, M., {et~al.} 2025{\natexlab{a}}, \apj, 989, 112

\bibitem[{{Gross} {et~al.}(2025{\natexlab{b}}){Gross}, {Chen}, {Oguri}, {Nolan}, {Liu}, {Shen}, {Zhuang}, {Li}, {Zakamska}, {Hwang}, \& {Ishikawa}}]{Gross2024}
{Gross}, A.~C., {Chen}, Y.-C., {Oguri}, M., {et~al.} 2025{\natexlab{b}}, \apj, 989, 112

\bibitem[{{Hamann} {et~al.}(2011){Hamann}, {Kanekar}, {Prochaska}, {Murphy}, {Ellison}, {Malec}, {Milutinovic}, \& {Ubachs}}]{Hamann11}
{Hamann}, F., {Kanekar}, N., {Prochaska}, J.~X., {et~al.} 2011, \mnras, 410, 1957

\bibitem[{{Hamann} {et~al.}(2012){Hamann}, {Simon}, {Rodriguez Hidalgo}, \& {Capellupo}}]{Hamann12}
{Hamann}, F., {Simon}, L., {Rodriguez Hidalgo}, P., \& {Capellupo}, D. 2012, in Astronomical Society of the Pacific Conference Series, Vol. 460, AGN Winds in Charleston, ed. G.~{Chartas}, F.~{Hamann}, \& K.~M. {Leighly}, 47

\bibitem[{{Hennawi} {et~al.}(2010){Hennawi}, {Myers}, {Shen}, {Strauss}, {Djorgovski}, {Fan}, {Glikman}, {Mahabal}, {Martin}, {Richards}, {Schneider}, \& {Shankar}}]{Hennawi10}
{Hennawi}, J.~F., {Myers}, A.~D., {Shen}, Y., {et~al.} 2010, \apj, 719, 1672

\bibitem[{{Hennawi} {et~al.}(2006){Hennawi}, {Strauss}, {Oguri}, {Inada}, {Richards}, {Pindor}, {Schneider}, {Becker}, {Gregg}, {Hall}, {Johnston}, {Fan}, {Burles}, {Schlegel}, {Gunn}, {Lupton}, {Bahcall}, {Brunner}, \& {Brinkmann}}]{Hennawi06}
{Hennawi}, J.~F., {Strauss}, M.~A., {Oguri}, M., {et~al.} 2006, \aj, 131, 1

\bibitem[{{Husemann} {et~al.}(2018){Husemann}, {Worseck}, {Arrigoni Battaia}, \& {Shanks}}]{Husemann18}
{Husemann}, B., {Worseck}, G., {Arrigoni Battaia}, F., \& {Shanks}, T. 2018, \aap, 610, L7

\bibitem[{{Hwang} {et~al.}(2020){Hwang}, {Shen}, {Zakamska}, \& {Liu}}]{Hwang20}
{Hwang}, H.-C., {Shen}, Y., {Zakamska}, N., \& {Liu}, X. 2020, \apj, 888, 73

\bibitem[{{Inada} {et~al.}(2005){Inada}, {Burles}, {Gregg}, {Becker}, {Schechter}, {Eisenstein}, {Oguri}, {Castander}, {Hall}, {Johnston}, {Pindor}, {Richards}, {Schneider}, {White}, {Brinkmann}, {Szalay}, \& {York}}]{Inada05}
{Inada}, N., {Burles}, S., {Gregg}, M.~D., {et~al.} 2005, \aj, 130, 1967

\bibitem[{{Inada} {et~al.}(2008){Inada}, {Oguri}, {Becker}, {Shin}, {Richards}, {Hennawi}, {White}, {Pindor}, {Strauss}, {Kochanek}, {Johnston}, {Gregg}, {Kayo}, {Eisenstein}, {Hall}, {Castander}, {Clocchiatti}, {Anderson}, {Schneider}, {York}, {Lupton}, {Chiu}, {Kawano}, {Scranton}, {Frieman}, {Keeton}, {Morokuma}, {Rix}, {Turner}, {Burles}, {Brunner}, {Sheldon}, {Bahcall}, \& {Masataka}}]{Inada08}
{Inada}, N., {Oguri}, M., {Becker}, R.~H., {et~al.} 2008, \aj, 135, 496

\bibitem[{{Inada} {et~al.}(2012){Inada}, {Oguri}, {Shin}, {Kayo}, {Strauss}, {Morokuma}, {Rusu}, {Fukugita}, {Kochanek}, {Richards}, {Schneider}, {York}, {Bahcall}, {Frieman}, {Hall}, \& {White}}]{Inada12}
{Inada}, N., {Oguri}, M., {Shin}, M.-S., {et~al.} 2012, \aj, 143, 119

\bibitem[{{Jim{\'e}nez-Vicente} {et~al.}(2015){Jim{\'e}nez-Vicente}, {Mediavilla}, {Kochanek}, \& {Mu{\~n}oz}}]{jimenezvicente2015}
{Jim{\'e}nez-Vicente}, J., {Mediavilla}, E., {Kochanek}, C.~S., \& {Mu{\~n}oz}, J.~A. 2015, \apj, 806, 251

\bibitem[{{Jing} {et~al.}(2025){Jing}, {Chen}, {Deng}, {Zhu}, {Zou}, \& {Wu}}]{Jing25_desidual}
{Jing}, L., {Chen}, Q., {Deng}, Z., {et~al.} 2025, arXiv e-prints, arXiv:2505.03103

\bibitem[{{Junkkarinen} {et~al.}(2001){Junkkarinen}, {Shields}, {Beaver}, {Burbidge}, {Cohen}, {Hamann}, \& {Lyons}}]{Junkkarinen01}
{Junkkarinen}, V., {Shields}, G.~A., {Beaver}, E.~A., {et~al.} 2001, \apjl, 549, L155

\bibitem[{{Kelley} {et~al.}(2017){Kelley}, {Blecha}, \& {Hernquist}}]{Kelley17}
{Kelley}, L.~Z., {Blecha}, L., \& {Hernquist}, L. 2017, \mnras, 464, 3131

\bibitem[{{Kennicutt}(1998)}]{Kennicutt98}
{Kennicutt}, Jr., R.~C. 1998, \apj, 498, 541

\bibitem[{{Kochanek} {et~al.}(2006){Kochanek}, {Mochejska}, {Morgan}, \& {Stanek}}]{Kochanek06}
{Kochanek}, C.~S., {Mochejska}, B., {Morgan}, N.~D., \& {Stanek}, K.~Z. 2006, \apjl, 637, L73

\bibitem[{{Koss} {et~al.}(2012){Koss}, {Mushotzky}, {Treister}, {Veilleux}, {Vasudevan}, \& {Trippe}}]{koss12}
{Koss}, M., {Mushotzky}, R., {Treister}, E., {et~al.} 2012, \apjl, 746, L22

\bibitem[{{Koss} {et~al.}(2018){Koss}, {Blecha}, {Bernhard}, {Hung}, {Lu}, {Trakhtenbrot}, {Treister}, {Weigel}, {Sartori}, {Mushotzky}, {Schawinski}, {Ricci}, {Veilleux}, \& {Sanders}}]{Koss2018}
{Koss}, M.~J., {Blecha}, L., {Bernhard}, P., {et~al.} 2018, \nat, 563, 214

\bibitem[{{Krone-Martins} {et~al.}(2019){Krone-Martins}, {Graham}, {Stern}, {Djorgovski}, {Delchambre}, {Ducourant}, {Teixeira}, {Drake}, {Scarano}, {Surdej}, {Galluccio}, {Jalan}, {Wertz}, {Kl{\"u}ter}, {Mignard}, {Spindola-Duarte}, {Dobie}, {Slezak}, {Sluse}, {Murphy}, {Boehm}, {Nierenberg}, {Bastian}, {Wambsganss}, \& {LeCampion}}]{KroneMartins19}
{Krone-Martins}, A., {Graham}, M.~J., {Stern}, D., {et~al.} 2019, arXiv e-prints, arXiv:1912.08977

\bibitem[{{Kuns{\'a}gi-M{\'a}t{\'e}} {et~al.}(2022){Kuns{\'a}gi-M{\'a}t{\'e}}, {Beck}, {Szapudi}, \& {Csabai}}]{wise_ps1}
{Kuns{\'a}gi-M{\'a}t{\'e}}, S., {Beck}, R., {Szapudi}, I., \& {Csabai}, I. 2022, \mnras, 516, 2662

\bibitem[{{Lamareille}(2010)}]{Lamareille10}
{Lamareille}, F. 2010, \aap, 509, A53

\bibitem[{{Lemon} {et~al.}(2024){Lemon}, {Courbin}, {More}, {Schechter}, {Ca{\~n}ameras}, {Delchambre}, {Leung}, {Shu}, {Spiniello}, {Hezaveh}, {Kl{\"u}ter}, \& {McMahon}}]{Lemon23}
{Lemon}, C., {Courbin}, F., {More}, A., {et~al.} 2024, \ssr, 220, 23

\bibitem[{{Lemon} {et~al.}(2017){Lemon}, {Auger}, {McMahon}, \& {Koposov}}]{Lemon17}
{Lemon}, C.~A., {Auger}, M.~W., {McMahon}, R.~G., \& {Koposov}, S.~E. 2017, \mnras, 472, 5023

\bibitem[{{Liu} {et~al.}(2011){Liu}, {Shen}, {Strauss}, \& {Hao}}]{Liu11_sdsspair}
{Liu}, X., {Shen}, Y., {Strauss}, M.~A., \& {Hao}, L. 2011, \apj, 737, 101

\bibitem[{{Lu} \& {Lin}(2019)}]{WeiJian19}
{Lu}, W.-J. \& {Lin}, Y.-R. 2019, \apj, 881, 105

\bibitem[{{Lyke} {et~al.}(2020){Lyke}, {Higley}, {McLane}, {Schurhammer}, {Myers}, {Ross}, {Dawson}, {Chabanier}, {Martini}, {Busca}, {Mas des Bourboux}, {Salvato}, {Streblyanska}, {Zarrouk}, {Burtin}, {Anderson}, {Bautista}, {Bizyaev}, {Brandt}, {Brinkmann}, {Brownstein}, {Comparat}, {Green}, {de la Macorra}, {Mu{\~n}oz Guti{\'e}rrez}, {Hou}, {Newman}, {Palanque-Delabrouille}, {P{\^a}ris}, {Percival}, {Petitjean}, {Rich}, {Rossi}, {Schneider}, {Smith}, {Vivek}, \& {Weaver}}]{Lyke20_SDSS}
{Lyke}, B.~W., {Higley}, A.~N., {McLane}, J.~N., {et~al.} 2020, \apjs, 250, 8

\bibitem[{{Maiolino} {et~al.}(2008){Maiolino}, {Nagao}, {Grazian}, {Cocchia}, {Marconi}, {Mannucci}, {Cimatti}, {Pipino}, {Ballero}, {Calura}, {Chiappini}, {Fontana}, {Granato}, {Matteucci}, {Pastorini}, {Pentericci}, {Risaliti}, {Salvati}, \& {Silva}}]{Maiolino08}
{Maiolino}, R., {Nagao}, T., {Grazian}, A., {et~al.} 2008, \aap, 488, 463

\bibitem[{{Maiolino} {et~al.}(2024){Maiolino}, {Scholtz}, {Curtis-Lake}, {Carniani}, {Baker}, {de Graaff}, {Tacchella}, {{\"U}bler}, {D'Eugenio}, {Witstok}, {Curti}, {Arribas}, {Bunker}, {Charlot}, {Chevallard}, {Eisenstein}, {Egami}, {Ji}, {Jones}, {Lyu}, {Rawle}, {Robertson}, {Rujopakarn}, {Perna}, {Sun}, {Venturi}, {Williams}, \& {Willott}}]{Maiolino2024dual}
{Maiolino}, R., {Scholtz}, J., {Curtis-Lake}, E., {et~al.} 2024, \aap, 691, A145

\bibitem[{{Mandelker} {et~al.}(2019){Mandelker}, {van den Bosch}, {Springel}, \& {van de Voort}}]{mandalker19}
{Mandelker}, N., {van den Bosch}, F.~C., {Springel}, V., \& {van de Voort}, F. 2019, \apjl, 881, L20

\bibitem[{{Mannucci} {et~al.}(2022){Mannucci}, {Pancino}, {Belfiore}, {Cicone}, {Ciurlo}, {Cresci}, {Lusso}, {Marasco}, {Marconi}, {Nardini}, {Pinna}, {Severgnini}, {Saracco}, {Tozzi}, \& {Yeh}}]{Mannucci22}
{Mannucci}, F., {Pancino}, E., {Belfiore}, F., {et~al.} 2022, Nature Astronomy, 6, 1185

\bibitem[{{Mannucci} {et~al.}(2023){Mannucci}, {Scialpi}, {Ciurlo}, {Yeh}, {Marconcini}, {Tozzi}, {Cresci}, {Marconi}, {Amiri}, {Belfiore}, {Carniani}, {Cicone}, {Nardini}, {Pancino}, {Rubinur}, {Severgnini}, {Ulivi}, {Venturi}, {Vignali}, {Volonteri}, {Pinna}, {Rossi}, {Puglisi}, {Agapito}, {Plantet}, {Ghose}, {Carbonaro}, {Xompero}, {Grani}, {Esposito}, {Power}, {Guerra Ramon}, {Lefebvre}, {Cavallaro}, {Davies}, {Riccardi}, {Macintosh}, {Taylor}, {Dolci}, {Baruffolo}, {Feuchtgruber}, {Kravchenko}, {Rau}, {Sturm}, {Wiezorrek}, {Dallilar}, \& {Kenworthy}}]{Mannucci23}
{Mannucci}, F., {Scialpi}, M., {Ciurlo}, A., {et~al.} 2023, \aap, 680, A53

\bibitem[{{Marconcini} {et~al.}(2023){Marconcini}, {Marconi}, {Cresci}, {Venturi}, {Ulivi}, {Mannucci}, {Belfiore}, {Tozzi}, {Ginolfi}, {Marasco}, {Carniani}, {Amiri}, {Di Teodoro}, {Scialpi}, {Tomicic}, {Mingozzi}, {Brazzini}, \& {Moreschini}}]{Marconcini23}
{Marconcini}, C., {Marconi}, A., {Cresci}, G., {et~al.} 2023, \aap, 677, A58

\bibitem[{{Martin} {et~al.}(2010){Martin}, {Scannapieco}, {Ellison}, {Hennawi}, {Djorgovski}, \& {Fournier}}]{Martin+10}
{Martin}, C.~L., {Scannapieco}, E., {Ellison}, S.~L., {et~al.} 2010, \apj, 721, 174

\bibitem[{{Massey} {et~al.}(2010){Massey}, {Kitching}, \& {Richard}}]{Massey10}
{Massey}, R., {Kitching}, T., \& {Richard}, J. 2010, Reports on Progress in Physics, 73, 086901

\bibitem[{{Matsuoka} {et~al.}(2024){Matsuoka}, {Izumi}, {Onoue}, {Strauss}, {Iwasawa}, {Kashikawa}, {Akiyama}, {Aoki}, {Arita}, {Imanishi}, {Ishimoto}, {Kawaguchi}, {Kohno}, {Lee}, {Nagao}, {Silverman}, \& {Toba}}]{Matsuoka2024}
{Matsuoka}, Y., {Izumi}, T., {Onoue}, M., {et~al.} 2024, \apjl, 965, L4

\bibitem[{{McMahon} {et~al.}(2013){McMahon}, {Banerji}, {Gonzalez}, {Koposov}, {Bejar}, {Lodieu}, {Rebolo}, \& {VHS Collaboration}}]{McMahon13}
{McMahon}, R.~G., {Banerji}, M., {Gonzalez}, E., {et~al.} 2013, The Messenger, 154, 35

\bibitem[{{Mediavilla} \& {Jim{\'e}nez-Vicente}(2021)}]{Mediavilla21}
{Mediavilla}, E. \& {Jim{\'e}nez-Vicente}, J. 2021, \apj, 914, 112

\bibitem[{{Misawa} {et~al.}(2007){Misawa}, {Eracleous}, {Charlton}, {Ganguly}, {Tytler}, {Kirkman}, {Suzuki}, \& {Lubin}}]{Misawa07}
{Misawa}, T., {Eracleous}, M., {Charlton}, J.~C., {et~al.} 2007, in Astronomical Society of the Pacific Conference Series, Vol. 373, The Central Engine of Active Galactic Nuclei, ed. L.~C. {Ho} \& J.~W. {Wang}, 291

\bibitem[{{Mosquera} {et~al.}(2013){Mosquera}, {Kochanek}, {Chen}, {Dai}, {Blackburne}, \& {Chartas}}]{mosquera2013}
{Mosquera}, A.~M., {Kochanek}, C.~S., {Chen}, B., {et~al.} 2013, \apj, 769, 53

\bibitem[{{Newman} {et~al.}(2017){Newman}, {Smith}, {Conroy}, {Villaume}, \& {van Dokkum}}]{Newman17}
{Newman}, A.~B., {Smith}, R.~J., {Conroy}, C., {Villaume}, A., \& {van Dokkum}, P. 2017, \apj, 845, 157

\bibitem[{{Osterbrock} \& {Ferland}(2006)}]{Osterbrock06}
{Osterbrock}, D.~E. \& {Ferland}, G.~J. 2006, {Astrophysics of gaseous nebulae and active galactic nuclei}

\bibitem[{{Ostrovski} {et~al.}(2017){Ostrovski}, {McMahon}, {Connolly}, {Lemon}, {Auger}, {Banerji}, {Hung}, {Koposov}, {Lidman}, {Reed}, {Allam}, {Benoit-L{\'e}vy}, {Bertin}, {Brooks}, {Buckley-Geer}, {Carnero Rosell}, {Carrasco Kind}, {Carretero}, {Cunha}, {da Costa}, {Desai}, {Diehl}, {Dietrich}, {Evrard}, {Finley}, {Flaugher}, {Fosalba}, {Frieman}, {Gerdes}, {Goldstein}, {Gruen}, {Gruendl}, {Gutierrez}, {Honscheid}, {James}, {Kuehn}, {Kuropatkin}, {Lima}, {Lin}, {Maia}, {Marshall}, {Martini}, {Melchior}, {Miquel}, {Ogando}, {Plazas Malag{\'o}n}, {Reil}, {Romer}, {Sanchez}, {Santiago}, {Scarpine}, {Sevilla-Noarbe}, {Soares-Santos}, {Sobreira}, {Suchyta}, {Tarle}, {Thomas}, {Tucker}, \& {Walker}}]{Ostrovski17}
{Ostrovski}, F., {McMahon}, R.~G., {Connolly}, A.~J., {et~al.} 2017, \mnras, 465, 4325

\bibitem[{{Perna} {et~al.}(2025){Perna}, {Arribas}, {Lamperti}, {Circosta}, {Bertola}, {P{\'e}rez-Gonz{\'a}lez}, {D'Eugenio}, {{\"U}bler}, {Cresci}, {Volonteri}, {Mannucci}, {Maiolino}, {Rodr{\'\i}guez Del Pino}, {B{\"o}ker}, {Bunker}, {Charlot}, {Willott}, {Carniani}, {Curti}, {Jones}, {Kumari}, {Marshall}, {Venturi}, {Saxena}, {Scholtz}, \& {Witstok}}]{Perna2025dual}
{Perna}, M., {Arribas}, S., {Lamperti}, I., {et~al.} 2025, \aap, 696, A59

\bibitem[{{Pfeifle} {et~al.}(2019){Pfeifle}, {Satyapal}, {Manzano-King}, {Cann}, {Sexton}, {Rothberg}, {Canalizo}, {Ricci}, {Blecha}, {Ellison}, {Gliozzi}, {Secrest}, {Constantin}, \& {Harvey}}]{Pfeifle19_triple}
{Pfeifle}, R.~W., {Satyapal}, S., {Manzano-King}, C., {et~al.} 2019, \apj, 883, 167

\bibitem[{{Pfeifle} {et~al.}(2023){Pfeifle}, {Weaver}, {Satyapal}, {Ricci}, {Secrest}, {Gliozzi}, {Blecha}, \& {Rothberg}}]{pfeifle23_dual}
{Pfeifle}, R.~W., {Weaver}, K., {Satyapal}, S., {et~al.} 2023, \apj, 954, 116

\bibitem[{{Pfeifle} {et~al.}(2025){Pfeifle}, {Weaver}, {Secrest}, {Rothberg}, \& {Patton}}]{Bigmac25}
{Pfeifle}, R.~W., {Weaver}, K.~A., {Secrest}, N.~J., {Rothberg}, B., \& {Patton}, D.~R. 2025, \apjs, 281, 25

\bibitem[{{Planck Collaboration} {et~al.}(2020){Planck Collaboration}, {Aghanim}, {Akrami}, {Ashdown}, {Aumont}, {Baccigalupi}, {Ballardini}, {Banday}, {Barreiro}, {Bartolo}, {Basak}, {Battye}, {Benabed}, {Bernard}, {Bersanelli}, {Bielewicz}, {Bock}, {Bond}, {Borrill}, {Bouchet}, {Boulanger}, {Bucher}, {Burigana}, {Butler}, {Calabrese}, {Cardoso}, {Carron}, {Challinor}, {Chiang}, {Chluba}, {Colombo}, {Combet}, {Contreras}, {Crill}, {Cuttaia}, {de Bernardis}, {de Zotti}, {Delabrouille}, {Delouis}, {Di Valentino}, {Diego}, {Dor{\'e}}, {Douspis}, {Ducout}, {Dupac}, {Dusini}, {Efstathiou}, {Elsner}, {En{\ss}lin}, {Eriksen}, {Fantaye}, {Farhang}, {Fergusson}, {Fernandez-Cobos}, {Finelli}, {Forastieri}, {Frailis}, {Fraisse}, {Franceschi}, {Frolov}, {Galeotta}, {Galli}, {Ganga}, {G{\'e}nova-Santos}, {Gerbino}, {Ghosh}, {Gonz{\'a}lez-Nuevo}, {G{\'o}rski}, {Gratton}, {Gruppuso}, {Gudmundsson}, {Hamann}, {Handley}, {Hansen}, {Herranz}, {Hildebrandt}, {Hivon}, {Huang}, {Jaffe}, {Jones}, {Karakci}, {Keih{\"a}nen},
  {Keskitalo}, {Kiiveri}, {Kim}, {Kisner}, {Knox}, {Krachmalnicoff}, {Kunz}, {Kurki-Suonio}, {Lagache}, {Lamarre}, {Lasenby}, {Lattanzi}, {Lawrence}, {Le Jeune}, {Lemos}, {Lesgourgues}, {Levrier}, {Lewis}, {Liguori}, {Lilje}, {Lilley}, {Lindholm}, {L{\'o}pez-Caniego}, {Lubin}, {Ma}, {Mac{\'\i}as-P{\'e}rez}, {Maggio}, {Maino}, {Mandolesi}, {Mangilli}, {Marcos-Caballero}, {Maris}, {Martin}, {Martinelli}, {Mart{\'\i}nez-Gonz{\'a}lez}, {Matarrese}, {Mauri}, {McEwen}, {Meinhold}, {Melchiorri}, {Mennella}, {Migliaccio}, {Millea}, {Mitra}, {Miville-Desch{\^e}nes}, {Molinari}, {Montier}, {Morgante}, {Moss}, {Natoli}, {N{\o}rgaard-Nielsen}, {Pagano}, {Paoletti}, {Partridge}, {Patanchon}, {Peiris}, {Perrotta}, {Pettorino}, {Piacentini}, {Polastri}, {Polenta}, {Puget}, {Rachen}, {Reinecke}, {Remazeilles}, {Renzi}, {Rocha}, {Rosset}, {Roudier}, {Rubi{\~n}o-Mart{\'\i}n}, {Ruiz-Granados}, {Salvati}, {Sandri}, {Savelainen}, {Scott}, {Shellard}, {Sirignano}, {Sirri}, {Spencer}, {Sunyaev}, {Suur-Uski}, {Tauber}, {Tavagnacco},
  {Tenti}, {Toffolatti}, {Tomasi}, {Trombetti}, {Valenziano}, {Valiviita}, {Van Tent}, {Vibert}, {Vielva}, {Villa}, {Vittorio}, {Wandelt}, {Wehus}, {White}, {White}, {Zacchei}, \& {Zonca}}]{Planck2020}
{Planck Collaboration}, {Aghanim}, N., {Akrami}, Y., {et~al.} 2020, \aap, 641, A6

\bibitem[{{Ricci} {et~al.}(2017){Ricci}, {Bauer}, {Treister}, {Schawinski}, {Privon}, {Blecha}, {Arevalo}, {Armus}, {Harrison}, {Ho}, {Iwasawa}, {Sanders}, \& {Stern}}]{Ricci17}
{Ricci}, C., {Bauer}, F.~E., {Treister}, E., {et~al.} 2017, \mnras, 468, 1273

\bibitem[{{Rosas-Guevara} {et~al.}(2019){Rosas-Guevara}, {Bower}, {McAlpine}, {Bonoli}, \& {Tissera}}]{Rosas-Guevara2019}
{Rosas-Guevara}, Y.~M., {Bower}, R.~G., {McAlpine}, S., {Bonoli}, S., \& {Tissera}, P.~B. 2019, \mnras, 483, 2712

\bibitem[{{Rubinur} {et~al.}(2019){Rubinur}, {Das}, \& {Kharb}}]{Rubinur19}
{Rubinur}, K., {Das}, M., \& {Kharb}, P. 2019, \mnras, 484, 4933

\bibitem[{{Salvato} {et~al.}(2025){Salvato}, {Wolf}, {Dwelly}, {Starck}, {Buchner}, {Shirley}, {Merloni}, {Georgakakis}, {Balzer}, {Brusa}, {Rau}, {Freund}, {Lang}, {Liu}, {Lamer}, {Schwope}, {Roster}, {Waddell}, {Scialpi}, {Igo}, {Kluge}, {Mannucci}, {Tiwari}, {Homan}, {Krumpe}, {Zenteno}, {Hernandez-Lang}, {Comparat}, {Fabricius}, {Snigula}, {Schlegel}, {Weaver}, {Zhou}, {Dey}, {Valdes}, {Myers}, {Juneau}, {Winkler}, {Marquez}, {di Mille}, {Ciroi}, {Schramm}, {Buckley}, {Brink}, {Gromadzki}, {Robrade}, \& {Nandra}}]{Salvato25}
{Salvato}, M., {Wolf}, J., {Dwelly}, T., {et~al.} 2025, \aap, 704, A344

\bibitem[{Satyapal {et~al.}(2017)Satyapal, Secrest, Ricci, Ellison, Rothberg, Blecha, Constantin, Gliozzi, McNulty, \& Ferguson}]{Satyapal17}
Satyapal, S., Secrest, N.~J., Ricci, C., {et~al.} 2017, The Astrophysical Journal, 848, 126

\bibitem[{{Schechter} {et~al.}(2017){Schechter}, {Morgan}, {Chehade}, {Metcalfe}, {Shanks}, \& {McDonald}}]{Scheckter17}
{Schechter}, P.~L., {Morgan}, N.~D., {Chehade}, B., {et~al.} 2017, \aj, 153, 219

\bibitem[{{Schmidt} {et~al.}(2023){Schmidt}, {Treu}, {Birrer}, {Shajib}, {Lemon}, {Millon}, {Sluse}, {Agnello}, {Anguita}, {Auger-Williams}, {McMahon}, {Motta}, {Schechter}, {Spiniello}, {Kayo}, {Courbin}, {Ertl}, {Fassnacht}, {Frieman}, {More}, {Schuldt}, {Suyu}, {Aguena}, {Andrade-Oliveira}, {Annis}, {Bacon}, {Bertin}, {Brooks}, {Burke}, {Carnero Rosell}, {Carrasco Kind}, {Carretero}, {Conselice}, {Costanzi}, {da Costa}, {Pereira}, {De Vicente}, {Desai}, {Doel}, {Everett}, {Ferrero}, {Friedel}, {Garc{\'\i}a-Bellido}, {Gaztanaga}, {Gruen}, {Gruendl}, {Gschwend}, {Gutierrez}, {Hinton}, {Hollowood}, {Honscheid}, {James}, {Kuehn}, {Lahav}, {Menanteau}, {Miquel}, {Palmese}, {Paz-Chinch{\'o}n}, {Pieres}, {Plazas Malag{\'o}n}, {Prat}, {Rodriguez-Monroy}, {Romer}, {Sanchez}, {Scarpine}, {Sevilla-Noarbe}, {Smith}, {Suchyta}, {Tarle}, {To}, {Varga}, \& {DES Collaboration}}]{Schmidt23}
{Schmidt}, T., {Treu}, T., {Birrer}, S., {et~al.} 2023, \mnras, 518, 1260

\bibitem[{{Schwartzman} {et~al.}(2024){Schwartzman}, {Clarke}, {Nyland}, {Secrest}, {Pfeifle}, {Schmitt}, {Satyapal}, \& {Rothberg}}]{Schwartzman24_vastrometry}
{Schwartzman}, E., {Clarke}, T.~E., {Nyland}, K., {et~al.} 2024, \apj, 961, 233

\bibitem[{{Schwartzman} {et~al.}(2025){Schwartzman}, {Fudolig}, {Clarke}, {Nyland}, {Secrest}, {Pfeifle}, {Schmitt}, {Satyapal}, \& {Rothberg}}]{Schwartzman25_vastrometry}
{Schwartzman}, E., {Fudolig}, P., {Clarke}, T.~E., {et~al.} 2025, \apj, 987, 200

\bibitem[{{Scialpi} {et~al.}(2024){Scialpi}, {Mannucci}, {Marconcini}, {Venturi}, {Pancino}, {Marconi}, {Cresci}, {Belfiore}, {Amiri}, {Bertola}, {Carniani}, {Cicone}, {Ciurlo}, {D'Amato}, {Ginolfi}, {Lusso}, {Marasco}, {Nardini}, {Rubinur}, {Severgnini}, {Tozzi}, {Ulivi}, {Vignali}, \& {Volonteri}}]{Scialpi24}
{Scialpi}, M., {Mannucci}, F., {Marconcini}, C., {et~al.} 2024, \aap, 690, A57

\bibitem[{Sedda {et~al.}(2023)Sedda, Naoz, \& Kocsis}]{Sedda23}
Sedda, M.~A., Naoz, S., \& Kocsis, B. 2023, Universe, 9, 138

\bibitem[{{Shajib} {et~al.}(2024){Shajib}, {Vernardos}, {Collett}, {Motta}, {Sluse}, {Williams}, {Saha}, {Birrer}, {Spiniello}, \& {Treu}}]{Shajib24}
{Shajib}, A.~J., {Vernardos}, G., {Collett}, T.~E., {et~al.} 2024, \ssr, 220, 87

\bibitem[{{Shen} {et~al.}(2023{\natexlab{a}}){Shen}, {Hwang}, {Oguri}, {Chen}, {Di Matteo}, {Ni}, {Bird}, {Zakamska}, {Liu}, {Chen}, \& {Kratter}}]{Shen23b}
{Shen}, Y., {Hwang}, H.-C., {Oguri}, M., {et~al.} 2023{\natexlab{a}}, \apj, 943, 38

\bibitem[{{Shen} {et~al.}(2023{\natexlab{b}}){Shen}, {Hwang}, {Oguri}, {Chen}, {Di Matteo}, {Ni}, {Bird}, {Zakamska}, {Liu}, {Chen}, \& {Kratter}}]{Shen23_deconv}
{Shen}, Y., {Hwang}, H.-C., {Oguri}, M., {et~al.} 2023{\natexlab{b}}, \apj, 943, 38

\bibitem[{{Shen} {et~al.}(2019){Shen}, {Hwang}, {Zakamska}, \& {Liu}}]{Shen19}
{Shen}, Y., {Hwang}, H.-C., {Zakamska}, N., \& {Liu}, X. 2019, \apjl, 885, L4

\bibitem[{{Skrutskie} {et~al.}(2006){Skrutskie}, {Cutri}, {Stiening}, {Weinberg}, {Schneider}, {Carpenter}, {Beichman}, {Capps}, {Chester}, {Elias}, {Huchra}, {Liebert}, {Lonsdale}, {Monet}, {Price}, {Seitzer}, {Jarrett}, {Kirkpatrick}, {Gizis}, {Howard}, {Evans}, {Fowler}, {Fullmer}, {Hurt}, {Light}, {Kopan}, {Marsh}, {McCallon}, {Tam}, {Van Dyk}, \& {Wheelock}}]{Skrutskie06}
{Skrutskie}, M.~F., {Cutri}, R.~M., {Stiening}, R., {et~al.} 2006, \aj, 131, 1163

\bibitem[{{Sluse} {et~al.}(2007){Sluse}, {Claeskens}, {Hutsemekers}, \& {Surdej}}]{Sluse07}
{Sluse}, D., {Claeskens}, J.~F., {Hutsemekers}, D., \& {Surdej}, J. 2007, \aap, 468, 885

\bibitem[{{Sluse} {et~al.}(2012){Sluse}, {Hutsem{\'e}kers}, {Courbin}, {Meylan}, \& {Wambsganss}}]{Sluse12}
{Sluse}, D., {Hutsem{\'e}kers}, D., {Courbin}, F., {Meylan}, G., \& {Wambsganss}, J. 2012, \aap, 544, A62

\bibitem[{{Sonnenfeld} {et~al.}(2019){Sonnenfeld}, {Jaelani}, {Chan}, {More}, {Suyu}, {Wong}, {Oguri}, \& {Lee}}]{Sonnenfeld19}
{Sonnenfeld}, A., {Jaelani}, A.~T., {Chan}, J., {et~al.} 2019, \aap, 630, A71

\bibitem[{{Sonnenfeld} {et~al.}(2013){Sonnenfeld}, {Treu}, {Gavazzi}, {Suyu}, {Marshall}, {Auger}, \& {Nipoti}}]{Sonnenfeld13}
{Sonnenfeld}, A., {Treu}, T., {Gavazzi}, R., {et~al.} 2013, \apj, 777, 98

\bibitem[{{Sonnenfeld} {et~al.}(2015){Sonnenfeld}, {Treu}, {Marshall}, {Suyu}, {Gavazzi}, {Auger}, \& {Nipoti}}]{Sonnenfeld15}
{Sonnenfeld}, A., {Treu}, T., {Marshall}, P.~J., {et~al.} 2015, \apj, 800, 94

\bibitem[{{Storey-Fisher} {et~al.}(2024){Storey-Fisher}, {Hogg}, {Rix}, {Eilers}, {Fabbian}, {Blanton}, \& {Alonso}}]{quaia}
{Storey-Fisher}, K., {Hogg}, D.~W., {Rix}, H.-W., {et~al.} 2024, \apj, 964, 69

\bibitem[{{Str{\"o}bele} {et~al.}(2012){Str{\"o}bele}, {La Penna}, {Arsenault}, {Conzelmann}, {Delabre}, {Duchateau}, {Dorn}, {Fedrigo}, {Hubin}, {Quentin}, {Jolley}, {Kiekebusch}, {Kirchbauer}, {Klein}, {Kolb}, {Kuntschner}, {Le Louarn}, {Lizon}, {Madec}, {Pettazzi}, {Soenke}, {Tordo}, {Vernet}, \& {Muradore}}]{museAO_strobele}
{Str{\"o}bele}, S., {La Penna}, P., {Arsenault}, R., {et~al.} 2012, in Society of Photo-Optical Instrumentation Engineers (SPIE) Conference Series, Vol. 8447, Adaptive Optics Systems III, ed. B.~L. {Ellerbroek}, E.~{Marchetti}, \& J.-P. {V{\'e}ran}, 844737

\bibitem[{{Temple} {et~al.}(2021){Temple}, {Hewett}, \& {Banerji}}]{Temple21}
{Temple}, M.~J., {Hewett}, P.~C., \& {Banerji}, M. 2021, \mnras, 508, 737

\bibitem[{{Treu}(2010)}]{Treu10b}
{Treu}, T. 2010, \araa, 48, 87

\bibitem[{{Treu} {et~al.}(2010){Treu}, {Auger}, {Koopmans}, {Gavazzi}, {Marshall}, \& {Bolton}}]{Treu10a}
{Treu}, T., {Auger}, M.~W., {Koopmans}, L. V.~E., {et~al.} 2010, \apj, 709, 1195

\bibitem[{{Treu} \& {Ellis}(2015)}]{treu15}
{Treu}, T. \& {Ellis}, R.~S. 2015, Contemporary Physics, 56, 17

\bibitem[{{{\"U}bler} {et~al.}(2024){{\"U}bler}, {Maiolino}, {P{\'e}rez-Gonz{\'a}lez}, {D'Eugenio}, {Perna}, {Curti}, {Arribas}, {Bunker}, {Carniani}, {Charlot}, {Rodr{\'\i}guez Del Pino}, {Baker}, {B{\"o}ker}, {Cresci}, {Dunlop}, {Grogin}, {Jones}, {Kumari}, {Lamperti}, {Laporte}, {Marshall}, {Mazzolari}, {Parlanti}, {Rawle}, {Scholtz}, {Venturi}, \& {Witstok}}]{Ubler2024dual}
{{\"U}bler}, H., {Maiolino}, R., {P{\'e}rez-Gonz{\'a}lez}, P.~G., {et~al.} 2024, \mnras, 531, 355

\bibitem[{{Ulivi} {et~al.}(2025){Ulivi}, {Mannucci}, {Scialpi}, {Marconcini}, {Cresci}, {Marconi}, {Feltre}, \& {Ginolfi}}]{Ulivi2025arXiv}
{Ulivi}, L., {Mannucci}, F., {Scialpi}, M., {et~al.} 2025, arXiv e-prints, arXiv:2508.19494

\bibitem[{{Vanden Berk} {et~al.}(2001){Vanden Berk}, {Richards}, {Bauer}, {Strauss}, {Schneider}, {Heckman}, {York}, {Hall}, {Fan}, {Knapp}, {Anderson}, {Annis}, {Bahcall}, {Bernardi}, {Briggs}, {Brinkmann}, {Brunner}, {Burles}, {Carey}, {Castander}, {Connolly}, {Crocker}, {Csabai}, {Doi}, {Finkbeiner}, {Friedman}, {Frieman}, {Fukugita}, {Gunn}, {Hennessy}, {Ivezi{\'c}}, {Kent}, {Kunszt}, {Lamb}, {Leger}, {Long}, {Loveday}, {Lupton}, {Meiksin}, {Merelli}, {Munn}, {Newberg}, {Newcomb}, {Nichol}, {Owen}, {Pier}, {Pope}, {Rockosi}, {Schlegel}, {Siegmund}, {Smee}, {Snir}, {Stoughton}, {Stubbs}, {SubbaRao}, {Szalay}, {Szokoly}, {Tremonti}, {Uomoto}, {Waddell}, {Yanny}, \& {Zheng}}]{Vandenberk01}
{Vanden Berk}, D.~E., {Richards}, G.~T., {Bauer}, A., {et~al.} 2001, \aj, 122, 549

\bibitem[{{Volonteri} {et~al.}(2003){Volonteri}, {Haardt}, \& {Madau}}]{volonteri03}
{Volonteri}, M., {Haardt}, F., \& {Madau}, P. 2003, \apj, 582, 559

\bibitem[{{Volonteri} {et~al.}(2022){Volonteri}, {Pfister}, {Beckmann}, {Dotti}, {Dubois}, {Massonneau}, {Musoke}, \& {Tremmel}}]{Volonteri22}
{Volonteri}, M., {Pfister}, H., {Beckmann}, R., {et~al.} 2022, \mnras, 514, 640

\bibitem[{{Wambsganss}(2006)}]{Wambsganss06}
{Wambsganss}, J. 2006, arXiv e-prints, 0604278

\bibitem[{{Weilbacher} {et~al.}(2020){Weilbacher}, {Palsa}, {Streicher}, {Bacon}, {Urrutia}, {Wisotzki}, {Conseil}, {Husemann}, {Jarno}, {Kelz}, {P{\'e}contal-Rousset}, {Richard}, {Roth}, {Selman}, \& {Vernet}}]{Weilbacher20}
{Weilbacher}, P.~M., {Palsa}, R., {Streicher}, O., {et~al.} 2020, \aap, 641, A28

\bibitem[{{White} \& {Frenk}(1991)}]{White91}
{White}, S. D.~M. \& {Frenk}, C.~S. 1991, \apj, 379, 52

\bibitem[{{White} \& {Rees}(1978)}]{White78}
{White}, S.~D.~M. \& {Rees}, M.~J. 1978, \mnras, 183, 341

\bibitem[{{Wong} {et~al.}(2020){Wong}, {Suyu}, {Chen}, {Rusu}, {Millon}, {Sluse}, {Bonvin}, {Fassnacht}, {Taubenberger}, {Auger}, {Birrer}, {Chan}, {Courbin}, {Hilbert}, {Tihhonova}, {Treu}, {Agnello}, {Ding}, {Jee}, {Komatsu}, {Shajib}, {Sonnenfeld}, {Blandford}, {Koopmans}, {Marshall}, \& {Meylan}}]{Wong20}
{Wong}, K.~C., {Suyu}, S.~H., {Chen}, G. C.~F., {et~al.} 2020, \mnras, 498, 1420

\bibitem[{{Yan} {et~al.}(2019){Yan}, {Chen}, {Lazarz}, {Bizyaev}, {Maraston}, {Stringfellow}, {McCarthy}, {Meneses-Goytia}, {Law}, {Thomas}, {Falcon Barroso}, {S{\'a}nchez-Gallego}, {Schlafly}, {Zheng}, {Argudo-Fern{\'a}ndez}, {Beaton}, {Beers}, {Bershady}, {Blanton}, {Brownstein}, {Bundy}, {Chambers}, {Cherinka}, {De Lee}, {Drory}, {Galbany}, {Holtzman}, {Imig}, {Kaiser}, {Kinemuchi}, {Liu}, {Luo}, {Magnier}, {Majewski}, {Nair}, {Oravetz}, {Oravetz}, {Pan}, {Sobeck}, {Stassun}, {Talbot}, {Tremonti}, {Waters}, {Weijmans}, {Wilhelm}, {Zasowski}, {Zhao}, \& {Zhao}}]{Yan19}
{Yan}, R., {Chen}, Y., {Lazarz}, D., {et~al.} 2019, \apj, 883, 175

\bibitem[{{Zhou} {et~al.}(2004){Zhou}, {Wang}, {Zhang}, {Dong}, \& {Li}}]{Zhou2004}
{Zhou}, H., {Wang}, T., {Zhang}, X., {Dong}, X., \& {Li}, C. 2004, \apjl, 604, L33

\end{thebibliography}

\appendix
\clearpage
\onecolumn
\section{Spectra of dual and lensed AGN systems}

In this section we present the spatially resolved spectra of the full sample of systems composed of at least two AGN. The spectra are extracted from the circular apertures shown in Fig.~\ref{fig:cubes_AGN}.

\begin{figure}[h!]
    \centering

        \includegraphics[width=1\linewidth]{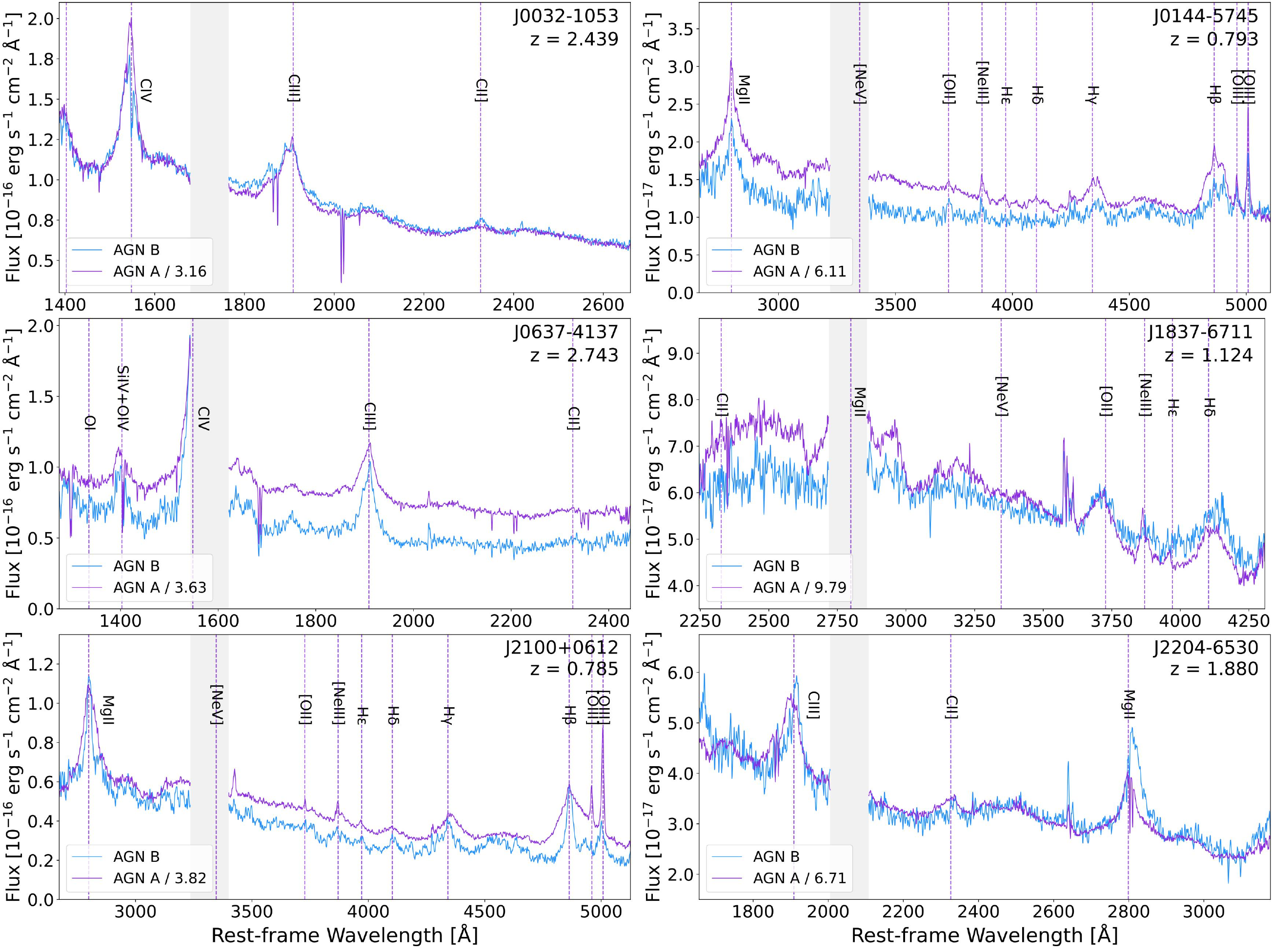}
    \caption{Final MUSE spectra of the dual AGN systems, with target names and redshifts indicated in each panel. The spectra were extracted from the circular apertures shown in Fig.~\ref{fig:cubes_AGN} (same color-coding), corrected for Galactic extinction, normalized to \textit{Gaia} magnitudes, and shifted to the rest frame. For clarity, the spectra of the primary AGN (component A) have been re-normalized to facilitate comparison with those of component B; the applied renormalization factor is indicated in each panel. Vertical dotted purple lines mark the expected positions of prominent emission lines at the redshift of AGN A. The gap around $\lambda_{\rm obs} = 6000~\AA$ (observed frame) corresponds to the New Generation Controllers (NGC) used in AO observations.}
    \label{fig:spectra_AGN}
\end{figure}

\begin{figure*}[h!]
\centering\includegraphics[width=1\linewidth]{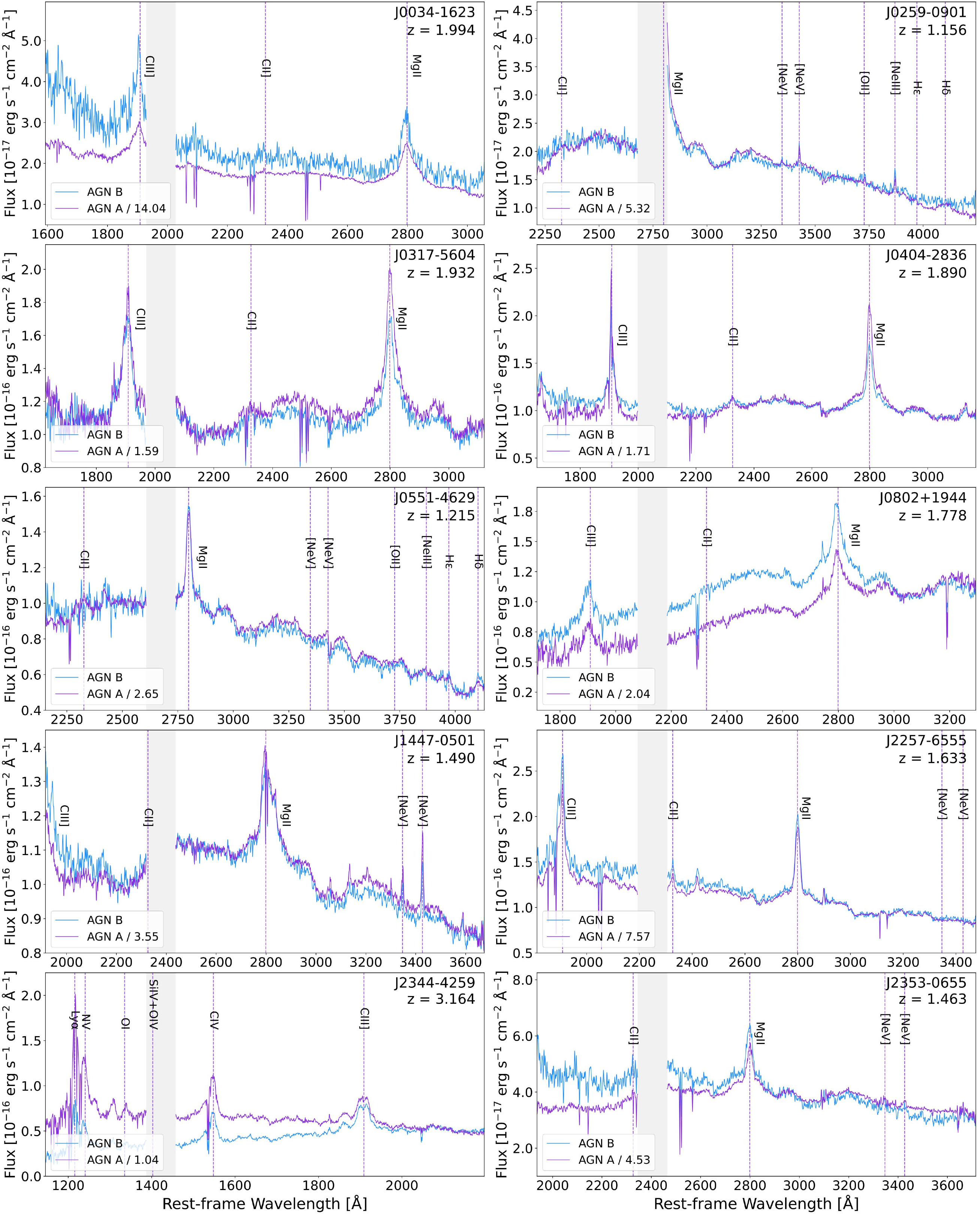}
        \vspace{-3mm}

    \caption{Final MUSE spectra of the double lensed AGN systems, with target names and redshifts indicated in each panel. The spectra were extracted from the circular apertures shown in the white-light images of Fig. \ref{fig:cubes_AGN} (same color-coding), corrected for Galactic extinction, normalized to \textit{Gaia} magnitudes, and shifted to the rest frame. Vertical dotted purple lines mark the expected positions of prominent emission lines at the redshift of AGN A. The gap around $\lambda_{\rm obs} = 6000~\AA$ corresponds to the NGC used in AO observations.}
    \label{fig:spectra_AGN_lensed}
\end{figure*}

Figure~\ref{fig:spectra_AGN} shows the spectra of the dual AGN systems, while Fig.~\ref{fig:spectra_AGN_lensed} presents the spectra of the doubly imaged lensed AGN systems. In both cases, the brightest AGN component is shown in purple, while the faintest component is shown in light blue.

\begin{figure*}[h!]
    \centering
    \includegraphics[width=1\linewidth]{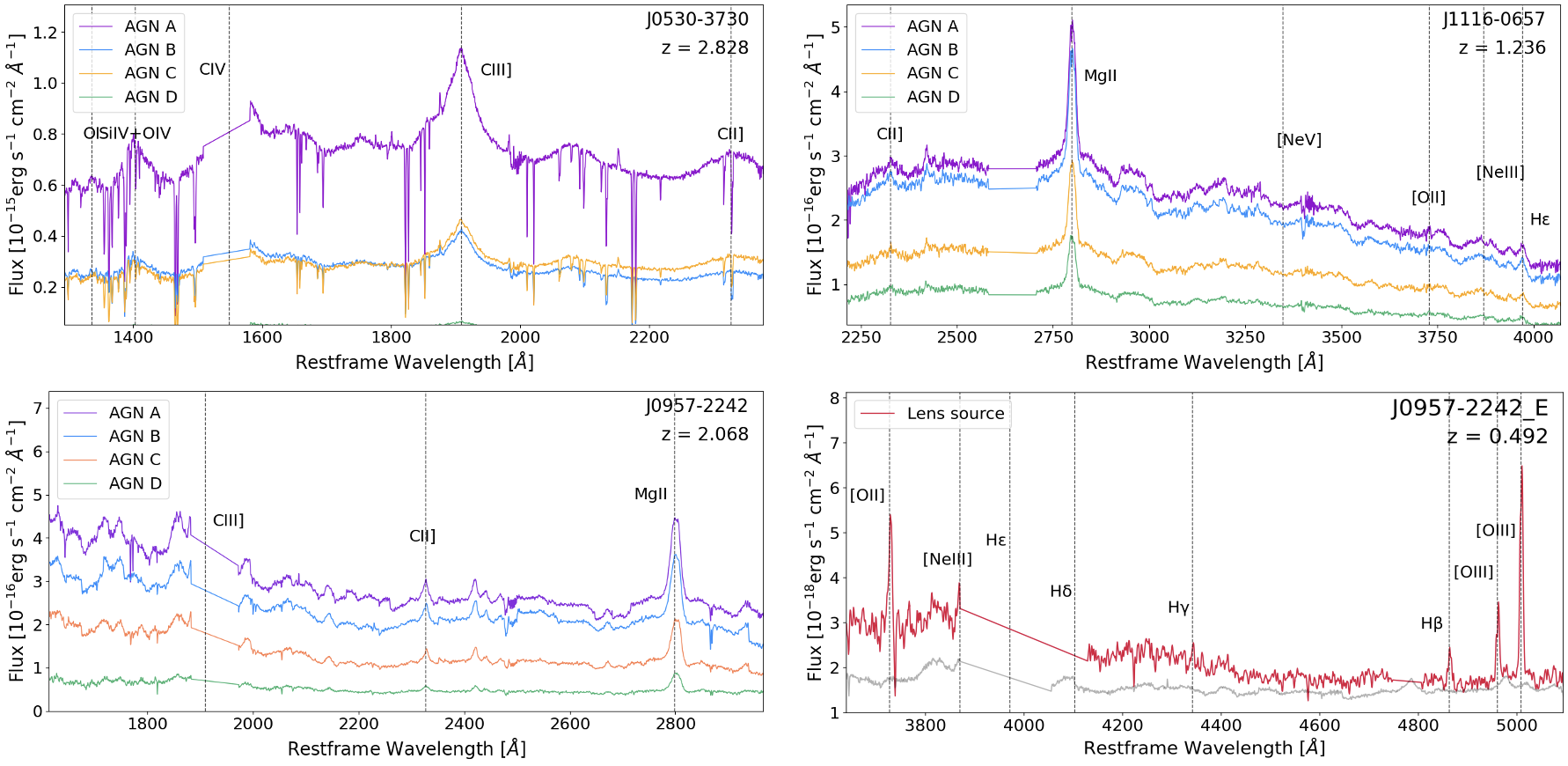}
    \caption{Final MUSE spectra for the three quadruply imaged AGN systems. 
    For each system, the four AGN images are shown in different colors, following the same color-coding adopted in Fig.~\ref{fig:cubes_quads}. 
    The top row shows the first two quadruple systems, while the bottom-left panel shows the four AGN spectra of J0957. 
    All spectra are corrected for Galactic extinction, normalized to \textit{Gaia} photometry, and shifted to the rest frame. 
    Vertical blue dashed lines mark the wavelengths of the main emission lines. 
    Since J0957 data also allowed us to identify the foreground lensing galaxy, its spectrum (component E) is shown separately in red in the bottom-right panel. 
    For reference, the AGN spectrum that contaminates the lens is shown in gray, while the emission lines of the lensing galaxy are indicated by black vertical dotted lines.}

    \label{fig:spectra_quads}
\end{figure*}

Finally, Fig.~\ref{fig:spectra_quads} shows the quadruply imaged lensed AGN systems. The third and fourth images are shown in orange and green, respectively. For the system J0957--2242, the spectrum of the foreground lensing galaxy is also shown in the bottom-right panel.

\vspace{0.5cm}

\clearpage
\newpage
\section{Lensed AGN distribution in our sample}
\label{app:lensed_dist}

To investigate the relative prevalence of lensed and dual AGN as a function of redshift, we define the fractions relative to the total number of AGN in each redshift bin:
\[
F_{\rm lens} = \frac{N_{\rm lensed}}{N_{\rm AGN}}, \qquad 
F_{\rm dual} = \frac{N_{\rm dual}}{N_{\rm AGN}},
\]
where $N_{\rm AGN} = N_{\rm lensed} + N_{\rm dual}$. 
These fractions describe the relative contributions within our observed sample and are useful to estimate the expected level of lensing "contamination" in dual-AGN studies. We stress that they should not be interpreted as intrinsic lens fractions suitable for lensing statistics, but only as indicators for selection purposes.

\begin{figure}[H]
    \centering
    \includegraphics[width=0.95\linewidth]{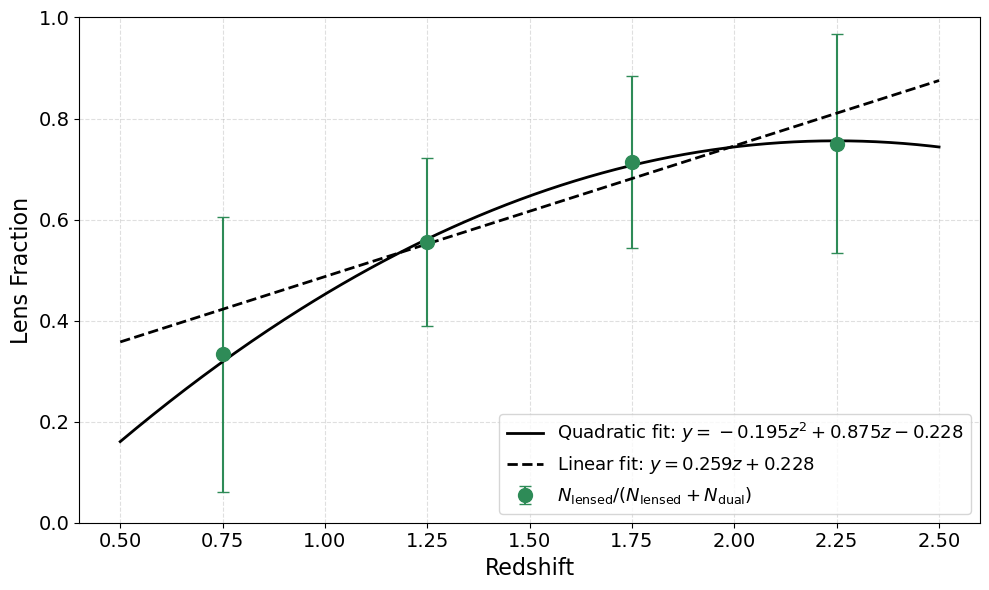}
    \caption{Fraction of lensed AGN within the lensed+dual sample, $F_{\rm lens} = N_{\rm lensed} / (N_{\rm lensed} + N_{\rm dual})$, as a function of redshift $z$. Data are binned in intervals of $\Delta z = 0.5$. Error bars represent $1\sigma$ binomial uncertainties. The solid black line shows the weighted quadratic fit, while the dashed line indicates the linear fit.}
    \label{fig:lens_fraction}
\end{figure}

Figure~\ref{fig:lens_fraction} shows $F_{\rm lens}$ in broad redshift bins ($\Delta z = 0.5$). We fit the redshift dependence of $F_{\rm lens}$ using both a weighted quadratic relation and a linear model, in order to provide a simple empirical description of the observed trend. The quadratic model is
$F_{\rm lens} = a z^2 + b z + c$,
with best-fit coefficients $a = -0.195$, $b = 0.875$, and $c = -0.229$, obtained using inverse variance weighting ($\sigma^{-2}$) to account for larger statistical uncertainties at high and low redshifts. For comparison, we also derive a linear fit of the form
$F_{\rm lens} = m z + q$,
with $m = 0.259$ and $q = 0.228$.The quadratic relation provides a better description of the data than the linear model in terms of $\chi^2$, but both parameterizations are shown to guide the eye and to facilitate comparison with future samples. We note that performing this fit is useful for future work, as it allowed us to estimate the expected dual-AGN fraction among GMP-selected systems based on the measured lensing fraction, providing a reference for selection effects in our sample. 
The resulting quadratic relation, $F_{\rm lens} = -0.195\, z^2 + 0.875\, z - 0.229$,
suggests a broad maximum in the lensing fraction at intermediate redshift, corresponding to the redshift range where survey selection effects and the quasar luminosity function are most favorably aligned.\\

\section{AGN+star system}
\label{app:star}
In this appendix we present the spectra of systems identified as fortuitous alignments of an AGN and a star along the same line of sight. We include the 11 systems observed in this Large Program, as well as an additional system, J1649+0812 (16:49:41.30 +08:12:33.5) with a projected separation of 0.59$''$, observed in the MUSE program (ID: 109.22W5, PI: Mannucci) and previously classified by \cite{Chen25_vastrometry}.

\begin{figure*}[]
\centering\includegraphics[width=0.84\linewidth]{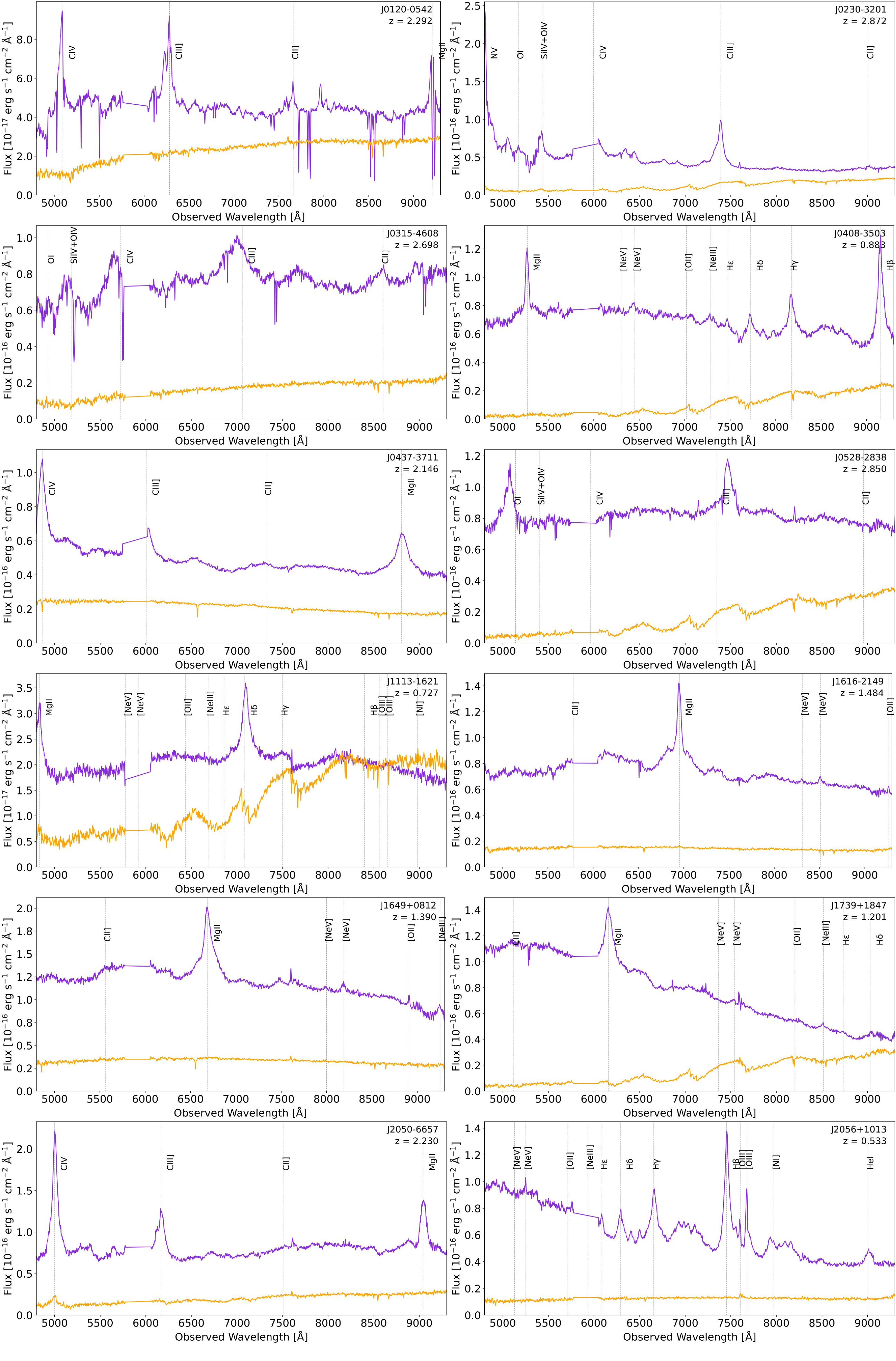}
        \vspace{-3mm}
    \caption{Final MUSE spectra of the AGN (purple) and star (orange) systems, with target names and redshifts indicated in each panel. The spectra were extracted from circular apertures of 5~px, corrected for Galactic extinction, normalized to \textit{Gaia} magnitudes, and shifted to the rest frame. Vertical dotted lines indicate the positions of the expected AGN emission lines. The gap around 6000~$\AA$ corresponds to the NGC used in AO observations.}

    \label{fig:AGN_STAR}
\end{figure*}

\end{document}